\documentclass[preprint,secnumarabic,amssymb, nobibnotes, aip, pop, superscriptaddress]{revtex4-2}
\usepackage{tikz}
\usepackage{upgreek}
\usepackage{amsmath} 
\usepackage{amsthm} 
\usepackage{amssymb}    
\usepackage{isomath}
\usepackage{graphicx} 
\usepackage[dvips,letterpaper,margin=1in,bottom=1in]{geometry}
\usepackage{tensor}
\usepackage{cancel}
\usepackage{color}
\usepackage{float}
\usepackage{blindtext}
\usepackage{enumerate}
\usepackage{subcaption}
\usepackage{chngcntr}
\usepackage[hidelinks]{hyperref}
\usepackage{cleveref}
\usepackage{fancyhdr}
\usepackage{lastpage}
\usepackage[utf8]{inputenc}
\usepackage[T1]{fontenc}
\usepackage{mathptmx}
\usepackage{etoolbox}

\usepackage{enumerate}
\renewcommand{\v}[1]{\ensuremath{\mathbf{#1}}} 
\newcommand{\gv}[1]{\ensuremath{\mbox{\boldmath$ #1 $}}} 
\newcommand{\abs}[1]{\left| #1 \right|} 
\newcommand{\avg}[1]{\left< #1 \right>} 
\renewcommand{\d}[2]{\frac{\text{d} #1}{\text{d} #2}} 
\newcommand{\pd}[2]{\frac{\partial #1}{\partial #2}} 
\newcommand{\pdd}[2]{\frac{\partial^2 #1}{\partial #2^2}} 

\newcommand{\grad}[1]{\nabla #1} 
\renewcommand{\div}[1]{\nabla\cdot \v{#1}} 
\newcommand{\curl}[1]{\nabla \times \v{#1}} 
\let\baraccent=\= 
\renewcommand{\=}[1]{\stackrel{#1}{=}} 

\providecommand{\fr}{\frac}

\newcommand{\vs}[1]{\vectorsym{#1}}

\usepackage{cancel}

\theoremstyle{definition}

\newcommand{\abpsi}{\abs{\grad\psi}}

\newcommand{\bhat}{\vs{\hat{b}}}
\newcommand{\nhat}{\vs{\hat{n}}}

\newcommand{\ibar}{{\mbox{$\,\iota\!\!$-}}}

\begin{document}
	\title{Suppressing Trapped-Electron-Mode-Driven Turbulence via Optimization of Three-Dimensional Shaping}
	\author{J.~M.~Duff}
	\email{jduff2@wisc.edu.}
	\affiliation{University of Wisconsin-Madison, Madison WI, 53706, USA}
	\author{B.~J.~Faber}
	\affiliation{University of Wisconsin-Madison, Madison WI, 53706, USA}
	\author{C.~C.~Hegna}
	\affiliation{University of Wisconsin-Madison, Madison WI, 53706, USA}
	\author{M.~J.~Pueschel}
	\affiliation{Dutch Institute for Fundamental Energy Research, 5612 AJ Eindhoven, The Netherlands}
	\affiliation{Eindhoven University of Technology, 5600 MB Eindhoven, The Netherlands}
	\author{P.~W.~Terry}
	\affiliation{University of Wisconsin-Madison, Madison WI, 53706, USA}
	\begin{abstract}
		Turbulent transport driven by trapped electron modes (TEMs) is believed to drive significant heat and particle transport in quasihelically symmetric stellarators. Two three-dimensionally-shaped magnetic configurations with suppressed trapped-electron-mode (TEM)-driven turbulence were generated through optimization that targeted quasihelical symmetry and the available energy of trapped electrons. Initial equilibria have flux surface shapes with a helically rotating negative triangularity (NT) and positive triangularity (PT). In gyrokinetic simulations, TEMs are suppressed in the reduced-TEM NT and PT configurations, showing that negative triangularity does not have the same beneficial turbulence properties over positive triangularity as seen in tokamaks. Heat fluxes from TEMs are also suppressed. Without temperature gradients and with a strong density gradient, the most unstable modes at low $k_y$ were consistent with toroidal universal instabilities (UIs) in the NT case and slab UIs in the PT case. Nonlinear simulations show that UIs drive substantial heat flux in both the NT and PT configurations. A moderate increase in $\beta$ halves the heat flux in the NT configuration, while suppressing the heat flux in the PT geometry. Based on the present work, future optimizations aimed at reducing electrostatic drift wave-driven turbulent transport will need to consider UIs if $\beta$ is sufficiently small.
	\end{abstract}

	\maketitle
	
	\section{Introduction}
	\label{sec:intro}
	Interest in optimizing stellarators to reduce turbulent transport has grown significantly in recent years. Many turbulence optimization efforts in stellarators have been focused on reducing turbulence driven by ion temperature gradient (ITG) modes\cite{rudakov_instability_1961,horton_toroidal_1981,rewoldt_comparison_2005,xanthopoulos_gyrokinetic_2007,xanthopoulos_nonlinear_2007,baumgaertel_simulating_2011,baumgaertel_gyrokinetic_2012,xanthopoulos_intrinsic_2016,plunk_distinct_2017,mckinney_comparison_2019}, which has been shown to have a substantial impact on ion temperature profiles in Wendelstein 7-X\cite{beidler_physics_1990,klinger_towards_2013,beurskens_ion_2021}. The importance of ITGs in stellarators for plasma confinement has been stressed and is a major consideration in future stellarator designs\cite{mckinney_comparison_2019,hegna_improving_2022}. Studies identified local shear and curvature as key quantities for ITG stability in stellarators\cite{mynick_geometry_2009} then used these quantities in a mixing-length estimate\cite{pueschel_stellarator_2016} for ITG-driven turbulent transport as an objective function to find stellarator equilibria with improved ITG turbulence properties\cite{mynick_optimizing_2010,mynick_reducing_2011,mynick_turbulent_2014,xanthopoulos_intrinsic_2016}. More recently, optimization based on the direct evaluation of the nonlinear heat flux from ITG turbulence with adiabatic electrons in a gyrokinetic model has been performed\cite{kim_optimization_2024}. 
	
	Turbulence driven by trapped electron modes\cite{kadomtsev_trapped_1971} (TEMs) is also considered to be a significant source of heat and particle loss in quasisymmetric stellarators\cite{canik_experimental_2007,faber_gyrokinetic_2015,hegna_improving_2022}. As a result, gyrokinetic TEM studies have been a major part of stellarator turbulence research\cite{baumgaertel_gyrokinetic_2012,proll_resilience_2012,proll_gyrokinetic_2012,helander_collisionless_2013,proll_collisionless_2013,proll_turbulence_2022}. Using the trapped particle fraction as a proxy for TEM stability, it has been possible to reduce linear growth rates and nonlinear heat fluxes in a Helically Symmetric eXperiment (HSX)-like configuration\cite{anderson_helically_1995,proll_tem_2016}. TEM stability has also been improved by examining a large database of perturbed HSX equilibria generated by changing currents in the shaping coils of HSX, finding that good quasihelical symmetry and co-rotating flux surface elongation are also associated with improved TEM stability \cite{gerard_optimizing_2023,gerard_effect_2024}. Optimizing for improved TEM stability and turbulence properties is a focus of the present work.
	
	However, even if TEMs are stabilized in low-shear stellarators, the universal instability (UI) can remain unstable and set the turbulence levels\cite{costello_universal_2023}. It has been suggested that UIs do not play a large role in driving turbulence in stellarators, because UIs tend to be subdominant to TEMs\cite{chowdhury_toroidal_2010,helander_universal_2015}. However, interest in UIs has grown recently because they can be the dominant instability in sheared-slab geometries\cite{landreman_universal_2015,brchnelova_density-gradient-driven_2024} and in low-shear stellarators where TEMs are stabilized\cite{costello_universal_2023}. In addition to the slab branch, a toroidal branch of UIs exists that is destabilized by the curvature and gradient of the magnetic field\cite{cheng_unstable_1980}. Methods for distinguishing UIs from TEMs involving cross-phases of the electrostatic potential with density fluctuations from the trapped and passing electron populations and artificially removing particle trapping and curvature drive in simulations to suppress TEMs have been employed in quasi-isodynamic and quasihelically symmetric equilibria\cite{costello_universal_2023}.
	
	One possible method of reducing UI-driven turbulence levels is by increasing plasma $\beta$, defined here as the ratio of electron kinetic to plasma pressure. The electromagnetic stabilization of UIs in slab-like geometry is attributed to coupling with Alfv\'enic perturbations at low $\beta$ ($\sim\mathcal{O}(m_\text{e}/m_\text{i})$, with $m_\text{i}$ and $m_\text{e}$ the ion and electron mass, respectively) and with the $\nabla B$ drift of the ions at $\beta\sim\mathcal{O}(10^{-2})$ \cite{mikhailovskaya_drift_1964,huba_finite-_1982,hastings_high-_1982}. However, increasing $\beta$ may lead to heat loss through electromagnetic channels by destabilizing electromagnetic modes or exite electromagnetic modes nonlinearly\cite{hatch_role_2011}. In the present work, it is shown that a $\beta$ can be chosen such that magnetic fluctuations significantly reduce electrostatic heat flux due to UIs without driving substantial electromagnetic heat flux in reduced-TEM configurations. 
	
	In the present work, two reduced-TEM equilibria are generated by optimizing for available energy of trapped electrons, whose stability and turbulence properties are then studied. First, details on the optimization techniques used to arrive at the reduced-TEM equilibria are given in Sec.~\ref{sec:opt}. Two scenarios are considered for each of the configurations: one with only density gradient drive and one with only electron temperature gradient drive. In Sec.~\ref{sec:linear}, a summary of the linear stability characteristics is given along with a procedure to identify which modes are the dominant instabilities. The detailed analysis and identification of linear modes is contained in Appendix~\ref{sec:lin_mode_appx}. Details on the properties of the turbulence are given in Sec.~\ref{sec:NL}. An examination of the stabilizing effects of plasma $\beta$ on UIs is performed in both reduced-TEM configurations in Sec.~\ref{sec:beta}, and the results of the paper are summarized in Sec.~\ref{sec:concl}.
	
	\section{Reduced-TEM Configurations}
	\label{sec:opt}
	Two local three-dimensional (3D) magnetohydrodynamic (MHD) equilibria\cite{hegna_local_2000,duff_effect_2022} are optimized, targeting quasihelical symmetry, the available energy of trapped electrons\cite{helander_available_2017,helander_available_2020,mackenbach_available_2022}, rotational transform, aspect ratio, global shear, poloidal curvature, distance to the cylindrical axis, and regularization of the gyrokinetic finite Larmor radius (FLR) terms. This local 3D equilibrium generates solutions to the MHD equilibrium equations near a single flux surface that satisfy the MHD force balance equation $\v{J} \times \v{B} =\nabla p$ and quasineutrality $\div{J} = 0$ consistently with Amp\`ere’s Law $\curl{B} = \mu_0\v{J}$, where $\v{B}$ is the magnetic field, $\v{J}$ is the current, $p$ the plasma pressure, and $\mu_0$ the vacuum permeability. Additionally, the MHD equilibrium conditions require the components of the magnetic field and current normal to the flux surface to vanish. Fully specifying a local MHD equilibrium requires the flux surface shape, a rotational transform $\ibar$, and two of three profile quantities on the flux surface: pressure gradient $p'=\text{d}p/\text{d}\psi$ ($\psi$ being the toroidal flux), flux-surface-averaged magnetic shear $\ibar'$, or flux-surface-averaged parallel current $\sigma=\mu_0\avg{\v{J}\cdot\v{B}/B^2}$, with $\avg{\cdot}$ denoting the flux-surface average.
	
	The magnetic coordinates satisfying the local 3D equilibrium $(\psi,\theta,\zeta)$ are used when describing local equilibrium quantities on the flux surface, and flux-tube coordinates $(x,y,z)$ are used to describe quantities along a field line, such that
	\begin{equation}
		\v{B} = \nabla\psi\times\nabla(\theta-\ibar\zeta)=\nabla\psi\times\nabla\alpha=\d{\psi}{x}\nabla x\times\nabla y,
	\end{equation}
	where $\theta$ is the poloidal straight-field-line angle, $\zeta=-\phi$ is the negative cylindrical toroidal angle, and $\alpha$ is the field-line label. The flux-tube coordinates are centered on a field line at $(\psi_0,\alpha_0,\theta=0)$ and are described by normalized radial, binormal, and parallel coordinates as
	\begin{align}
		&x = \fr{1}{B_0\rho_0}(\psi-\psi_0),\\
		&y = \rho_0(\alpha-\alpha_0),\\
		&z = \theta,
	\end{align}
	where $\rho_0$ is the minor radius of the magnetic surface at the center of the domain, and $B_0$ the on-axis magnetic field.
	
	\subsection{Optimization Setup}
	The optimization problem is set up to solve the unconstrained minimization problem
	\begin{equation}
		\min\limits_\v{x}f^2(\v{x}),
	\end{equation}
	where a local 3D equilibrium was varied to find satisfactory values for each term in the objective function $f$. The optimization variables include the Fourier coefficients $R_{m,n}$ and $Z_{m,n}$ of mode numbers $0\le m\le 3$ and $-3\le n\le 3$ of the inverse mapping of the flux surface such that
	\begin{align}
		&R(\theta,\zeta) = \sum\limits_m\sum\limits_nR_{m,n}\cos(m\theta+nN_\text{fp}\zeta),\\
		&Z(\theta,\zeta) = \sum\limits_m\sum\limits_nZ_{m,n}\sin(m\theta+nN_\text{fp}\zeta),
	\end{align}
	where $N_\text{fp}$ is the number of field periods. Additional variables used in the optimization are the rotational transform $\ibar$ and the flux-surface-averaged parallel current $\sigma$, which are necessary to specify the local 3D equilibrium. To fully specify the local equilibria, the pressure gradient is held fixed at $p'=0$. The squared objective function thus reads,
	\begin{align}
		f^2 =& w_\text{QS}f^2_\text{QS} + w_\text{TEM}f^2_\text{TEM} + w_{\hat{s}}(\hat{s}-\hat{s}_0)^2 + w_\ibar (\ibar-\ibar_0)^2 + w_\sigma (\sigma-\sigma_0)^2 + w_A (A-A_0)^2 \nonumber\\&+ w_{\kappa p}f^2_{\kappa p} + w_{R}f^2_R + w_{\text{FLR}}f^2_\text{FLR},
	\end{align}
	where the target values are denoted by the ``0" subscript and relative weights are constants given by $w$. The first two terms target quasisymmetry and the available energy, which will be discussed later in this section. Subsequent terms, in order, target the global shear $\hat{s}=-\text{d}(\ln\ibar)/\text{d}(\ln x)$, rotational transform, flux-surface-averaged parallel current, and flux-surface-averaged aspect ratio $A$ with a simple quadratic penalty function. These penalty functions were included to keep the respective quantities near values that would be considered reasonable for an experimental design. A term penalizing the distance of the flux surface to the cylindrical axis,
	\begin{equation}
		f_R = \avg{H(R_{\min}-R)(R_{\min}-R)^2},
	\end{equation}
	was included, where $R_\text{min}$ is the minimum major radius before penalization, and $H$ is the Heaviside function. The term penalizing the poloidal curvature of the flux surface,
	\begin{align}
		&f_{\kappa p} = \fr{1}{N_\zeta}\sum\limits_{i=1}^{N_\zeta}\fr{1}{L_i}\sqrt{\left(\pd{R_i}{\theta}\right)^2+\left(\pd{Z_i}{\theta}\right)^2}H(\kappa_{pi}-\kappa_{p0})(\kappa_{pi}-\kappa_{p0})^2,\\
		&\kappa_{pi} = \abs{R_i\left(\pd{R_i}{\theta}\pdd{Z_i}{\theta} - \pdd{R_i}{\theta}\pd{Z_i}{\theta}\right)}\left[\left(\pd{R_i}{\theta}\right)^2+\left(\pd{Z_i}{\theta}\right)^2\right]^{-3/2},
	\end{align}
	 is inspired by coil optimization\cite{kruger_constrained_2021}, where $N_\zeta$ is the number of points used to discretize a single field period of the flux surface, and $L_i$ is the length of the boundary curve enclosing the flux surface at toroidal angle $\zeta_i$. The index $i$ indicates the terms $R$ and $Z$, and their poloidal derivatives, are evaluated at the angle $\zeta_i$. The term $f_{\kappa p}$ penalizes sharp features in the flux surface shape that would be difficult to realize in an experimental design. A term to regularize the perpendicular wavenumber arguments of the FLR terms in the gyrokinetic Vlasov-Maxwell equations\cite{brizard_foundations_2007},
	\begin{equation}
		\label{eq:FLR_pen}
		f_\text{FLR} = H\left(\max\limits_z\left(\fr{g^{yy}}{g^{xx}}\right)-g_0\right)\left(\max\limits_z\left(\fr{g^{yy}}{g^{xx}}\right)-g_0\right),
	\end{equation}
	was also used, where $g^{ij}=\nabla u^i\cdot \nabla u^j$ are the contravariant metric coefficients for coordinates $u^i$, $u^j\in\{x,y,z\}$. This ensures an equilibrium that, in gyrokinetic simulations, can be resolved at practically achievable numerical expense. If $g^{yy}\gg g^{xx}$, then the FLR terms in the gyrokinetic equations will not effectively damp modes at high radial wave number $k_x$, and energy may accumulate at high $k_x$ end of the spectral domain of the simulation, leading to nonphysical results. The number of radial Fourier modes needed to resolve the nonlinear physics could then be prohibitively large for the computational resources available.
	
	The objective function used to target QHS is a measure of the deviation from quasisymmetry and uses the helicity of symmetry $N/M$ and assumes either an irrational surface or $\v{B}\cdot\nabla\psi\times\nabla B=0$ on the flux surface\cite{rodriguez_measures_2022},
	\begin{align}
		&f_{\text{QS}} = \sqrt{\fr{\avg{f_C}^2}{\avg{B}^3}\abs{\fr{N}{M}-\ibar}^2},\\
		&f_C = \v{B}\cdot\nabla\psi\times\nabla B-C\v{B}\cdot\nabla B,\\
		&C = \fr{G-(N/M)I}{\ibar-N/M},
	\end{align}
	where $I$ and $G$ are the Boozer currents\cite{boozer_plasma_1981}. This measure of quasisymmetry is advantageous to optimization, because it is a local measure and does not require on a transformation to Boozer coordinates\cite{boozer_plasma_1981}. Additionally, using this formulation in an optimization problem has been successful in reducing the deviation from quasisymmetry to negligible levels\cite{landreman_magnetic_2022}. For the two reduced-TEM configurations, a quasihelically symmetric magnetic field with helicity of $N=N_\text{fp}=4$, $M=-1$ was targeted.
	
	The TEM target function used in the optimization to improve TEM stability was the Available Energy (AE) metric\cite{mackenbach_available_2022}. The AE is a measure of the free energy in the background density and temperature gradients that could be converted into collisionless TEM fluctuations. In other words, AE is the difference in energy between the ground state of the trapped electrons and the background Maxwellian\cite{helander_available_2017,helander_available_2020}. The AE is evaluated by integrating over velocity space and adding together the contributions of bounce-averaged drifts in each magnetic well within the flux tube domain, arriving at
	\begin{equation}
		f_\text{TEM} = \int\int\text{d}K\text{d}\lambda\sum\limits_{\text{wells}(\lambda)}\text{e}^{-K}K^{5/2}\left[\hat{\omega}_y^2\left(\fr{\hat{\omega}^T_*}{\hat{\omega}_y}-1+\hat{F}\right)+\hat{\omega}_x^2\left(\hat{F}-1\right)\right]\hat{G}^{1/2}.
	\end{equation}
	The normalized kinetic energy and pitch angle are given by
	\begin{align}
		&K=\fr{m_\text{e} v^2}{2T_{\text{e}0}},\\
		&\lambda=\fr{2\mu \bar{B}}{m_\text{e}v^2},
	\end{align}
	where $\bar{B}$ is the average magnetic field strength in the flux tube, $\mu=m_\text{e}(v^2-v^2_\parallel)/2B$ is the magnetic moment for an electron with speed $v$ and parallel velocity $v_\parallel$, and $T_{\text{e}0}$ is a normalizing temperature at the center of the flux tube. The AE Jacobian is given by
	\begin{equation}
		\hat{G}^{1/2} = v\fr{\tau_b}{L},
	\end{equation}
	where $L$ is the length of the field line, and the bounce time is defined as
	\begin{equation}
		\tau_b = \int B\sqrt{g}\fr{\text{d}z}{v_\parallel},
	\end{equation}
	where the flux-tube Jacobian is $\sqrt{g}=(\nabla x\cdot\nabla y\times\nabla x)^{-1}$, and the bounce points are dependent on the pitch angle and magnetic well. The background pressure gradients $\omega_{(T\text{e},n)}=-a \text{d}{\ln(T_\text{e},n)}/\text{d}x$ appear in the normalized diamagnetic drift frequency
	\begin{equation}
		\hat{\omega}_*^T = \fr{2e\mathrm{\Delta}x}{m_\text{e}v^2}\omega_*^T = \fr{2T_{\text{e}0}\mathrm{\Delta}x}{m\text{e}v^2}\left(\omega_n+\omega_{T\text{e}}\left[K-\fr{3}{2}\right]\right),
	\end{equation}
	where $a$ is the minor radius of a last closed flux surface. Although there is no last closed flux surface specified in a local equilibrium, for normalization purposes, the local equilibria used in the optimization were assumed to lie at $s=\psi/\psi_\text{ref}=0.5$, where the reference toroidal flux is chosen as the toroidal flux of a last closed flux surface $\psi_\text{ref}=a^2B_0/2$ with minor radius $a$. This normalization was chosen for consistency with the HSX geometry in Ref.~\onlinecite{faber_stellarator_2018}.
	
	For the present work, $\omega_{T\text{e}}=3$ and $\omega_n=2$ are chosen because TEMs can drive significant heat flux in low shear stellarators near these gradients \cite{faber_gyrokinetic_2015,gerard_effect_2024,costello_universal_2023}. The normalized bounce-averaged drift (BAD) frequencies are defined as
	\begin{align}
		&\hat{\omega}_y = \fr{2e\mathrm{\Delta}x}{m\text{e}v^2}\omega_y = \fr{2e\mathrm{\Delta}x}{m\text{e}v^2}\fr{1}{\tau_b}\int\v{v}_d\cdot\nabla yB\sqrt{g}\fr{\text{d}z}{v_\parallel},\\
		&\hat{\omega}_x = \fr{2e\mathrm{\Delta}y}{m\text{e}v^2}\omega_x = \fr{2e\mathrm{\Delta}y}{m\text{e}v^2}\fr{1}{\tau_b}\int\v{v}_d\cdot\nabla x B\sqrt{g}\fr{\text{d}z}{v_\parallel},
	\end{align}
	where $e$ is the elementary charge. The quantities $\mathrm{\Delta}x$ and $\mathrm{\Delta}y$ correspond to normalizing perpendicular widths and have the property $\mathrm{\Delta}x=\mathrm{\Delta}y=1$ in GENE coordinates. The quantity
	\begin{equation}
		\hat{F} = \fr{\sqrt{(\hat{\omega}_y-\hat{\omega}_*^T)^2+\hat{\omega}_x^2}}{\sqrt{\hat{\omega}_y^2+\hat{\omega}_x^2}}
	\end{equation}
	is a functional of $\hat{\omega}_x$, $\hat{\omega}_y$, and $\hat{\omega}_*^T$, which contains the resonance between the BAD frequencies and the diamagnetic drift frequency that is responsible for TEM destabilization. The quantities $\hat{\omega}_x$ and $\hat{\omega}_y$ are calculated using GENE flux tube geometry quantities \cite{mackenbach_bounce-averaged_2023}. Practical evaluation of the AE can be performed via
	\begin{align}
		&\hat{\omega}_y = \fr{2e\mathrm{\Delta}x}{mv^2}\fr{1}{\tau_b}\int \sqrt{g}\mathcal{K}^y\left(1-\fr{B}{2B_{t}}\right)\fr{\text{d}z}{v_\parallel},\\
		&\hat{\omega}_x = \fr{2e\mathrm{\Delta}y}{mv^2}\fr{1}{\tau_b}\int \sqrt{g}\mathcal{K}^x\left(1-\fr{B}{2B_{t}}\right)\fr{\text{d}z}{v_\parallel},\\
		&v_\parallel = v\sqrt{1-\fr{B}{B_t}},
	\end{align}
	where $B_t=\bar{B}/\lambda$ is the magnetic field strength at the bounce point\cite{duff_availableenergymetricjl_nodate}. In the absence of an equilibrium pressure gradient, the curvatures are, for $p'=0$,
	\begin{align}
		&\mathcal{K}^x = -a\left(\d{\psi}{x}\right)^{-1}\abpsi\kappa_g,\\
		&\mathcal{K}^y = \fr{a}{B_0}\d{\psi}{x}\fr{B^2}{\abpsi^2}(\kappa_n+\mathrm{\Lambda}\kappa_g),
	\end{align}
	where $\kappa_n$ and $\kappa_g$ are the normal and geodesic components of the curvature vector $\gv{\kappa}=\kappa_n\nhat+\kappa_g\bhat\times\nhat$ for $\bhat=\v{B}/B$ and $\nhat=\nabla\psi/\abpsi$, and $\mathrm{\Lambda}=-\nabla\psi\cdot\nabla\alpha/B$ is the integrated local shear.
	
	The AE metric is useful as an objective function in an optimization targeting TEM-driven turbulence for a two reasons. First, the quantity $f_{\text{TEM}}$ only contains geometric and background plasma gradient information. This means that the computational cost of evaluating $f_{\text{TEM}}$ is substantially lower than using linear or nonlinear gyrokinetic observables. Second, the AE metric has a power law scaling\cite{mackenbach_available_2022} with the nonlinear heat flux $Q\propto f_{\text{TEM}}^{3/2}$. While the correlation of this scaling with nonlinear gyrokinetic results may be far from perfect, reaching a solution with significantly reduced $f_\text{TEM}$ will likely produce reduced nonlinear TEM-driven heat flux.
	
	The initial configurations are local equilibria with analyticall-prescribed flux-surface shapes. The parameterization of the flux-surface shapes is based on Miller geometry\cite{miller_noncircular_1998} that rotates helically with the magnetic axis,
	\begin{align}
		\label{eq:R_h_axis}
		&R = R_0 + \mathrm{\Delta}\sin(N_\text{fp}\zeta) + r[\cos(N_\text{fp}\zeta-\theta +\arcsin\delta_\text{h}\sin(N_\text{fp}\zeta-\theta))\cos(N_\text{fp}\zeta)\nonumber \\&\quad\quad + (\kappa_\text{h}-1)\sin(N_\text{fp}\zeta-\theta)\sin(N_\text{fp}\zeta) + \cos(\theta) - \cos(N_\text{fp}\zeta-\theta)\cos(N_\text{fp}\zeta)],\\
		\label{eq:Z_h_axis}
		&Z = \mathrm{\Delta}\cos(N_\text{fp}\zeta) + r[\cos(N_\text{fp}\zeta-\theta +\arcsin\delta_\text{h}\sin(N_\text{fp}\zeta-\theta))\sin(N_\text{fp}\zeta)\nonumber \\&\quad\quad + (\kappa_\text{h}-1)\sin(N_\text{fp}\zeta-\theta)\cos(N_\text{fp}\zeta) + \sin(\theta) - \cos(N_\text{fp}\zeta-\theta)\sin(N_\text{fp}\zeta)],
	\end{align}
	where $\mathrm{\Delta}$ is the helical radius of the magnetic axis, $\kappa_\text{h}$ is the helical elongation, and $\delta_\text{h}$ is the helical triangularity. The two initial equilibria were chosen to have negative helical triangularity (NT) and positive helical triangularity (PT), motivated by the differential response of TEMs to NT vs. PT in tokamaks\cite{weisen_effect_1997,moret_influence_1997,pochelon_energy_1999,camenen_impact_2007,fontana_effect_2018,austin_achievement_2019,marinoni_h-mode_2019,pueschel_reducing_2024}. However, the results shown in the present work indicate that the available energy rather than the sign of the helical triangularity, primarily determines the TEM stability and saturation properties.
	
\begin{figure}[!b]
		\centering
		\begin{subfigure}[b]{0.49\textwidth}
			\centering
			\includegraphics[width=\textwidth]{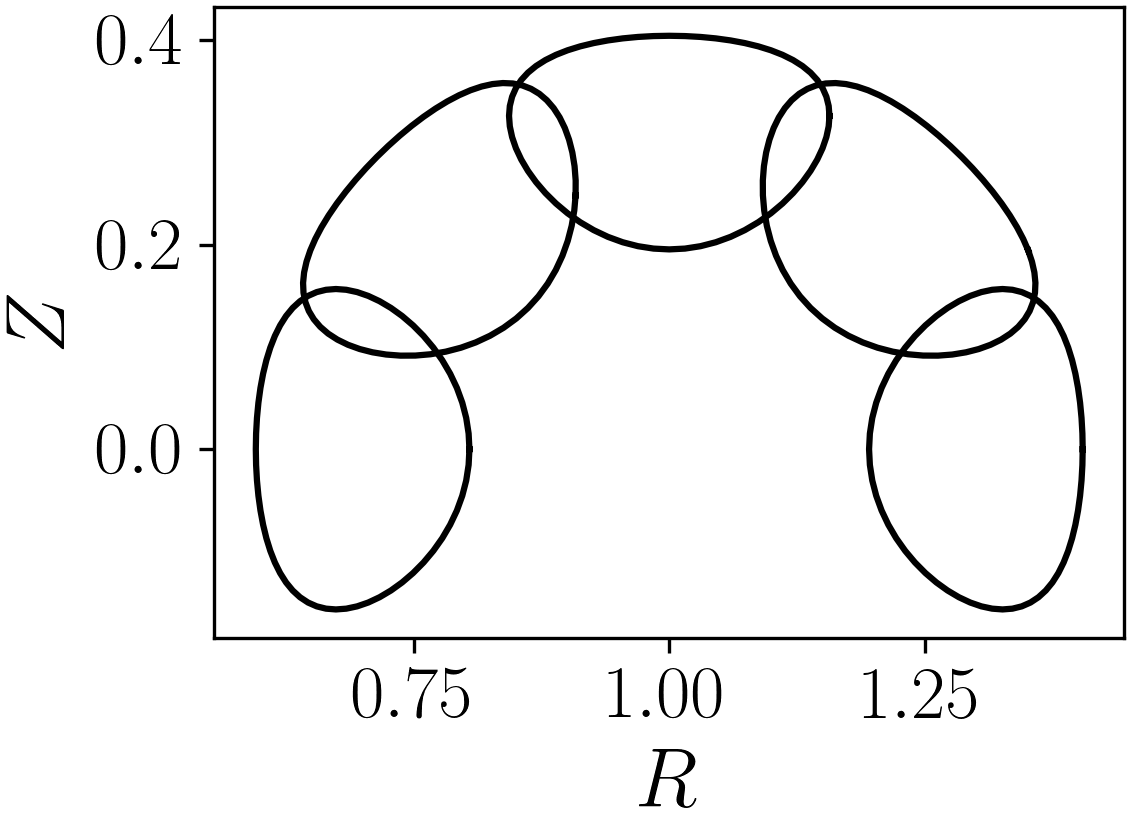}
			\caption{}
			\label{fig:NT_init}
		\end{subfigure}
		\begin{subfigure}[b]{0.49\textwidth}
			\centering
			\includegraphics[width=\textwidth]{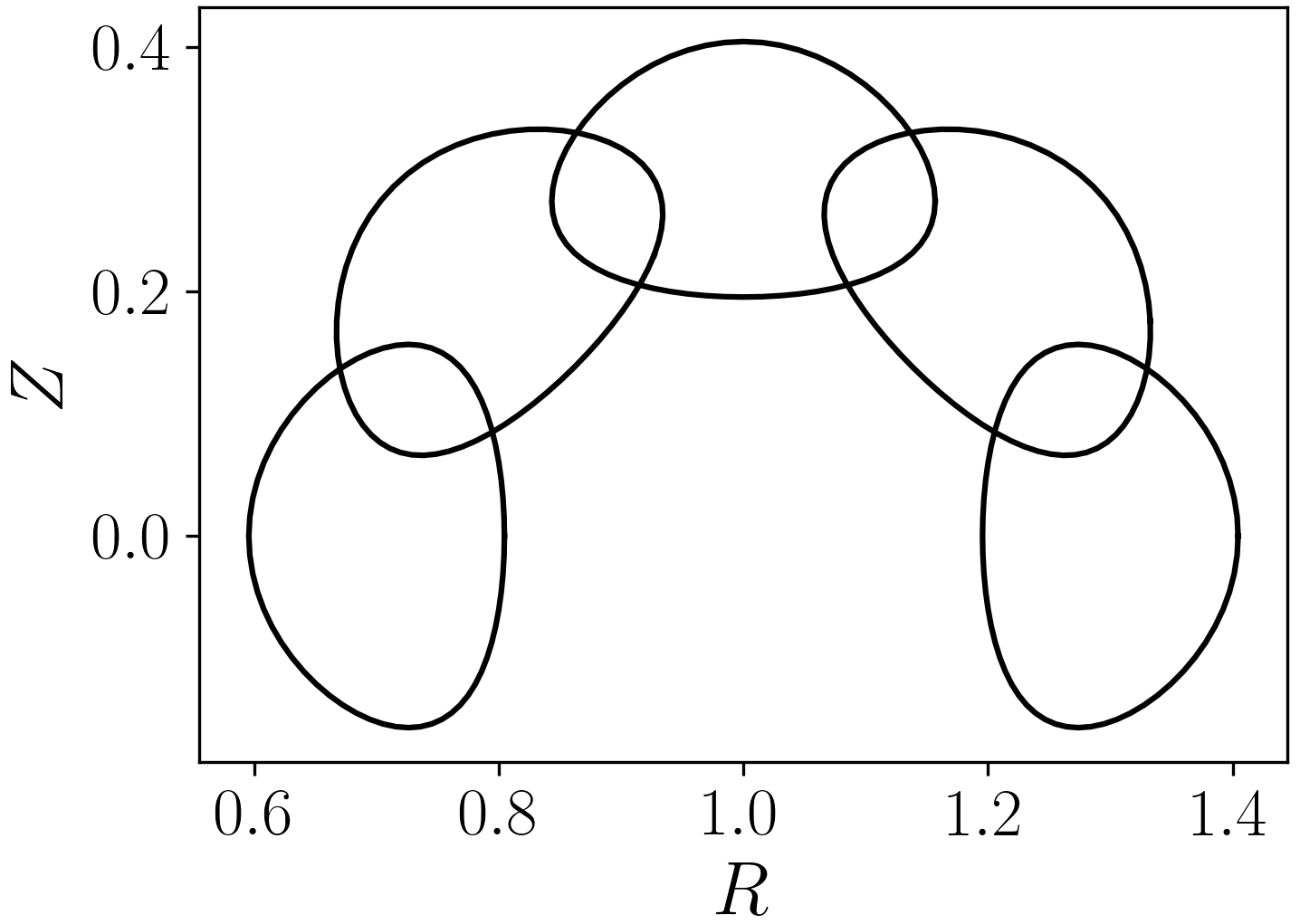}
			\caption{}
			\label{fig:PT_init}
		\end{subfigure}
		\caption{Toroidal cross sections over a half field period for the initial equilibrium used in the optimization for the NT case (a) and the PT case (b).}
		\label{fig:init_guess}
	\end{figure}

	The initial local equilibria have the flux surface shape parameterized by $R_0=1$, $r=0.1045$, $\mathrm{\Delta}=0.3$, $N=4$, $\kappa_h=1.5$ and $\delta_h=\pm0.25$. The initial equilibrium parameters are $\ibar=1.05013$ and $\sigma=0$, with pressure gradient is held fixed at $p'=0$. The minor radius $r$ is chosen to have a comparatively low aspect ratio, expecting the aspect ratio to increase in order to improve quantities like quasisymmetry. A relatively large $\mathrm{\Delta}$ is chosen because it was observed that quasisymmetry tends to improve with larger $\mathrm{\Delta}$. A rotational transform just above unity was chosen for two of reasons. First, a benefit is expected in terms of neoclassical confinement\cite{helander_theory_2014}. Second, in combination with a low global shear, a rotational transform just above unity avoid low-order rational surfaces, which can degrade plasma confinement through the formation of magnetic islands\cite{freidberg_ideal_2014}. Several toroidal cross-sections of each initial equilibrium over a half field period are depicted in Fig.~\ref{fig:init_guess}. The cross-section at the largest major radius $R$ corresponds to $\zeta=0$. As $\zeta$ increases, the cross-sections rotate counter clockwise, with the final depicted cross-section at $\zeta=\pi/N_\text{fp}$. The flux surfaces are symmetric about the $Z=0$ plane.
	
	\subsection{Optimization Results}
\begin{figure}[!b]
		\centering
		\begin{subfigure}[b]{0.49\textwidth}
			\centering
			\includegraphics[width=\textwidth]{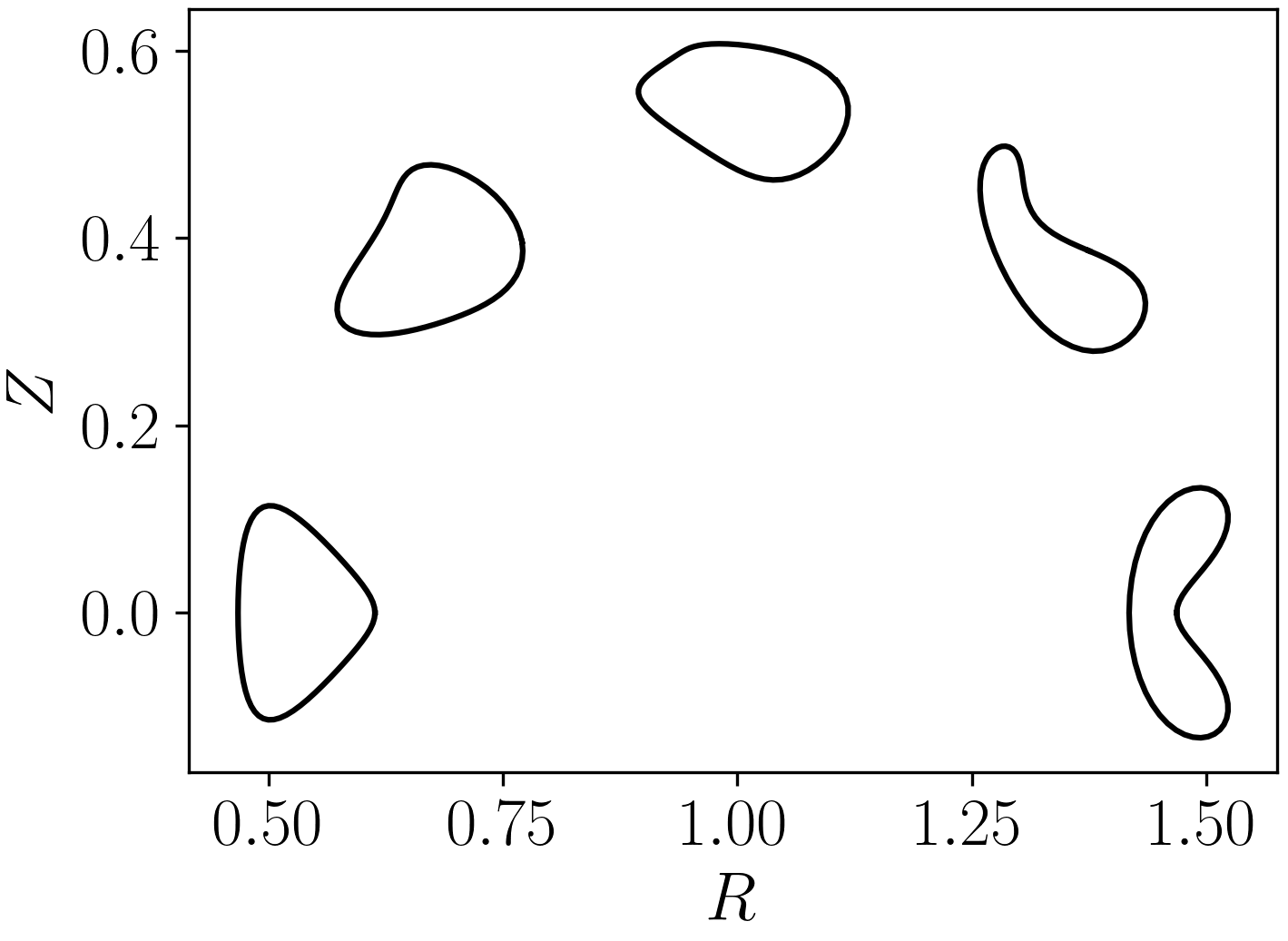}
			\caption{}
			\label{fig:NT_opt}
		\end{subfigure}
		\begin{subfigure}[b]{0.49\textwidth}
			\centering
			\includegraphics[width=\textwidth]{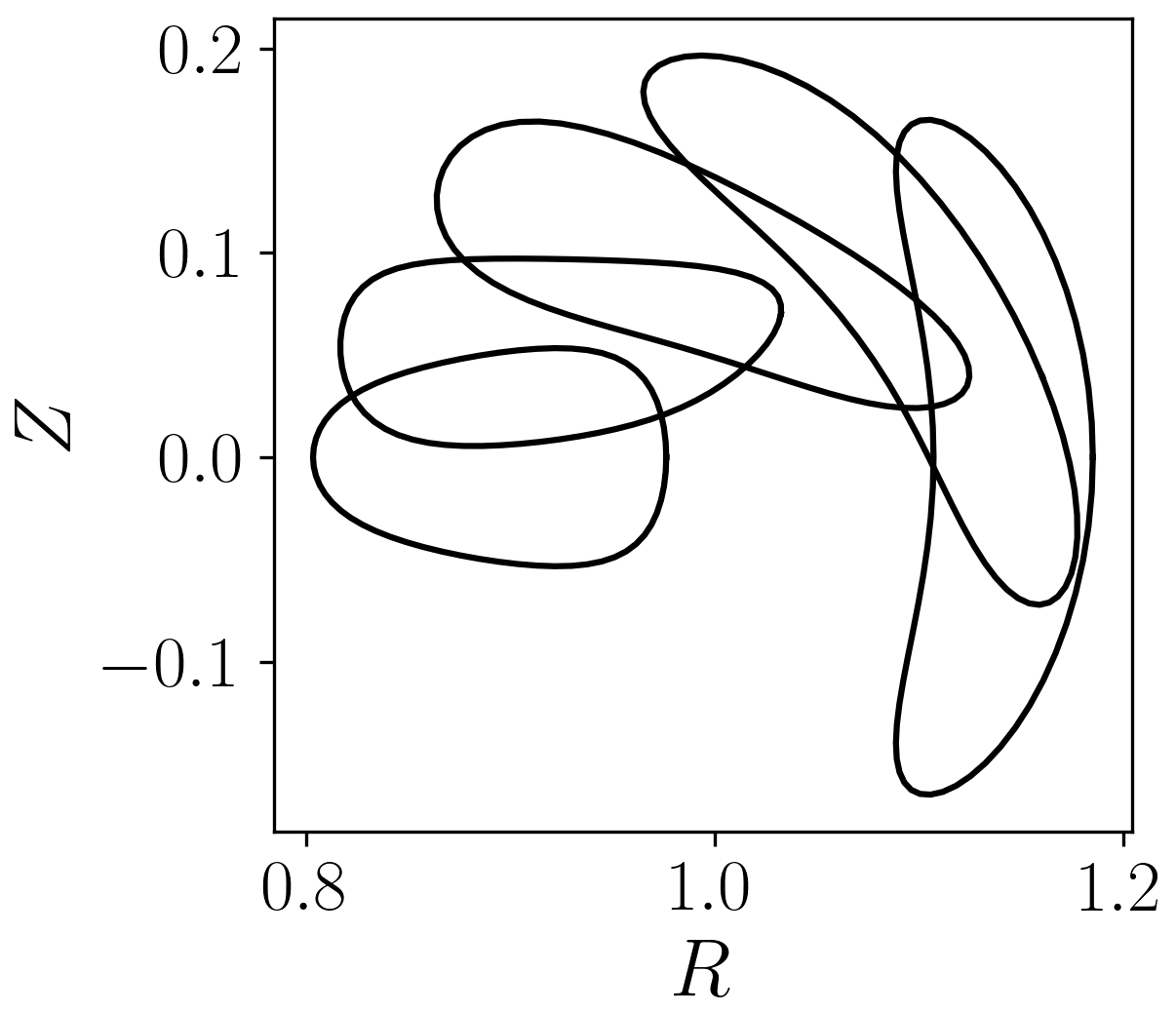}
			\caption{}
			\label{fig:PT_opt}
		\end{subfigure}\\
		\begin{subfigure}[b]{0.49\textwidth}
			\centering
			\includegraphics[width=\textwidth]{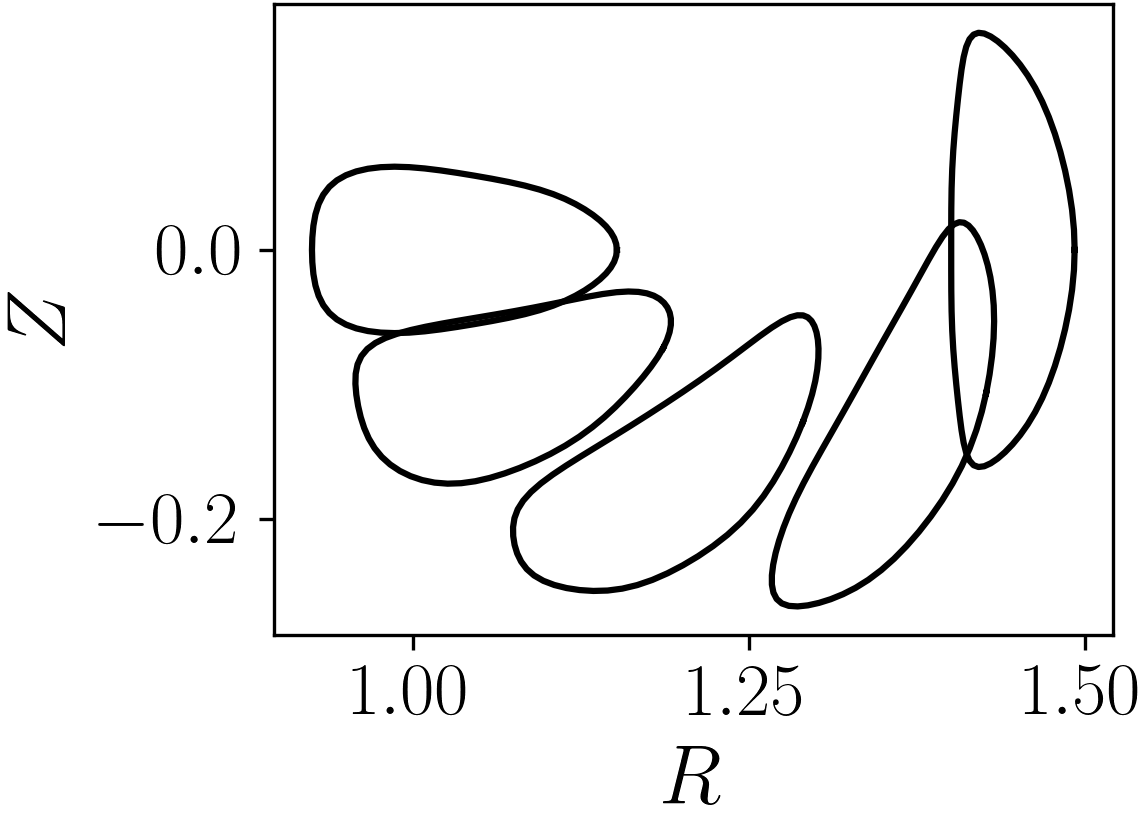}
			\caption{}
			\label{fig:HSX}
		\end{subfigure}
		\caption{Toroidal cross-sections over a half field period for the reduced-TEM equilibria for the NT case (a), the PT case (b), and the non-reduced-TEM HSX (c).}
		\label{fig:opt_xsection}
	\end{figure}

	Two reduced-TEM local equilibria were generated---one with NT and one with PT---using the Numerical Evaluation of 3D Local Equilibria (NE3DLE)\cite{duff_numerical_nodate} code to generate solutions to the local 3D MHD equilibrium equations and StellaratorOptimization.jl\cite{faber_stellaratoroptimizationjl_nodate} to perform the optimization. As a comparison, an $s=0.5$ flux surface of HSX in the QHS configuration\cite{faber_gyrokinetic_2015} was also considered. The toroidal cross-sections of the reduced-TEM flux surfaces and HSX are shown in Fig.~\ref{fig:opt_xsection}. Table~\ref{tab:opt_values} contains a list of quantities that were targeted in the optimization by the objective function. Optimized configurations lie well within acceptable ranges of $f_R$ and $f_{\kappa p}$, meaning the flux surface did not cross $R=0.4$ while maintaining poloidal curvature lower than a tokamak with elongation $\kappa=1$ and triangularity $\abs{\delta}=0.75$. The radius of the helical axis increased substantially for the final NT configuartion, as was preferred by the quasisymmetry metric, and the aspect ratio also increased but remains in the neighborhood of the aspect ratio of the HSX flux surface. The quasisymmetry was improved for the NT configuration as well but did not reach the levels of HSX. AE decreased substantially, arriving at a value lower by a factor of 2.5 than in HSX, which already has a relatively low AE compared to the initial equilibria. Similar observations were made in the PT case; however, the deviation from quasisymmetry was larger and the AE was lower than the optimized NT case. Based on values of $f_{\text{TEM}}$, HSX would be expected to be the most linearly unstable to TEMs and produce more TEM-driven heat flux than the reduced-TEM cases. Both the NT and PT reduced-TEM cases are expected to have similar TEM properties.
	
	\begin{table}[h]
		\centering
		\begin{tabular}{c|c|c|c|c|c}
			Quantity&NT initial&NT optimized&PT initial&PT optimized&HSX\\
			\hline
			$f_{\text{QS}}$&$1.117$&$0.2840$&$8.906$&$0.3791$&$0.1311$\\
			\hline
			$f_{\text{TEM}}$&$3.148$&$0.1495$&$8.683$&$0.1141$&$0.3702$\\
			\hline
			$\hat{s}$&$-1.146$&$-0.3107$&$-0.3725$&$0.3225$&$-0.0798$\\
			\hline
			$\ibar$&$1.050103$&$1.337202$&$1.050103$&$1.387012$&$1.060230$ \\
			\hline
			$\sigma$&$0$&$-0.0619$&$0$&$0.0136$&$-$\\
			\hline
			$A$&$9.576$&$15.18$&$9.576$&$15.12$&$14.10$\\
			\hline
			$f_{\kappa p}(\kappa_{p0}=0.01)$&0&0&0&0&0\\
			\hline
			$f_{R}(R_{\text{min}}=0.6)$&0&$9.605\times10^{-4}$&0&0&0\\
			\hline
			$\max\limits_z (g^{yy}/g^{xx})$&$213.6$&$7.959$&$25.96$&$19.67$&$21.91$
		\end{tabular}
		\caption{Quantities targeted by optimization for initial, reduced-TEM, and HSX configurations. A dash entries indicates this quantity could not be computed due technical reasons.}
		\label{tab:opt_values}
	\end{table}
	
\begin{figure}[h]
		\centering
		\begin{subfigure}[b]{0.49\textwidth}
			\includegraphics[width=\textwidth]{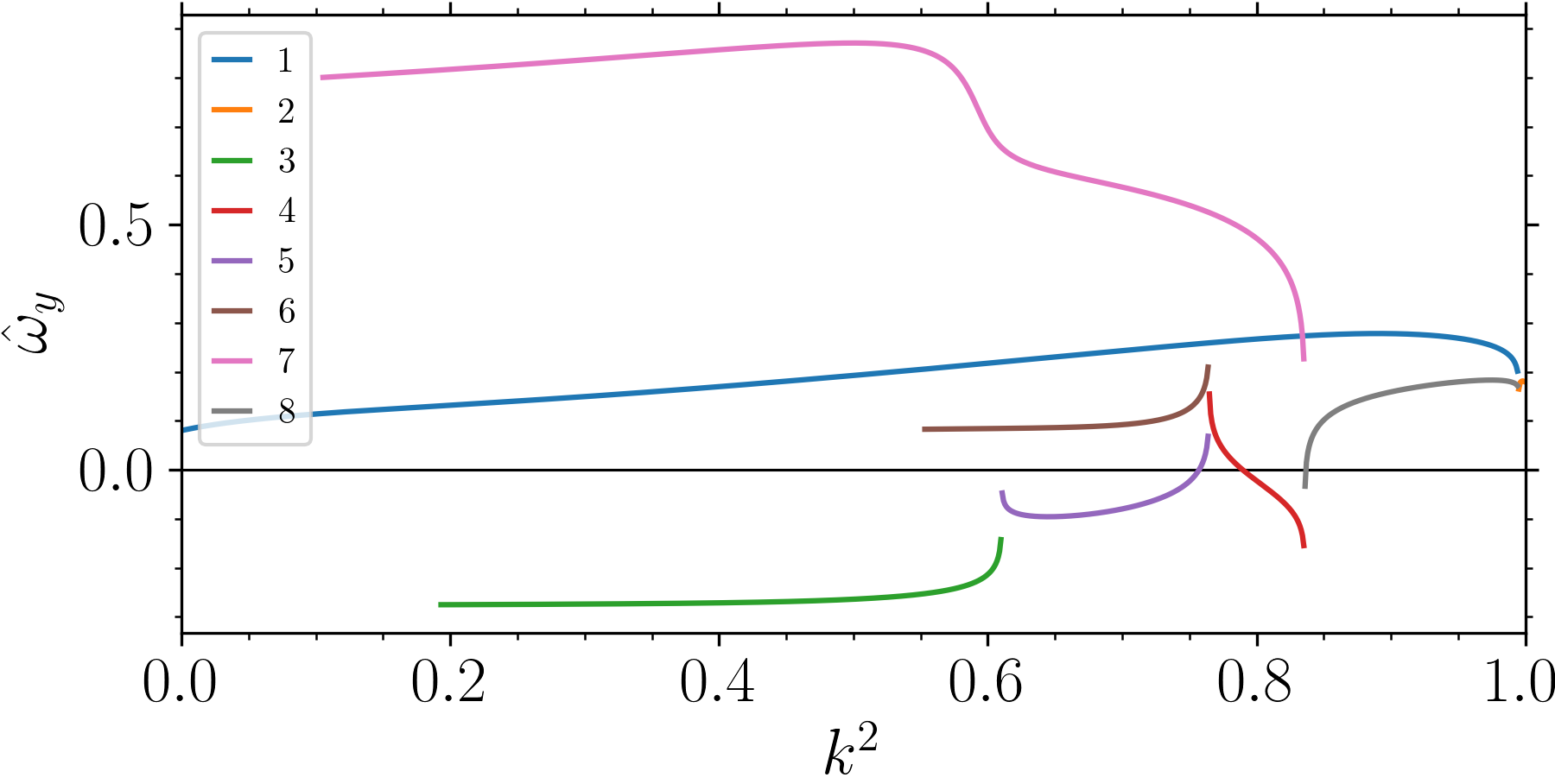}
			\caption{}
		\end{subfigure}
		\begin{subfigure}[b]{0.49\textwidth}
			\includegraphics[width=\textwidth]{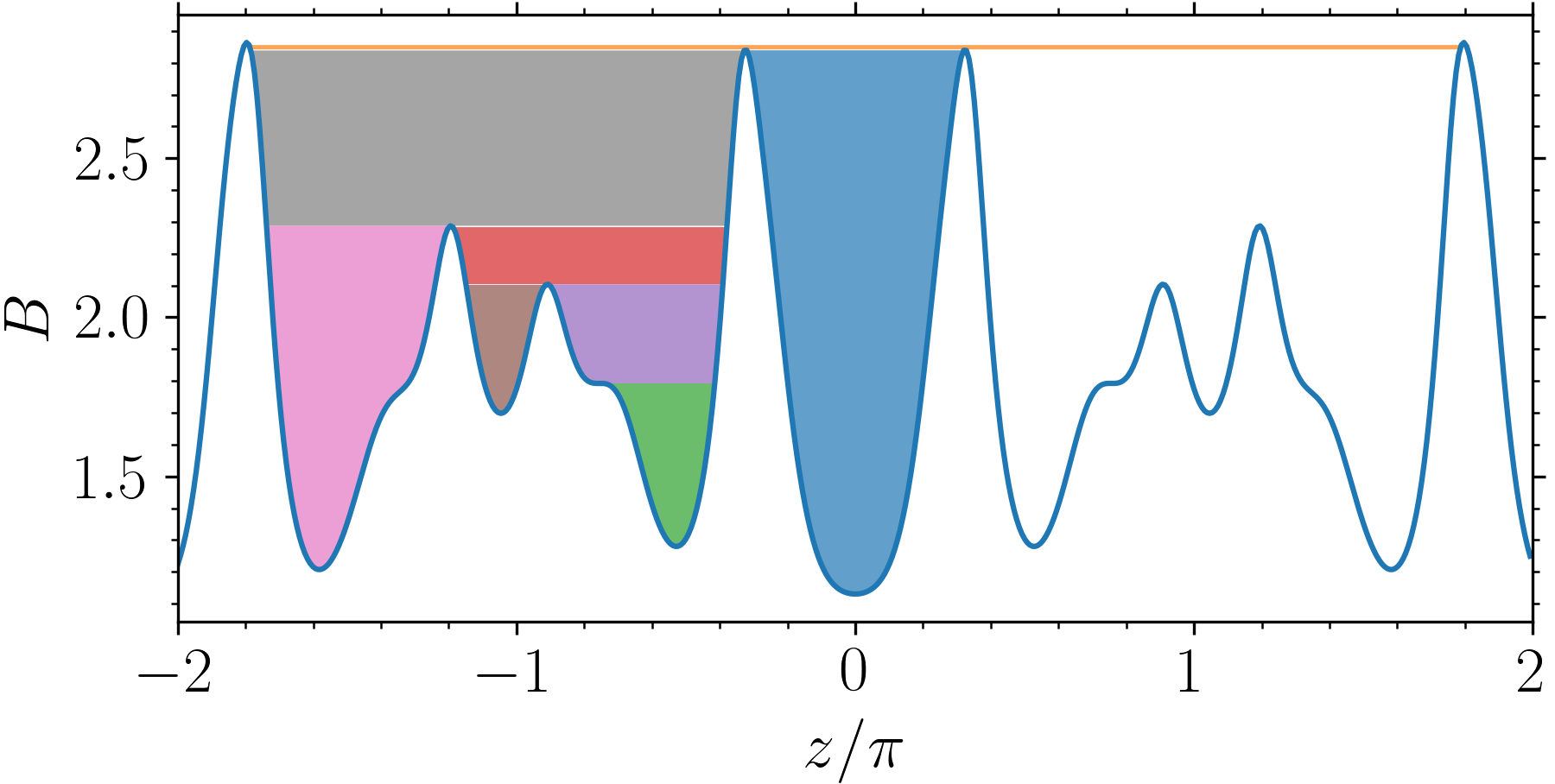}
			\caption{}
		\end{subfigure} \\
		\begin{subfigure}[b]{0.49\textwidth}
			\includegraphics[width=\textwidth]{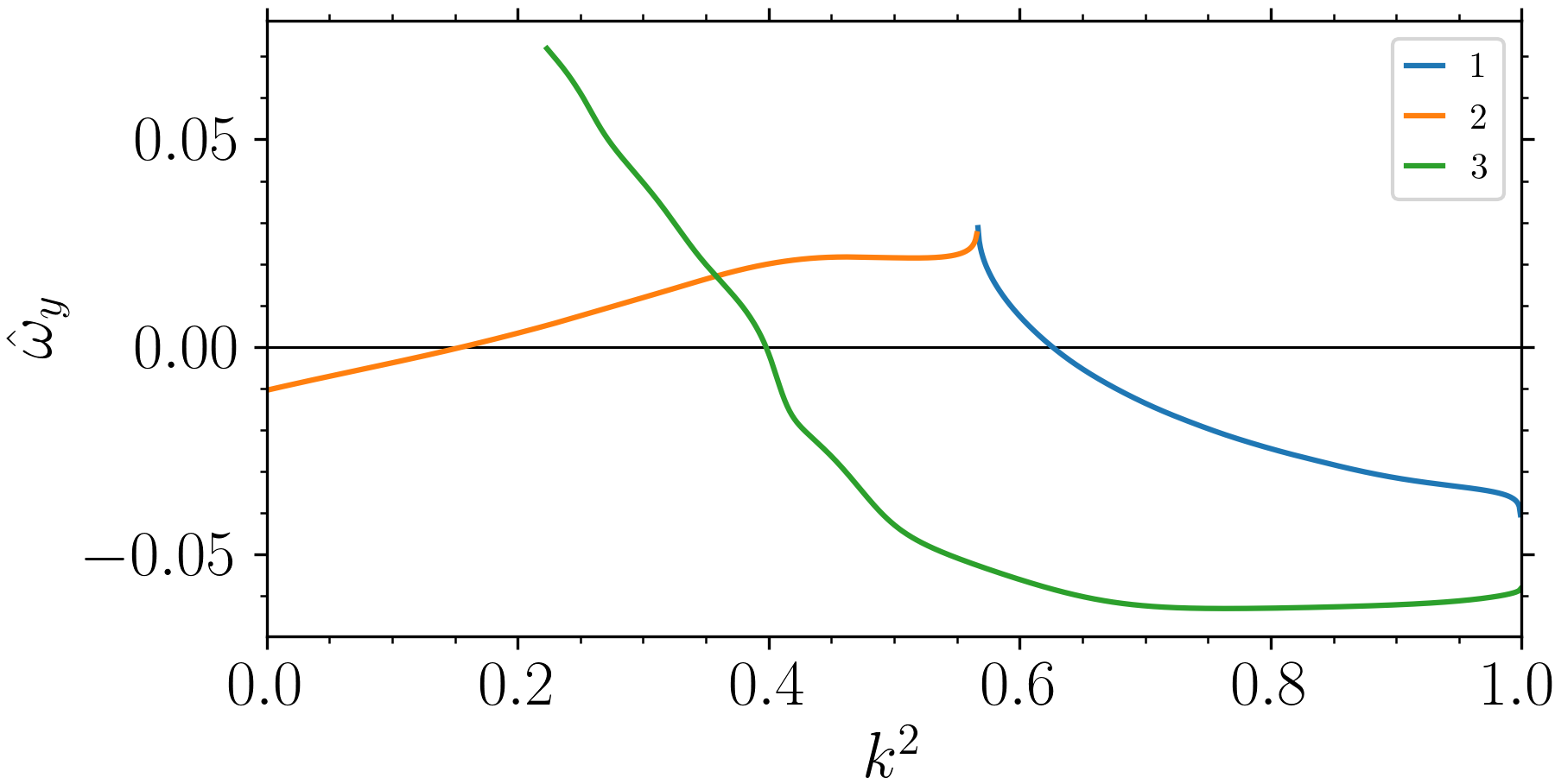}
			\caption{}
		\end{subfigure}
		\begin{subfigure}[b]{0.49\textwidth}
			\includegraphics[width=\textwidth]{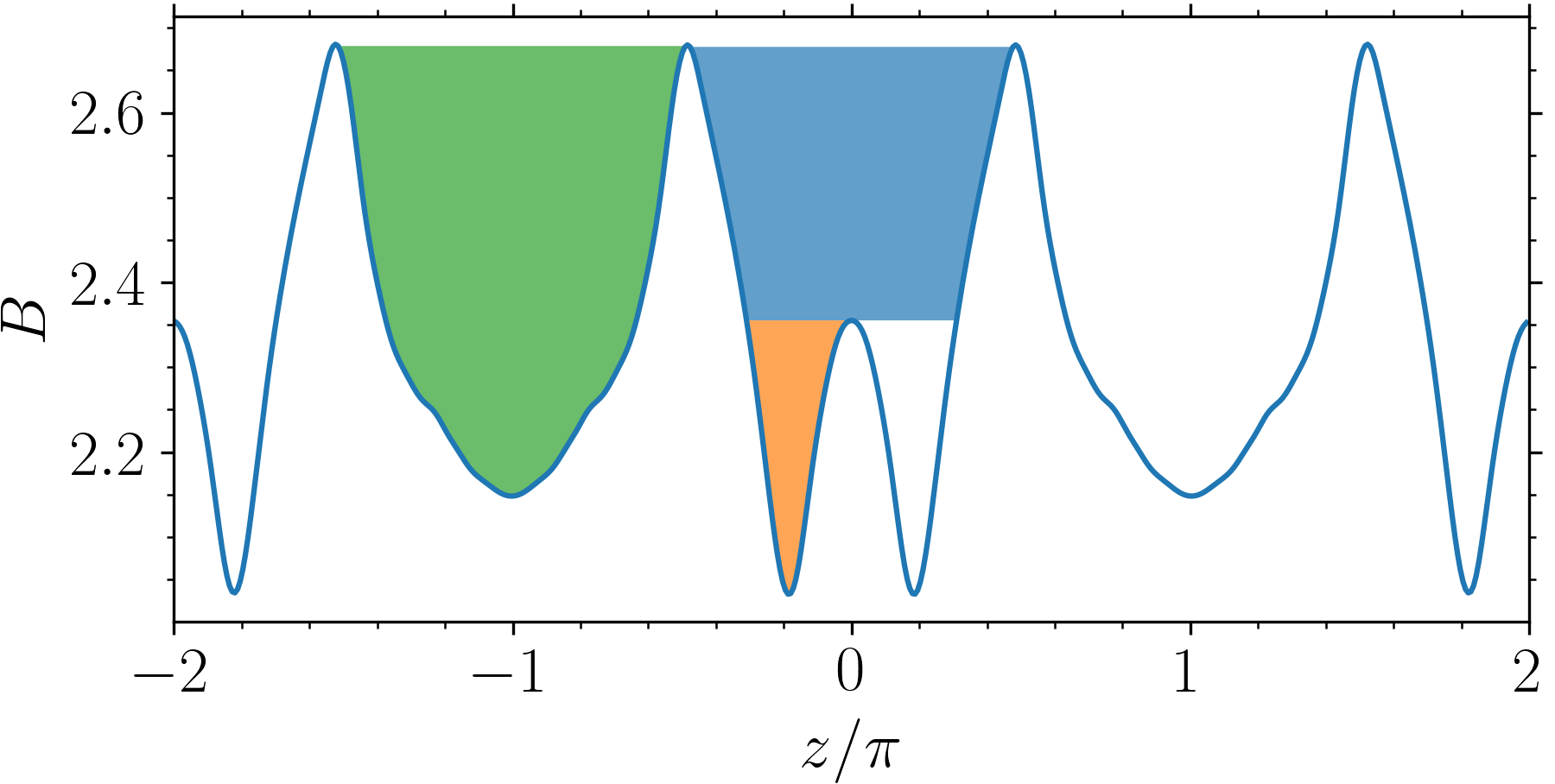}
			\caption{}
		\end{subfigure}
		\caption{Normalized bounce-averaged drift frequency as a function of normalized pitch angle variable $k^2$ for the initial NT (a) and optimized NT (c) equilibria. The colors correspond to the shaded magnetic wells, plotted along the field line in (b) and (d), respectively. Deeply trapped particles exist near $k^2=0$, and the trapped-passing boundary is at $k^2=1$. Optimizing for available energy reduces $\hat{\omega}_y$, and trapped particles in the wells centered at $z\approx\pm\pi$ (green) have the most unstable drifts for the optimized configuration.}
		\label{fig:NT_omega_alpha}
	\end{figure}

	The BAD frequency $\hat{\omega}_y$ is known to be destabilizing for TEMs\cite{connor_effect_1983,hegna_effect_2015} and was targeted through the available energy in the optimization. In Fig.~\ref{fig:NT_omega_alpha}, $\hat{\omega}_y$ for the initial and optimized NT configurations is plotted as a function of a normalized pitch angle variable
	\begin{equation}
		k^2 = \left(1-\lambda\fr{B_\text{min}}{B}\right)\fr{B_\text{max}}{B_\text{max}-B_\text{min}},
	\end{equation}
	where $B_\text{max}$ is the maximum value of the magnetic field along the field line and $B_\text{min}$ is the minimum\cite{beer_bounce_1996}. Deeply trapped particles correspond to $k=0$, and the trapped-passing boundary lies at $k=1$. The line colors correspond to the colors used to shade the magnetic wells in panel (b), and $\hat{\omega}_y>0$ is destabilizing. The optimization reduced the largest value of $\hat{\omega}_y$ by an order of magnitude and reversed $\hat{\omega}_y$ for weakly trapped and very deeply trapped particles in wells near $z=0$. However, the wells centered at $z\approx\pm\pi$ for the TEM-reduced NT configuration had the most destabilizing BAD frequency for deeply trapped particles.
	
\begin{figure}[h]
		\centering
		\begin{subfigure}[b]{0.49\textwidth}
			\includegraphics[width=\textwidth]{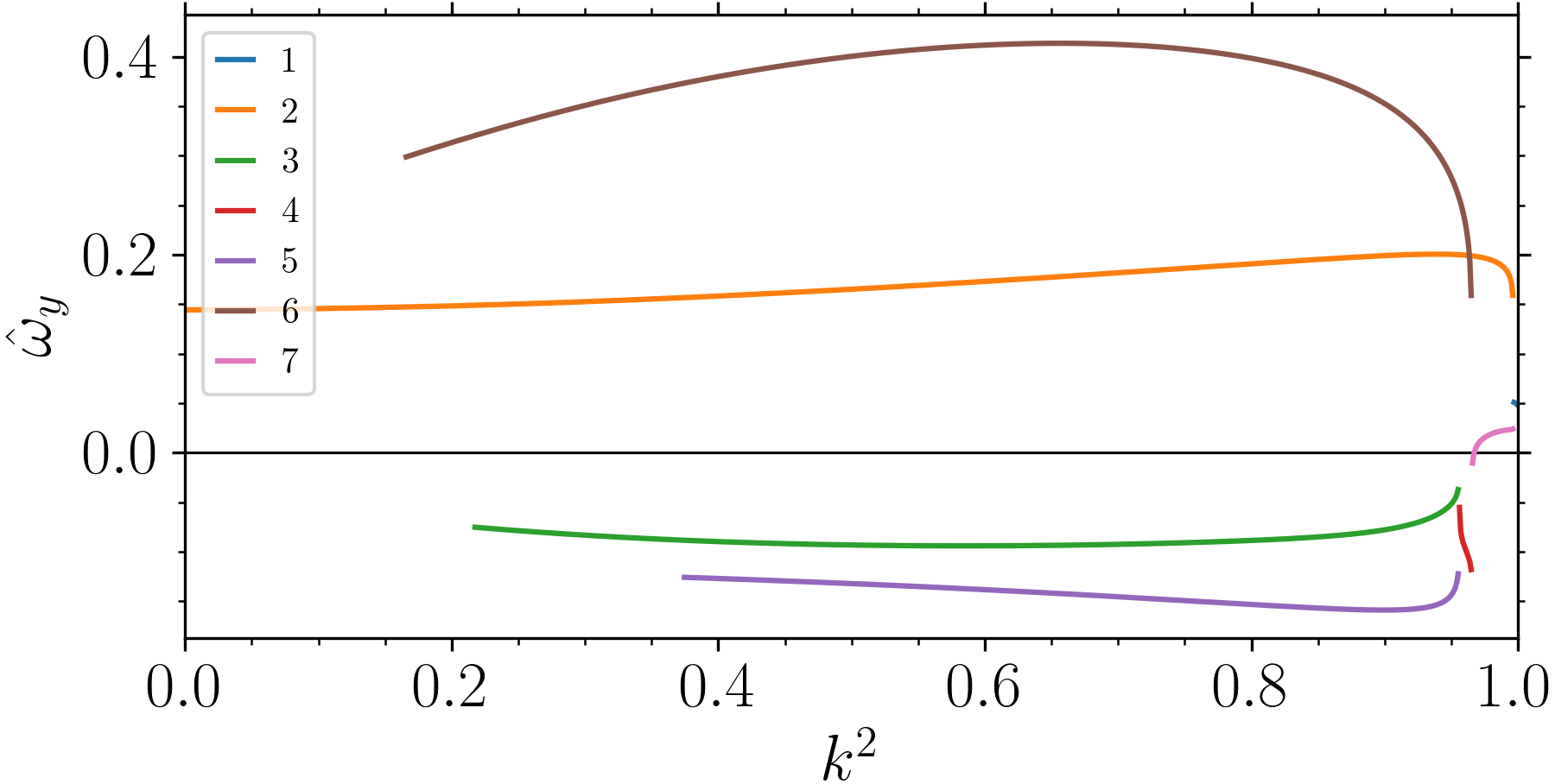}
			\caption{}
		\end{subfigure}
		\begin{subfigure}[b]{0.49\textwidth}
			\includegraphics[width=\textwidth]{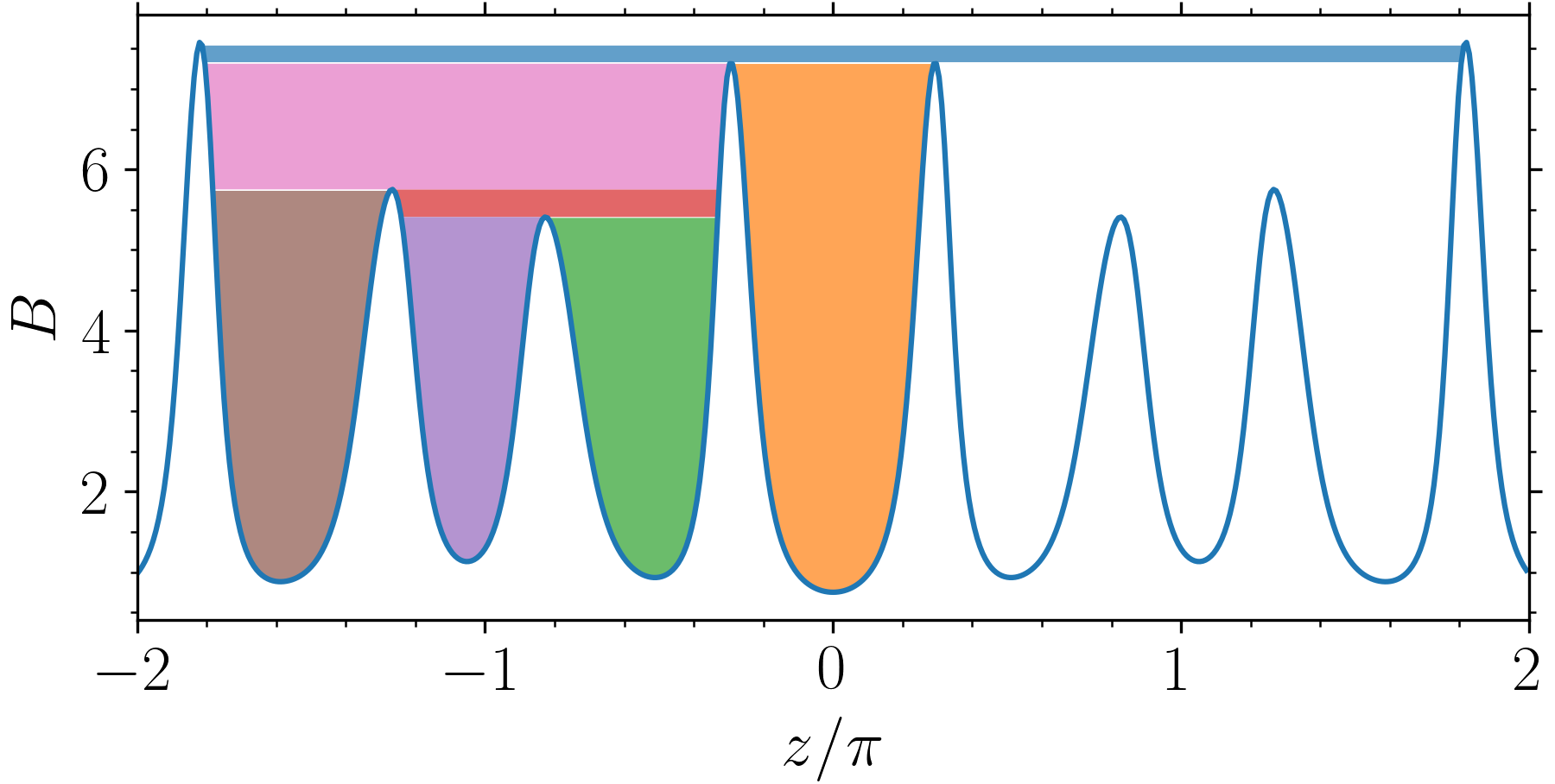}
			\caption{}
		\end{subfigure} \\
		\begin{subfigure}[b]{0.49\textwidth}
			\includegraphics[width=\textwidth]{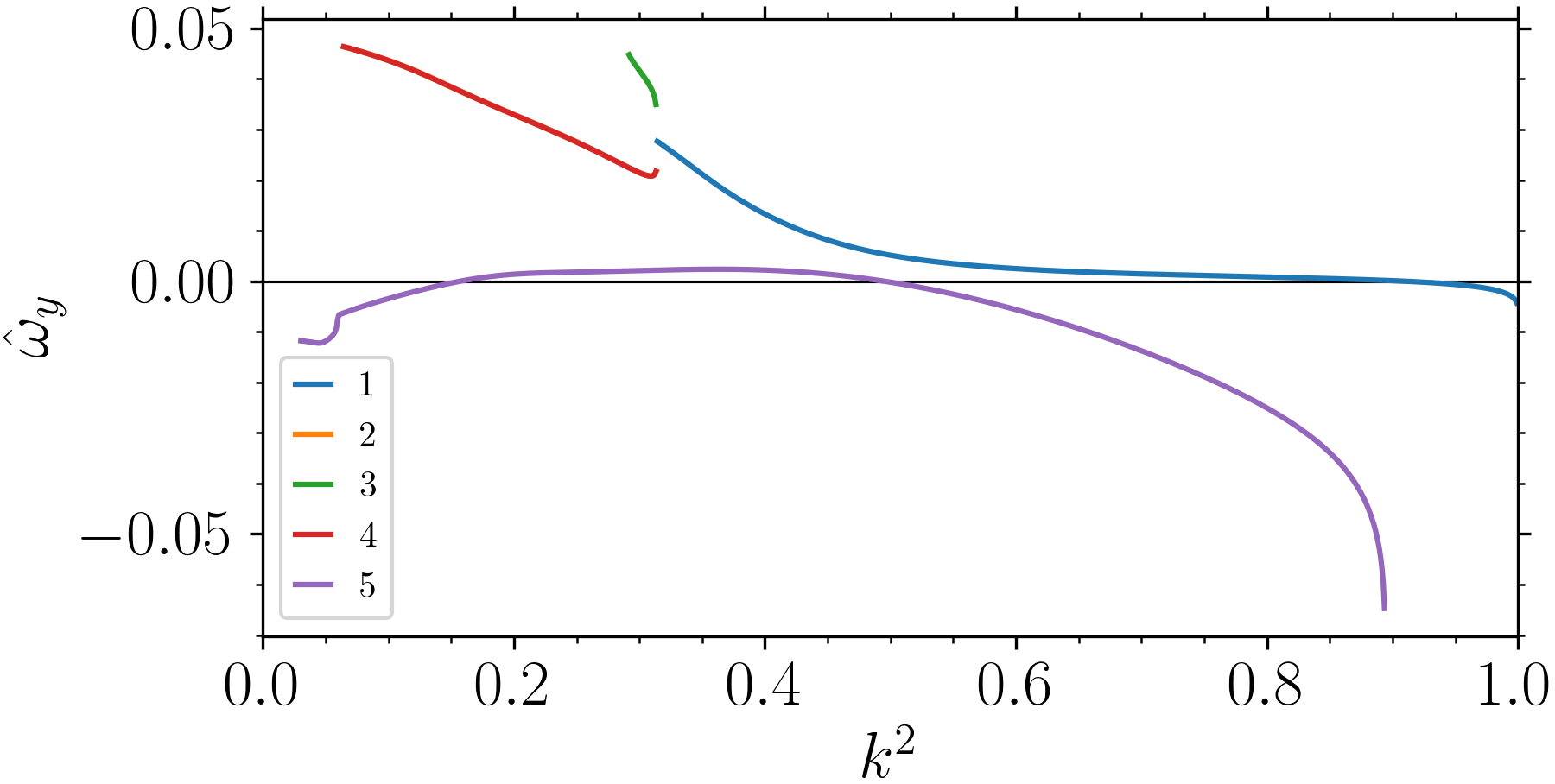}
			\caption{}
		\end{subfigure}
		\begin{subfigure}[b]{0.49\textwidth}
			\includegraphics[width=\textwidth]{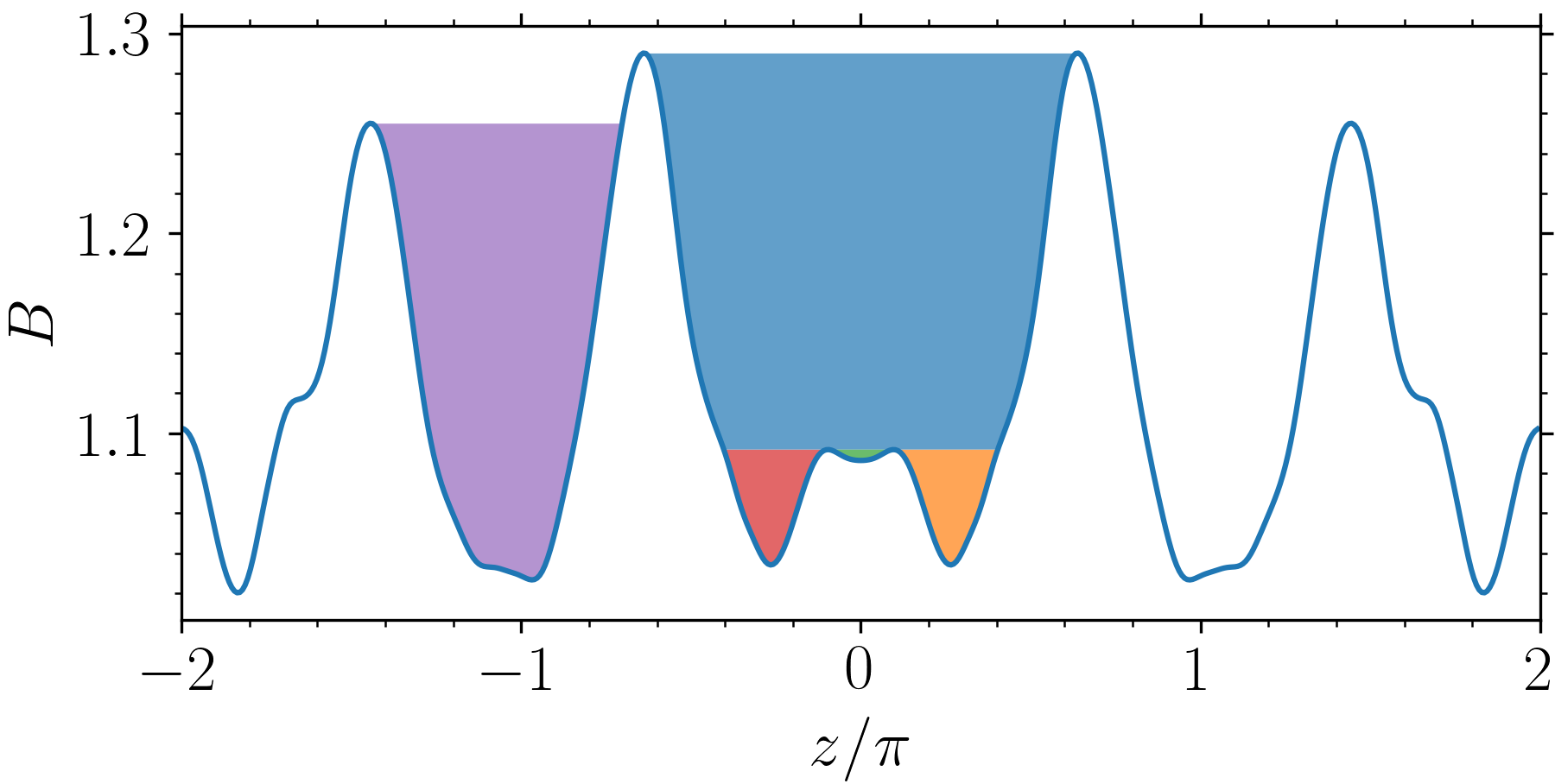}
			\caption{}
		\end{subfigure}
		\caption{Normalized bounce-averaged drift frequency as a function of normalized pitch angle variable $k^2$ for the initial PT (a) and optimized PT (c) equilibria. The colors correspond to the shaded magnetic wells, plotted along the field line in (b) and (d), respectively. Deeply trapped particles exist near $k^2=0$, and the trapped-passing boundary is at $k^2=1$. Optimizing for available energy reduced $\hat{\omega}_y$, and trapped particles in the wells centered at $z\approx\pm0.15\pi$ and $z=0$ (orange, red, and green) have the most unstable drifts for the optimized configuration.}
		\label{fig:PT_omega_alpha}
	\end{figure}

	These analyses are repeated for the PT initial and optimized equilibria, see Fig.~\ref{fig:PT_omega_alpha}. Once again, the optimization reduced $\hat{\omega}_y$ by an order of magnitude. Here, the wells centered at $z=0$ and $z\approx\pm0.15\pi$ in the optimized PT configuration have the most destabilizing $\hat{\omega}_y$. The most destabilizing BAD frequencies for the reduced-TEM PT and NT cases are similar in magnitude but are located in different wells.
	
\begin{figure}[h]
		\centering
		\begin{subfigure}[b]{0.49\textwidth}
			\includegraphics[width=\textwidth]{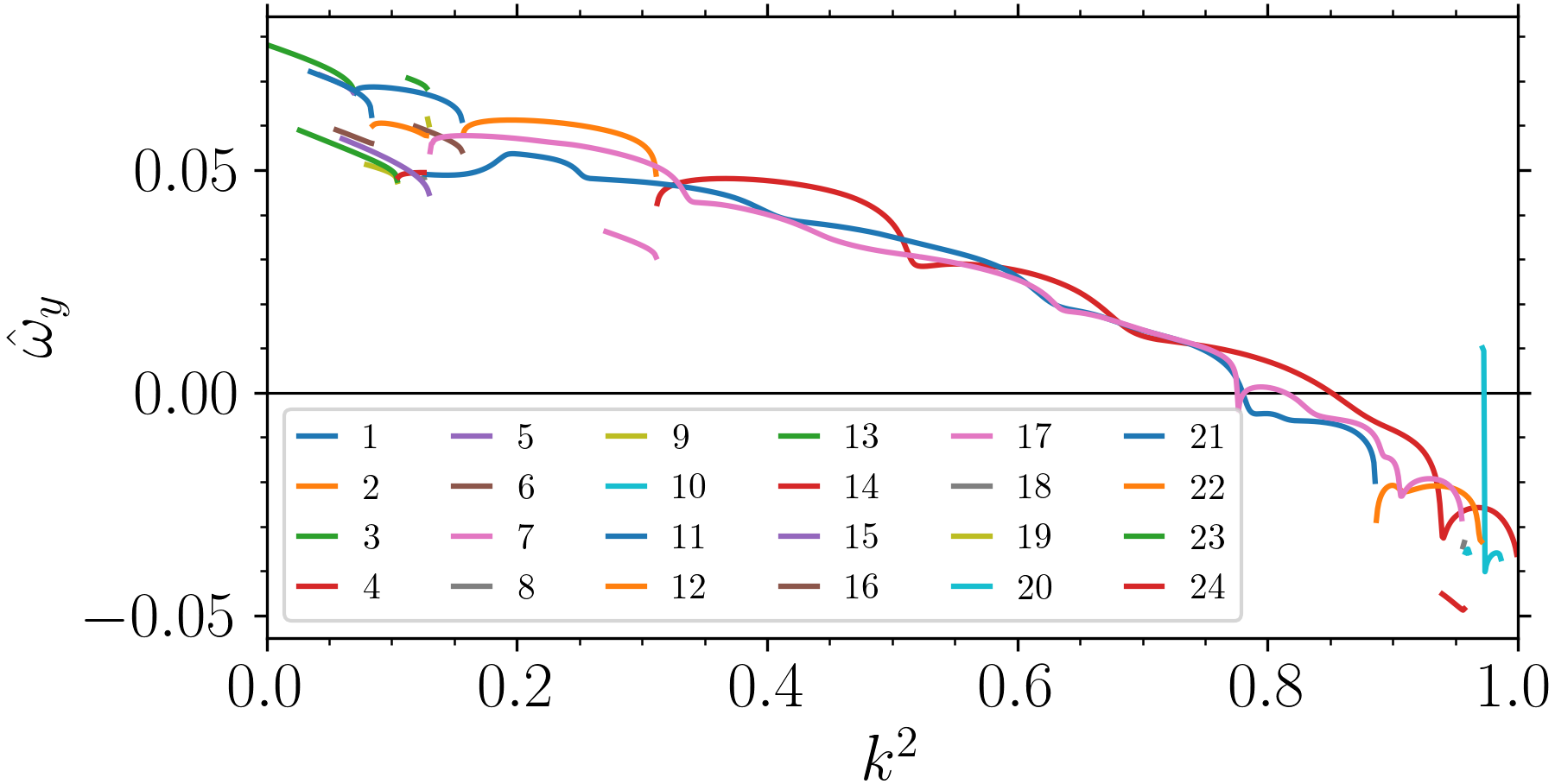}
			\caption{}
		\end{subfigure}
		\begin{subfigure}[b]{0.49\textwidth}
			\includegraphics[width=\textwidth]{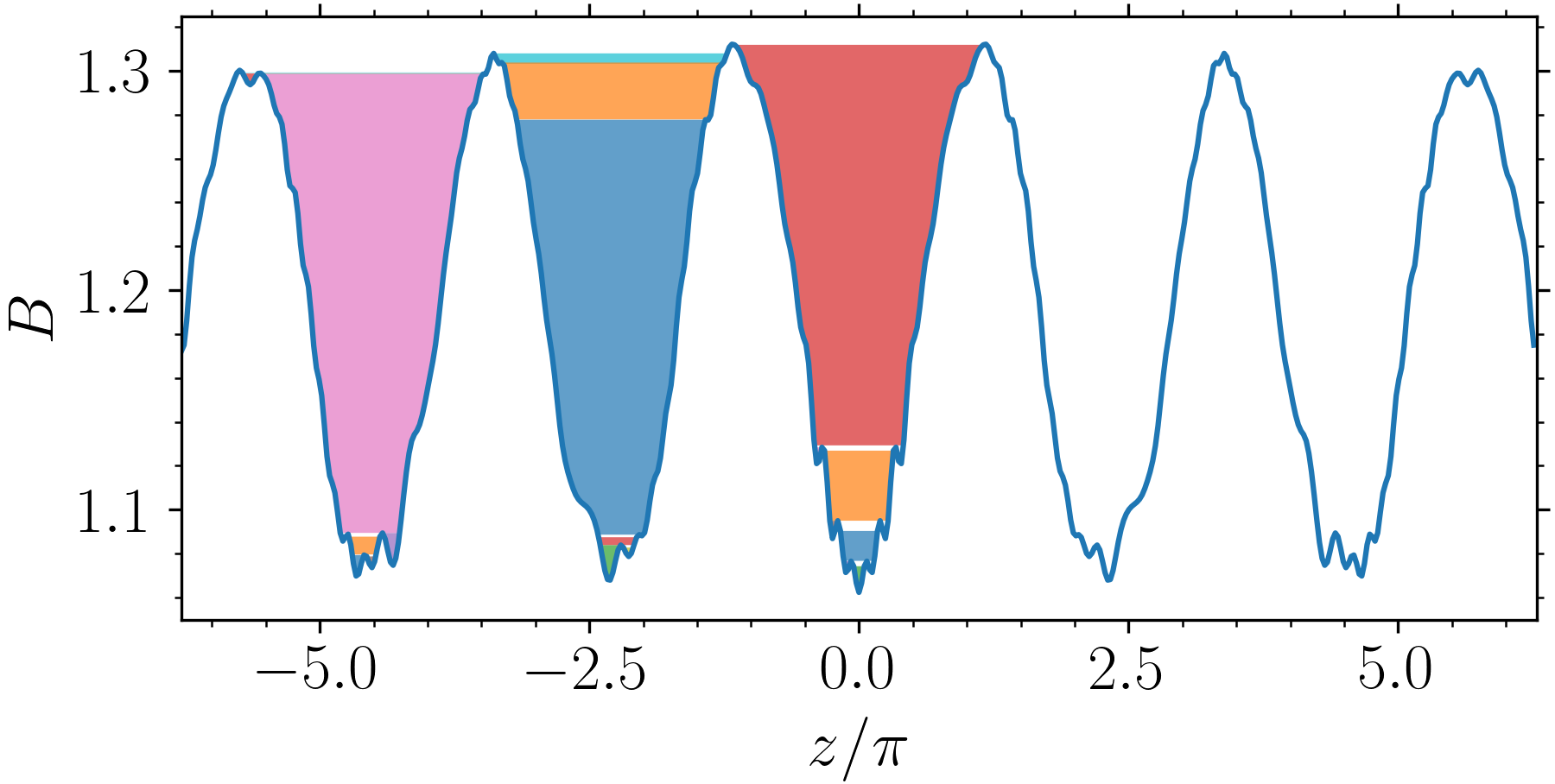}
			\caption{}
		\end{subfigure} 
		\caption{Normalized bounce-averaged drift frequency as a function of normalized pitch angle variable $k^2$ for HSX (a). The colors correspond to the shaded magnetic wells, plotted along the field line in (b). Deeply trapped particles exist near $k^2=0$ and the trapped-passing boundary is at $k^2=1$. All wells have similarly destabilizing $\hat{\omega}_y$ up to the trapped-passing boundary.}
		\label{fig:HSX_omega_alpha}
	\end{figure}

	Equivalent data are shown for HSX in Fig.~\ref{fig:HSX_omega_alpha}. In HSX, TEMs are the most unstable modes and turbulence is driven by TEMs\cite{faber_gyrokinetic_2015}. The magnitude of $\hat{\omega}_y$ in HSX is similar to that of the reduced-TEM NT and PT cases. However, the BAD frequency is similarly destabilizing in each well. The fact that HSX has a destabilizing BAD frequency in each well leads to stronger expected TEM growth than for the reduced-TEM equilibria.
	
	\section{Linear Stability and Mode Identification}
	\label{sec:linear}
	The linear stability and mode characteristics were studied for each reduced-TEM equilibrium and the HSX equilibrium. The linear modes were computed with initial value calculations using the gyrokinetic\cite{abel_multiscale_2013,brizard_foundations_2007} turbulence code \textsc{Gene}\cite{jenko_electron_2000,jenko_massively_2000,noauthor_gene_nodate}, where the linearized gyrokinetic equations were evolved in time to find the most unstable mode at a given wavenumber $(k_x,k_y)$. Only modes centered at $k_x=0$ were computed, as these modes tend to dominate the turbulent spectrum. Growth rates $\gamma$ and real frequencies $\omega$ are given in units of $c_\text{s}/a$, where $c_s=\sqrt{T_{\text{i}0}/m_\text{i}}$ is the ion sound speed with background ion temperature $T_{\text{i}0}$. Linear simulations were computed with a numerical grid size of $N_x\times N_z\times N_{v_\parallel}\times N_\mu = 7\times 512\times 32\times 8$ with four poloidal turns ($n_\text{pol}=4$), where $N_x$ is the number of radial grid points. $N_z$ is for parallel spatial, $N_{v_\parallel}$ is for parallel velocity, and $N_\mu$ is for magnetic moment grid points. The present scenario is collisionless, has no ion temperature gradient, uses a hydrogen mass ratio $m_\text{i}/m_\text{e} = 1836$, a background temperature ratio of $T_{\text{i}0}/T_{\text{e}0}=1$, and was effectively electrostatic with $\beta=10^{-4}$. The hyperdiffusion coefficients\cite{pueschel_role_2010} are set to $\varepsilon_z=4$ and $\varepsilon_{v}=0.2$. A scenario with $\omega_n=4$ and $\omega_{T\text{e}}=0$ was investigated to determine if $\nabla n$-driven TEMs were stabilized, and a scenario with $\omega_n=0$ and $\omega_{T\text{e}}=4$ examined $\nabla T$-driven TEMs.
	
	To identify which modes are the dominant instability, several different mode properties were determined. First, the cross phases $\alpha$ between electrostatic potential fluctuations $\mathrm{\Phi}$ and the fluctuations of density $n$, parallel electron temperature $T_\parallel$, and perpendicular electron temperature $T_\perp$ for both the trapped and passing populations of electrons were evaluated. The cross phases between $\mathrm{\Phi}$ and the other quantities indicate how efficiently trapped or passing electron populations drive instability via a given gradient. Nonlinearly, $0<\alpha<\pi$ is associated with outward radial transport, with the highest efficiency occurring at $\pi/2$. For $-\pi<\alpha<0$, a pinch is indicated. An efficient drive from the density and temperature fluctuations of the trapped electrons $n_{\text{t}}$ and $T_{\text{t}}$, respectively, the indicates a TEM\cite{faber_gyrokinetic_2015}, while an inefficient drive from $n_{\text{t}}$ and $T_{\text{t}}$ may point to a UI\cite{costello_universal_2023}.
	
	In cases where the phase picture is inconclusive, the magnetic geometry can be artificially altered to remove the particle trapping by setting $B_0(z)=\bar{B}_0$ and $\partial B_0/\partial z=0$. If the shape of $\mathrm{\Phi}$ and the eigenvalues are not substantially different between the physical magnetic geometry and the constant-$B$ geometry, then the mode is consistent with a UI. Further, a toroidal UI and a slab UI can be distinguished from one another by zeroing the curvature $\mathcal{K}^x=\mathcal{K}^y=0$ to produce a slab-like geometry. Again, if $\mathrm{\Phi}$, $\gamma$, and $\omega$ do not change substantially from the real geometry to the slab-like geometry, then the mode is consistent with a slab UI. If $\mathrm{\Phi}$ does not change from the real geometry to the constant-$B$ geometry but does in the slab-like geometry, then the mode is consistent with a toroidal UI. However, a mode may be a hybrid\cite{kammerer_exceptional_2008,pueschel_gyrokinetic_2008} TEM-UI\cite{helander_universal_2015}, with cross phases of a typical TEM and mode localization consistent with UI or both UI and TEM. Also, it should be noted that by removing the non-resonant trapped particles, growth rates of UIs can increase in the constant-$B$ geometry compared to the real geometry \cite{costello_universal_2023}. The change in growth rates can cause one branch of UIs to become the dominant instability at a certain $k_y$ in the constant-$B$ or slab-like geometries that was subdominant in the physical geometry. The branch change can be observed by a jump in frequency. In this case, comparing the mode structure at this $k_y$ with $\mathrm{\Phi}$ at a nearby $k_y$ on the same branch can help to identify the type of mode. A summary of the results of the linear physics and mode identification procedure detailed is given next, with a more detailed analysis given in Appendix~\ref{sec:lin_mode_appx}.
	
	\subsection{Density Gradient Drive}
\begin{figure}[!b]
		\centering
		\includegraphics[width=0.5\textwidth]{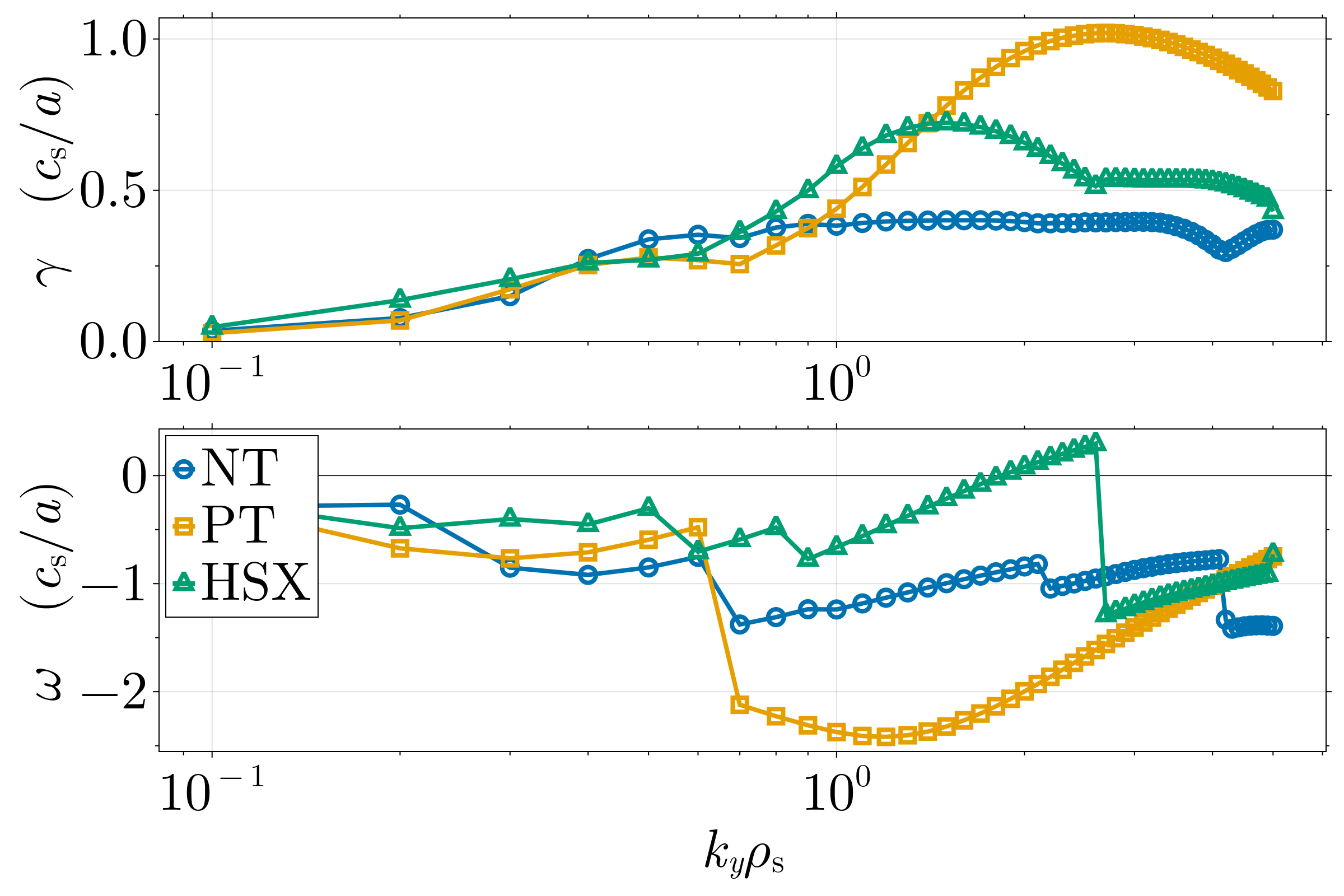}
		\caption{Growth rates (top) and frequencies (bottom) of the fastest-growing modes with $k_x=0$ for the NT (blue) and PT (orange) reduced-TEM configurations compared to HSX (green) for $\omega_n=4$, $\omega_{T\text{e}}=0$. The frequencies are in the electron diamagnetic direction, except for $1.9\le k_y\le2.6$ in HSX. The growth rates at low $k_y$, which drive the most turbulent transport, are similar for each configuration.}
		\label{fig:omn4_gamma_omega}
	\end{figure}

	In the scenario with $\omega_n=4$ and $\omega_{T\text{e}}=0$, the growth rates and frequencies are compared for the reduced-TEM configurations and HSX in Fig.~\ref{fig:omn4_gamma_omega}. The signs of the real frequencies are in the electron diamagnetic direction, which means the dominant instabilities were electron-driven. The modes propagating in the ion direction at $1.9\le k_y\le2.6$ in HSX hare not the focus of the present work \cite{faber_gyrokinetic_2015,plunk_collisionless_2017}. Because large-scale structures are more efficient at driving turbulent transport, the focus is mainly on the low-$k_y$ modes. At low $k_y$, each of the configurations has similar growth rates.
	
	The dominant modes are consistent with TEMs in HSX. For $k_y>1$, the density cross phases of the trapped electrone population indicates a substantial contribution to the instability drive. While the fastest growing modes for $k_y\le 1$ have cross phases that do not conclusively point to TEMs, the parallel structure of $\mathrm{\Phi}$ aligns with regions of particle trapping and destabilizing curvature. With this information, the fastest-growing modes in HSX with $\omega_n=4$ and $\omega_{T\text{e}}=0$ are consistent with TEMs.
	
	In the reduced-TEM NT configuration, the dominant modes are consistent with toroidal UIs for $k_y\le0.7$ and hybridized TEM-UIss for $k_y\ge0.8$ and have stronger TEM characteristics at higher $k_y$. The cross phases for $\mathrm{\Phi}$ and $n_\text{t}$ in modes with $k_y\gtrsim1.5$ shows a substantial contribution to the instability drive, which is consistent with TEMs. However, removing particle trapping from the magnetic geometry reveals UI-like characteristics for $k_y\ge0.8$. For $k_y\le0.7$, removing the particle trapping did not substantially change $\mathrm{\Phi}$, but removing curvature did. Similarly, growth rates slightly increased when trapped particles were removed, possibly due to the removal of stabilizing nonresonant electrons\cite{costello_universal_2023,helander_universal_2015}. The frequency trends indicate that the dominant branch of UI can change at a given $k_y$, but the mode still remains toroidal in nature. Combined with the fact that the cross phases of both trapped and passing electrons indicate inefficient drive, the mode response to varying the geometry leads to the conclusion that for $k_y\le0.7$, the most unstable modes in the NT configuration in the density gradient drive scenario are UIs.
	
	In the reduced-TEM PT configuration, the fastest growing modes were consistent with slab UIs for $k_y\le0.6$ and TEMs for $k_y\ge0.7$. While the cross phases indicated that the $n_\text{t}$ channel is not an efficient mechanism, removing particle trapping in the geometry showed that for $k_y\ge0.7$, $\mathrm{\Phi}(z)$ changed substantially, along with the growth rates and frequencies of the dominant mode. In contrast, for $k_y\le0.6$, $\abs{\mathrm{\Phi}}(z)$, $\gamma$, and $\omega$ did not substantially change when the trapping and curvature were suppressed in the magnetic geometry. Together, the cross phases, mode structure, and eigenvalues of the fastest growing modes are consistent with slab UIs dominating for $k_y\le0.6$ and TEMs for $k_y\ge0.7$.
	
	\subsection{Electron Temperature Gradient Drive}
\begin{figure}[h]
		\centering
		\includegraphics[width=0.5\textwidth]{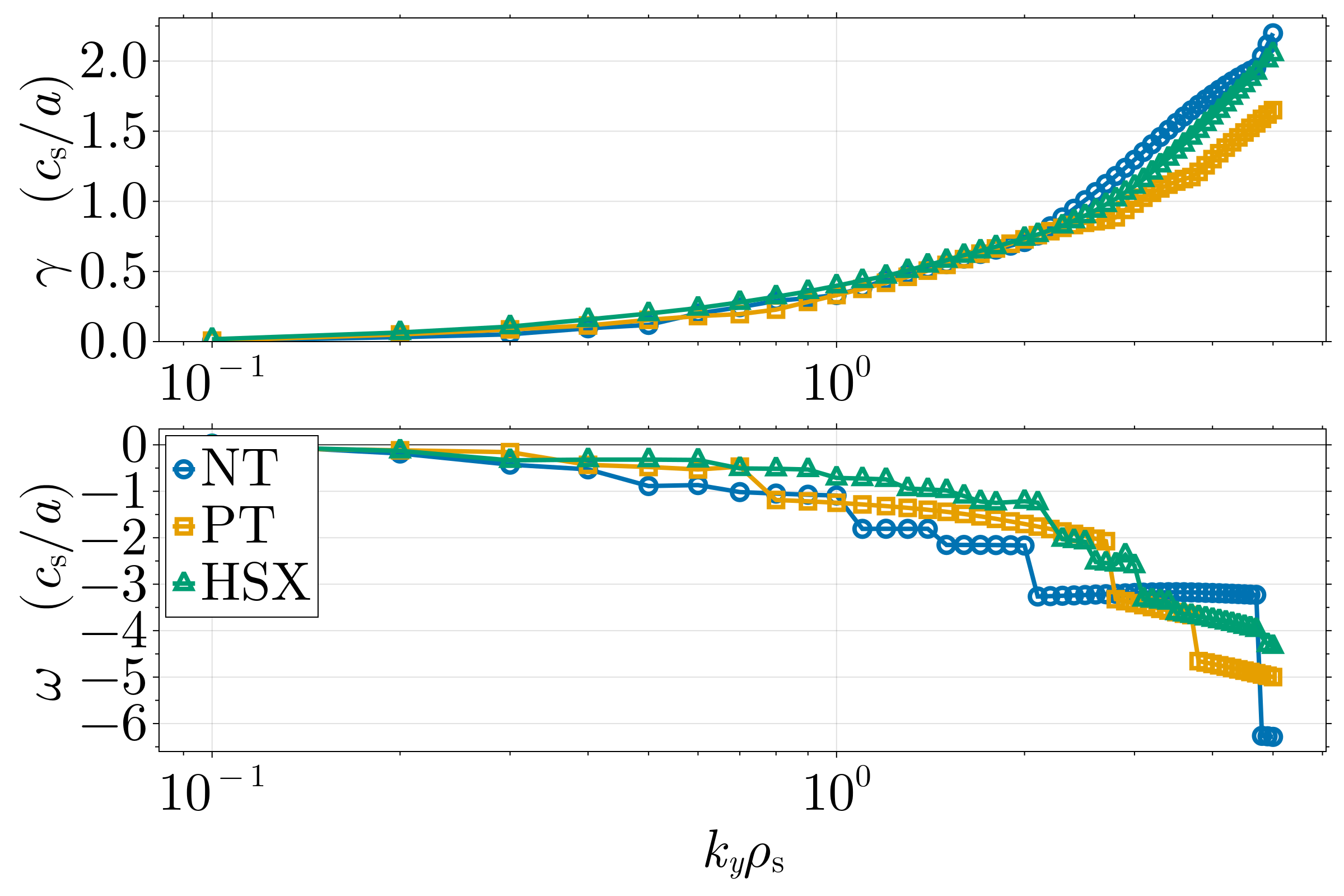}
		\caption{Growth rates (top) and frequencies (bottom) of the fastest growing modes with $k_x=0$ for the NT (blue) and PT (orange) reduced-TEM configurations compared to HSX (green) for $\omega_n=0$, $\omega_{T\text{e}}=4$. The frequencies are in the electron diamagnetic direction. The growth rates for ion-scale modes are similar for each configuration.}
		\label{fig:omte4_gamma_omega}
	\end{figure}

	For the scenario with $\omega_n=0$ and $\omega_{T\text{e}}=4$, the growth rates and frequencies for the NT, PT, and HSX configurations are shown in Fig.~\ref{fig:omte4_gamma_omega}. The signs of the real frequencies were in the electron diamagnetic direction. At all $k_y$, each of the configurations had similar growth rates. These modes may be either $\nabla T$-driven TEMs or electron temperature gradient modes (ETGs).
	
	For this scenario, if either parallel or perpendicular temperature fluctuations for passing electrons $T_{\parallel,\text{p}}$ or $T_{\perp,\text{p}}$, respectively, are efficient drives, then the cross phases are consistent with an ETG. If either of the trapped electron temperature cross phases are efficient, then the mode is a TEM. Commonly, at intermediate $k_y$, TEM-ETG hybrids are observed for $\omega_{T\text{e}}$ drive\cite{pueschel_multi-scale_2020}.
	
	In HSX and both reduced-TEM configurations, the dominant mode at every $k_y$ investigated is consistent with an ETG except for $k_y=0.1$ in the NT configuration. Consistently, $T_{\parallel,\text{p}}$ is the most efficient drive, with the noted exception of $k_y=0.1$ in the NT configuration, where $T_{\parallel,\text{t}}\approx \pi/2$. An analysis of the parallel structure of $\mathrm{\Phi}$ shows all modes with ETG-like cross phases do not show localization of the electrostatic potential where destabilizing curvature overlapping regions of deeply trapped particles. For the fastest growing mode at $k_y=0.1$ in the NT configuration, $\mathrm{\Phi}$ is localized to those regions of deep trapping and destabilizing curvature. 
	
	While only the most unstable mode at a given $k_y$ was computed, there may exist unstable subdominant modes. Such subdominant modes could be a mix of TEMs and UIs in the density gradient scenario or TEMs and ETGs in the electron temperature gradient scenario. It is important to keep these subdominant modes in mind, as it has been shown that subdominant modes can play a major role in setting turbulent amplitudes in stellarators \cite{pueschel_stellarator_2016,faber_stellarator_2018,mulholland_enhanced_2023}. Nonetheless, targeting available energy in an optimization was successful in substantially reducing $\nabla n$- and $\nabla T_\text{e}$-TEM instability at low $k_y$.
	
	\section{Nonlinear Physics}
	\label{sec:NL}
	Studying the properties of linear eigenmodes can be insightful; however, the nonlinear properties are not guaranteed to be the same as the linear characteristics. Nonlinear simulations are performed for the scenario $\omega_n=4$, $\omega_{T\text{e}}=0$ with converged grid resolutions. For NT, the resolution is $N_x\times N_{k_y}\times N_z\times N_{v_\parallel}\times N_\mu = 128\times 64\times 128\times32\times 8$ with $n_\text{pol}=4$, where $N_{k_y}$ is the number of binormal wavenumbers used. The minimum binormal wavenumber is $k_y^{\text{min}}=0.1$, and radial box size $L_x=56.32$. The PT case has a resolution of $N_x\times N_{k_y}\times N_z\times N_{v_\parallel}\times N_\mu = 256\times 128\times 128\times32\times 8$ with $n_\text{pol}=4$, $k_y^{\text{min}}=0.05$, and $L_x=124$. The nonlinear simulations are collisionless, have no ion temperature gradient, use a hydrogen mass ratio $m_\text{i}/m_\text{e} = 1836$, have a background temperature ratio $T_{0\text{i}}/T_{0\text{e}}=1$, have $\beta=10^{-4}$, and hyperdiffusivities $\varepsilon_z=4$ and $\varepsilon_{v}=0.2$. Nonlinear HSX data is taken from Ref.~\onlinecite{faber_stellarator_2018}. All nonlinear data in this section time-averaged over the quasistationary state.
	
\begin{figure}[!htp]
		\centering
		\begin{subfigure}[b]{0.49\textwidth}
			\centering
			\includegraphics[width=\textwidth]{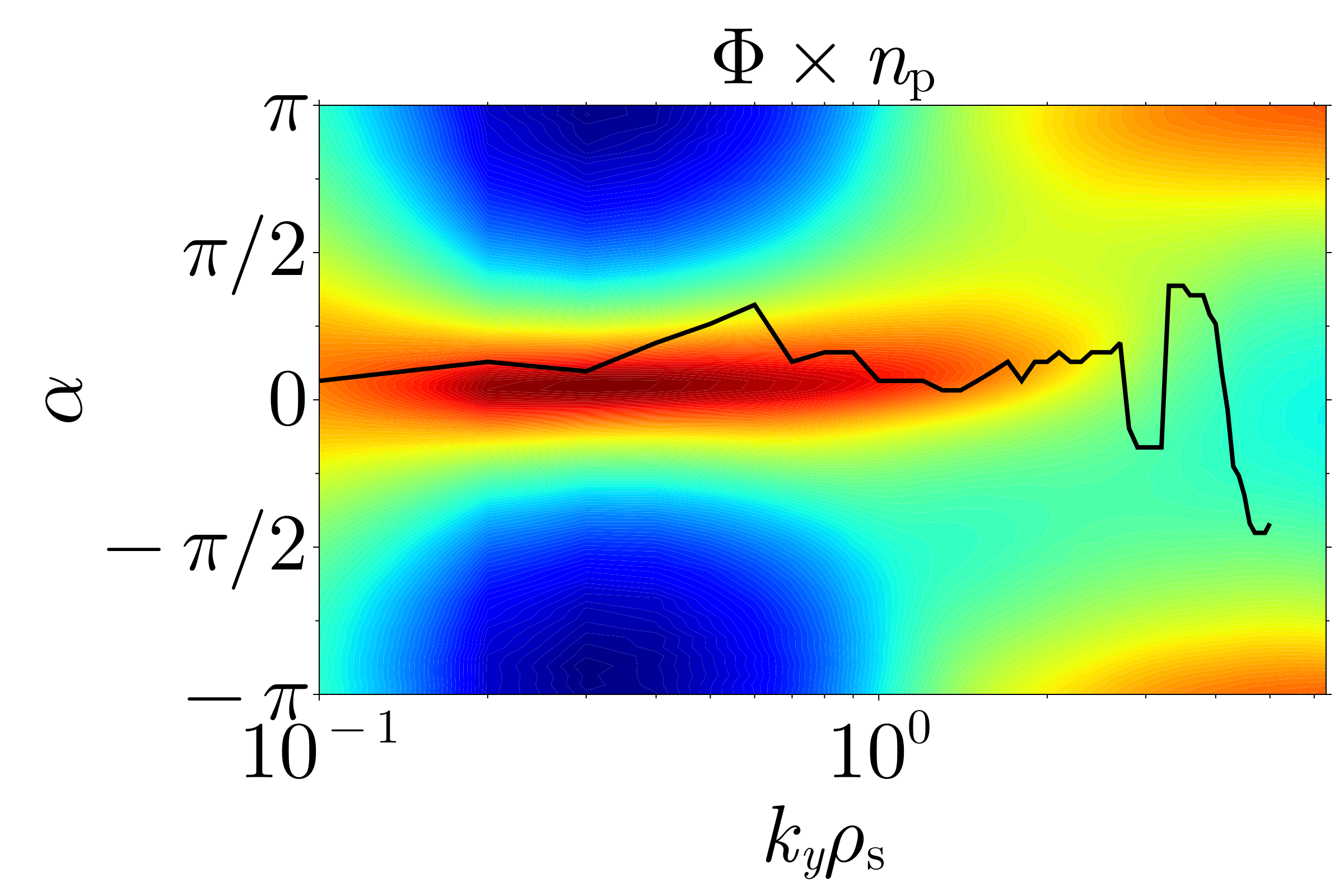}
			\caption{}
			\label{fig:NT_dens_pass}
		\end{subfigure}
		\begin{subfigure}[b]{0.49\textwidth}
			\centering
			\includegraphics[width=\textwidth]{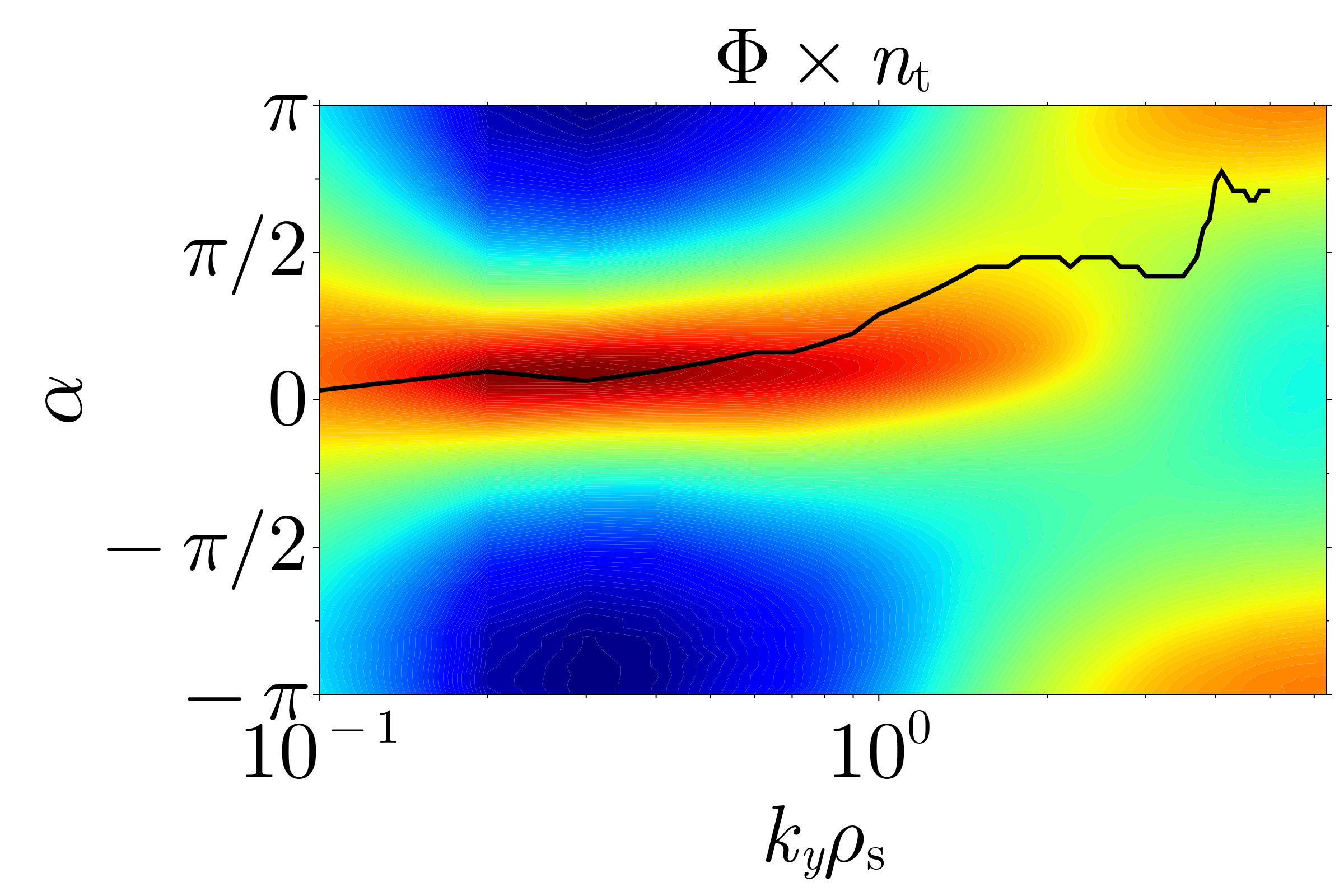}
			\caption{}
			\label{fig:NT_dens_trap}
		\end{subfigure}\\
		\begin{subfigure}[b]{0.49\textwidth}
			\centering
			\includegraphics[width=\textwidth]{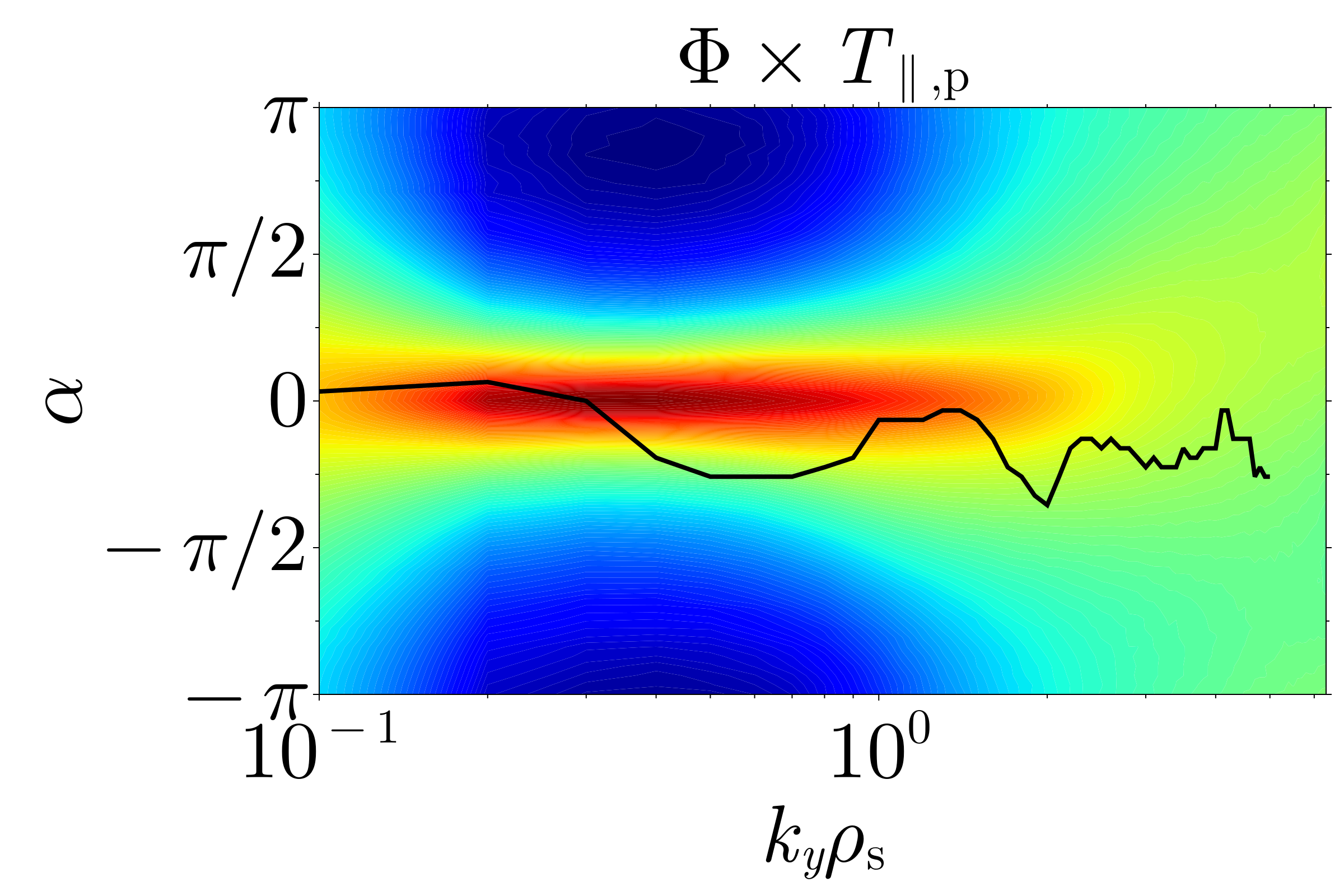}
			\caption{}
			\label{fig:NT_T_par_pass}
		\end{subfigure}
		\begin{subfigure}[b]{0.49\textwidth}
			\centering
			\includegraphics[width=\textwidth]{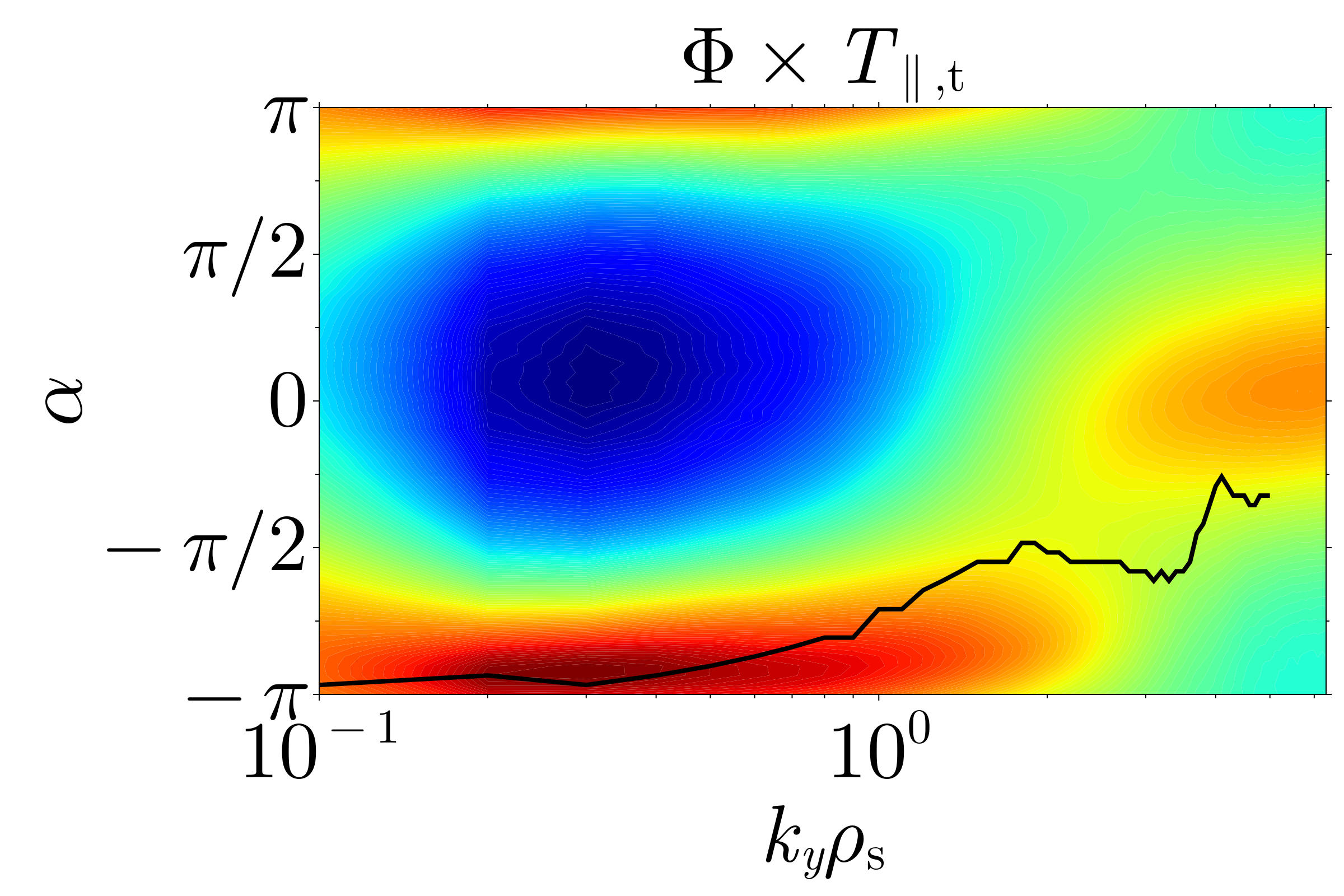}
			\caption{}
			\label{fig:NT_T_par_trap}
		\end{subfigure}\\
		\begin{subfigure}[b]{0.49\textwidth}
			\centering
			\includegraphics[width=\textwidth]{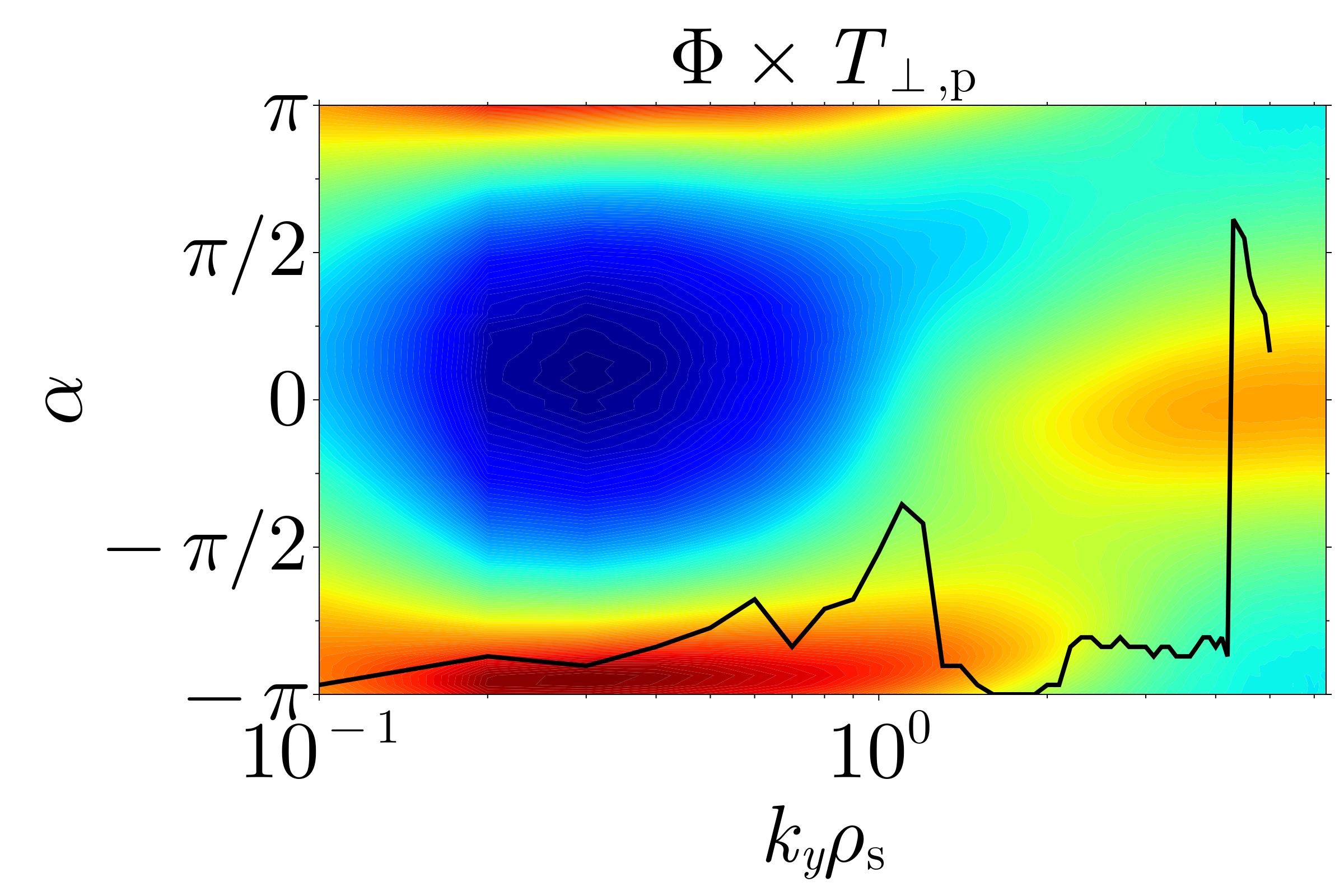}
			\caption{}
			\label{fig:NT_T_perp_pass}
		\end{subfigure}
		\begin{subfigure}[b]{0.49\textwidth}
			\centering
			\includegraphics[width=\textwidth]{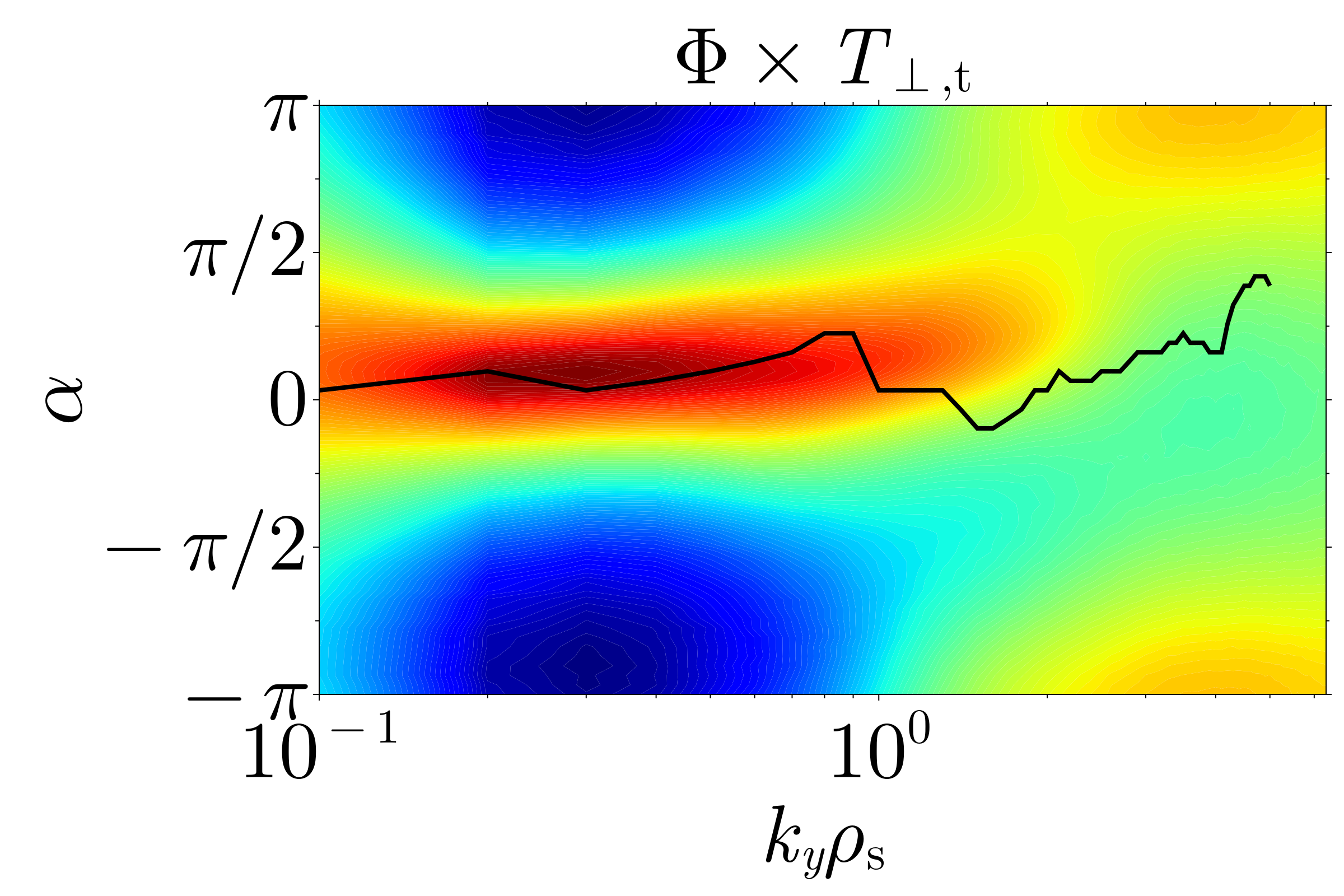}
			\caption{}
			\label{fig:NT_T_perp_trap}
		\end{subfigure}
		\caption{Histograms of nonlinear cross phases for the NT equilibrium (logarithmic color scale) and linear cross phases (black line) of the fastest growing modes versus $k_y$ with $\omega_n=4$, $\omega_{T\text{e}}=0$. Linear cross phases are in good agreement with nonlinear cross phases, showing the nonlinear heat flux is driven by UIs: (a) $\mathrm{\Phi}\times n_\text{p}$, (b) $\mathrm{\Phi}\times n_\text{t}$, (c) $\mathrm{\Phi}\times T_{\parallel,\text{p}}$, (d) $\mathrm{\Phi}\times T_{\parallel,\text{t}}$, (e) $\mathrm{\Phi}\times T_{\perp,\text{p}}$, (f) $\mathrm{\Phi}\times T_{\perp,\text{t}}$.}
		\label{fig:NT_cross_NL}
	\end{figure}

	To determine the role of UIs in driving turbulence in the NT and PT geometries when there is only density gradient drive, the nonlinear cross phases were computed and compared to the dominant linear cross phases. The cross phases for the NT geometry are shown in Fig.~\ref{fig:NT_cross_NL} as a function of $k_y$, with a logarithmic color scale representing a histogram of the nonlinear cross phases at every grid point in $x$-$z$ space and the black line representing the linear cross phases. The nonlinear cross phases are similar to the UI signatures where the trapped and passing densities had similar cross phases. Additionally, the linear cross phases closely lay on top of the peaks in nonlinear cross phases at low $k_y$, where UIs were the dominant instability. The differences between the linear and nonlinear cross phases at $0.4\le k_y\le0.7$ are suggestive of either clusters of subdominant UIs or UIs at finite $k_x$ playing a larger role in driving turbulence than the fastest growing linear modes at those wavenumbers.
	
\begin{figure}[!htp]
		\centering
		\begin{subfigure}[b]{0.49\textwidth}
			\centering
			\includegraphics[width=\textwidth]{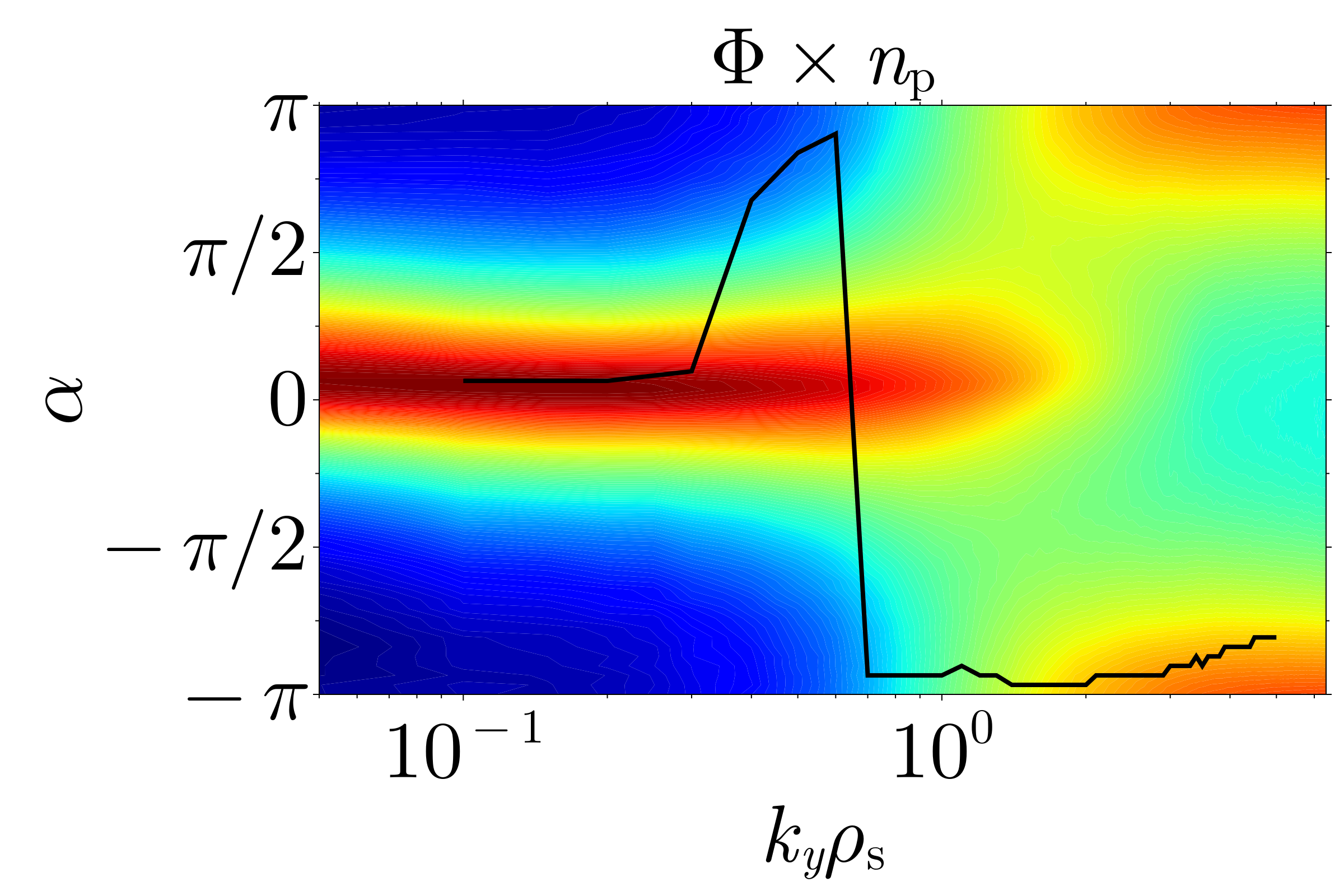}
			\label{fig:PT_dens_pass}
		\end{subfigure}
		\begin{subfigure}[b]{0.49\textwidth}
			\centering
			\includegraphics[width=\textwidth]{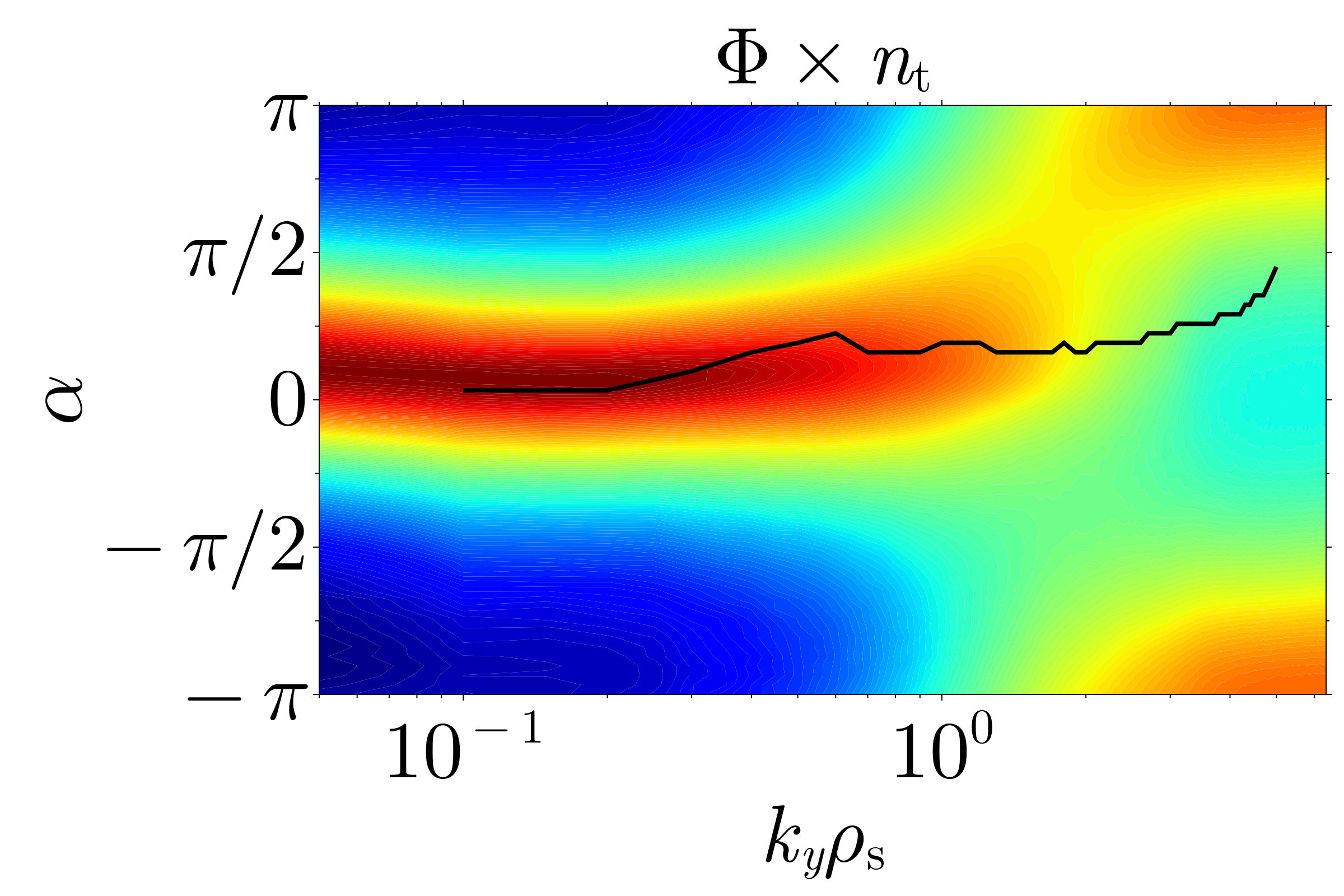}
			\label{fig:PT_dens_trap}
		\end{subfigure}\\
		\begin{subfigure}[b]{0.49\textwidth}
			\centering
			\includegraphics[width=\textwidth]{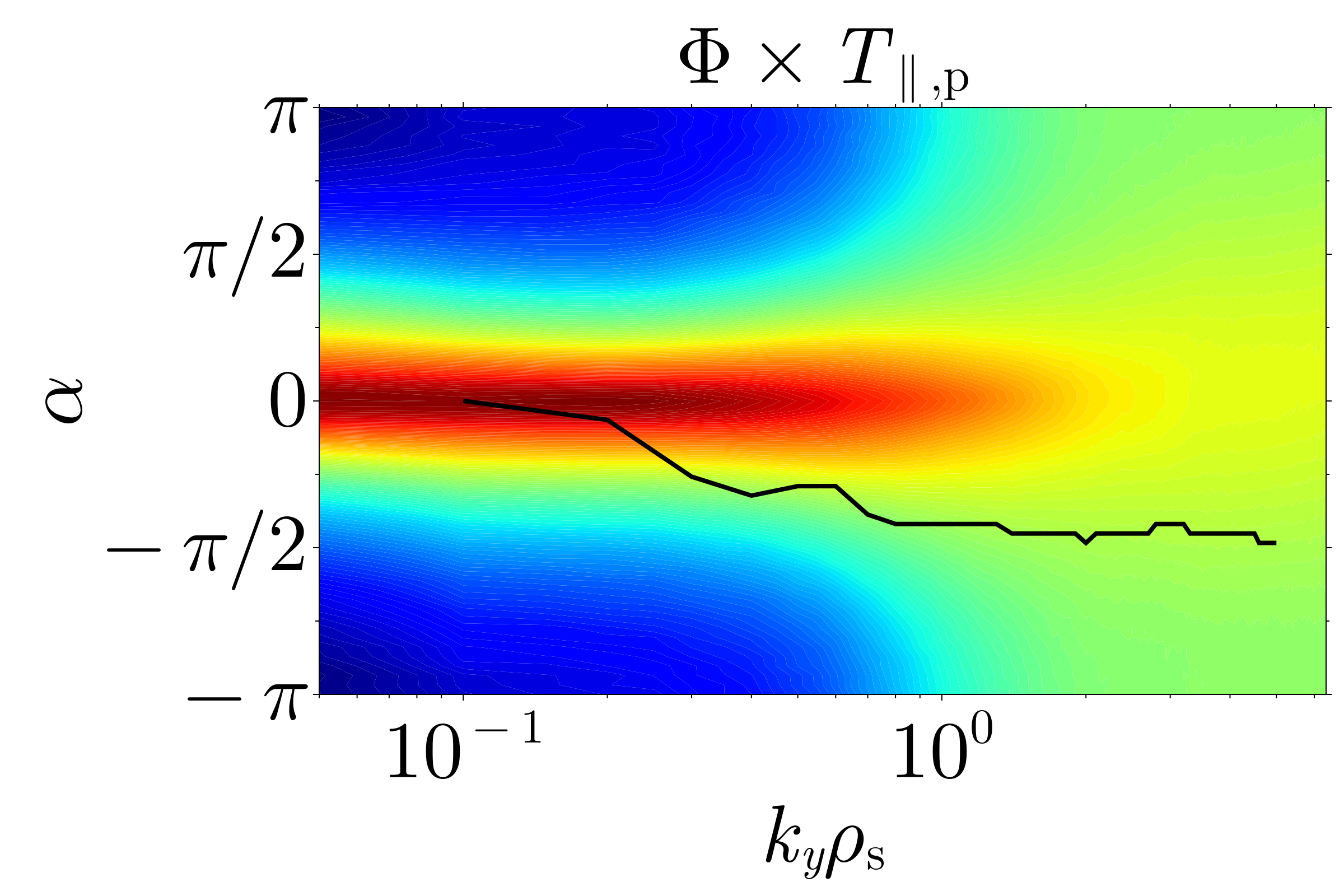}
			\label{fig:PT_T_par_pass}
		\end{subfigure}
		\begin{subfigure}[b]{0.49\textwidth}
			\centering
			\includegraphics[width=\textwidth]{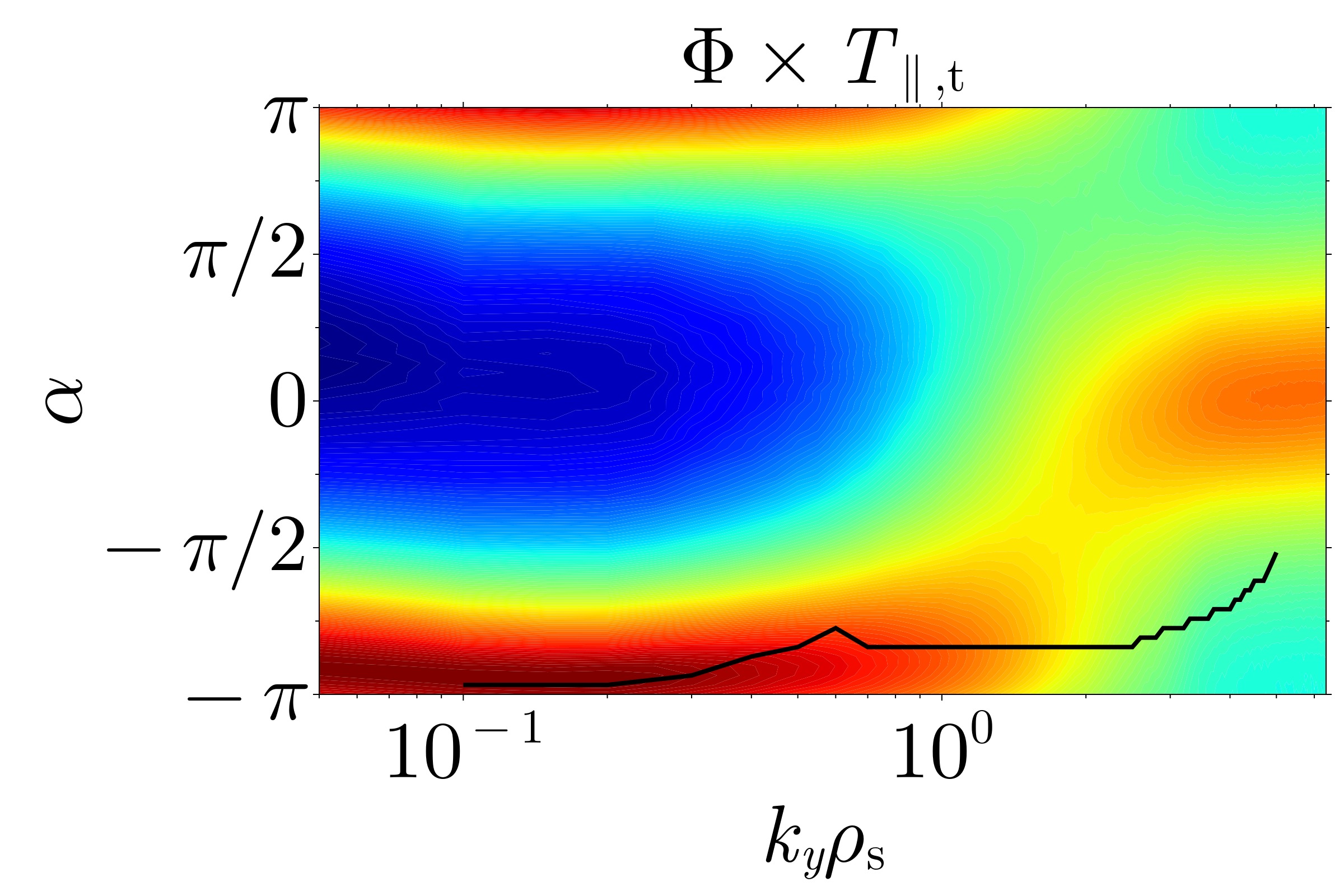}
			\label{fig:PT_T_par_trap}
		\end{subfigure}\\
		\begin{subfigure}[b]{0.49\textwidth}
			\centering
			\includegraphics[width=\textwidth]{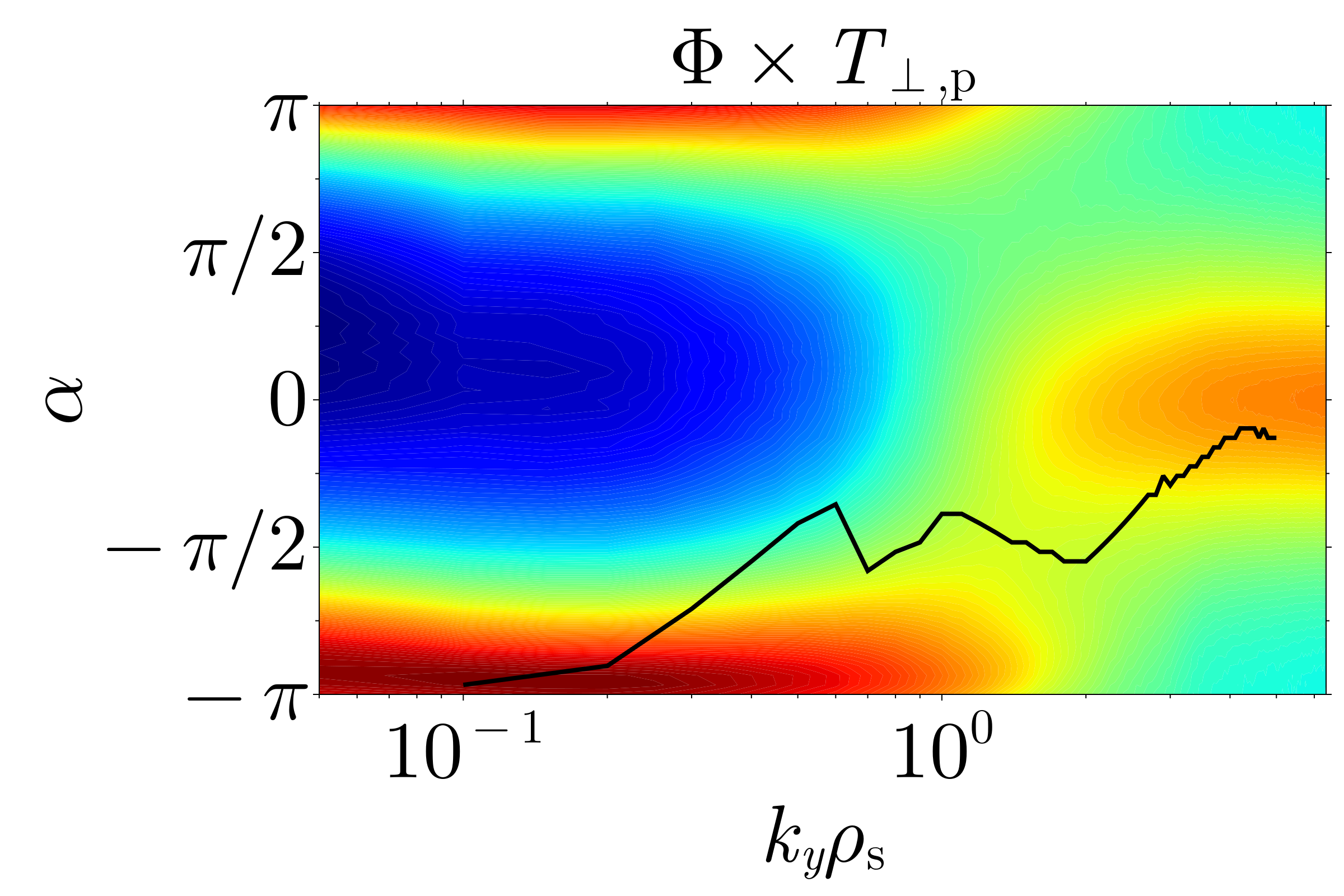}
			\label{fig:PT_T_perp_pass}
		\end{subfigure}
		\begin{subfigure}[b]{0.49\textwidth}
			\centering
			\includegraphics[width=\textwidth]{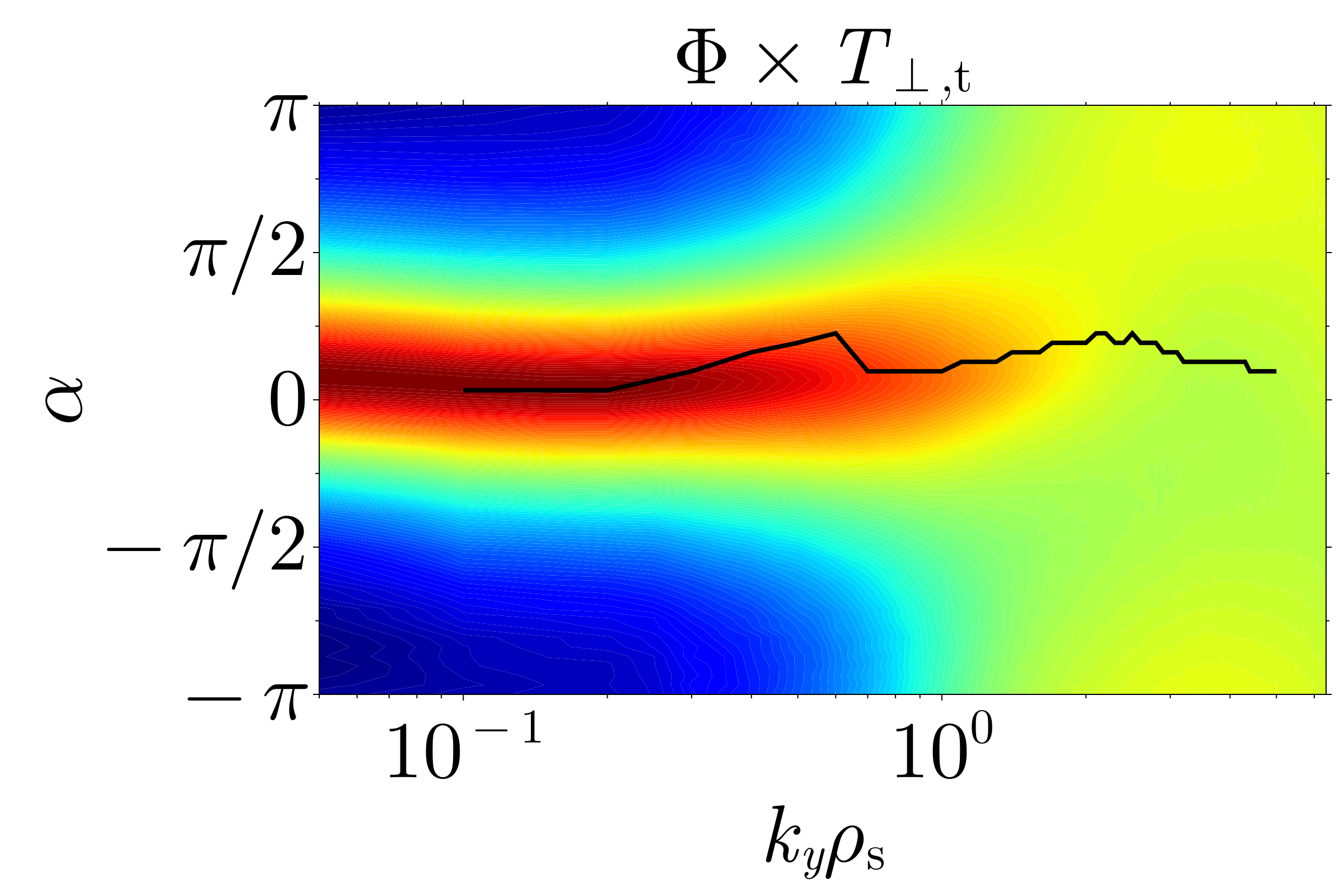}
			\label{fig:PT_T_perp_trap}
		\end{subfigure}
		\caption{Histograms of nonlinear cross phases for the PT equilibrium (logarithmic color scale) and linear cross phases (black line) of the fastest growing modes versus $k_y$ with $\omega_n=4$, $\omega_{T\text{e}}=0$. Linear cross phases are in mostly good agreement with nonlinear cross phases, showing the nonlinear heat flux is driven by UIs: (a) $\mathrm{\Phi}\times n_\text{p}$, (b) $\mathrm{\Phi}\times n_\text{t}$, (c) $\mathrm{\Phi}\times T_{\parallel,\text{p}}$, (d) $\mathrm{\Phi}\times T_{\parallel,\text{t}}$, (e) $\mathrm{\Phi}\times T_{\perp,\text{p}}$, (f) $\mathrm{\Phi}\times T_{\perp,\text{t}}$.}
		\label{fig:PT_cross_NL}
	\end{figure}

	In Fig.~\ref{fig:PT_cross_NL}, the cross phases for the PT geometry are shown as function of $k_y$. Again, the nonlinear cross phases were similar to the UI signatures where the trapped and passing densities had similar cross phases. The linear cross phases also closely lay on top of the peaks in nonlinear cross phases at low $k_y$, where UIs were the dominant instability. Like with the NT case, differences between the linear and nonlinear cross phases at $0.4\le k_y\le0.6$ may point to subdominant or finite $k_x$ UIs action.
	
\begin{figure}[!b]
		\centering
		\includegraphics[width=0.5\textwidth]{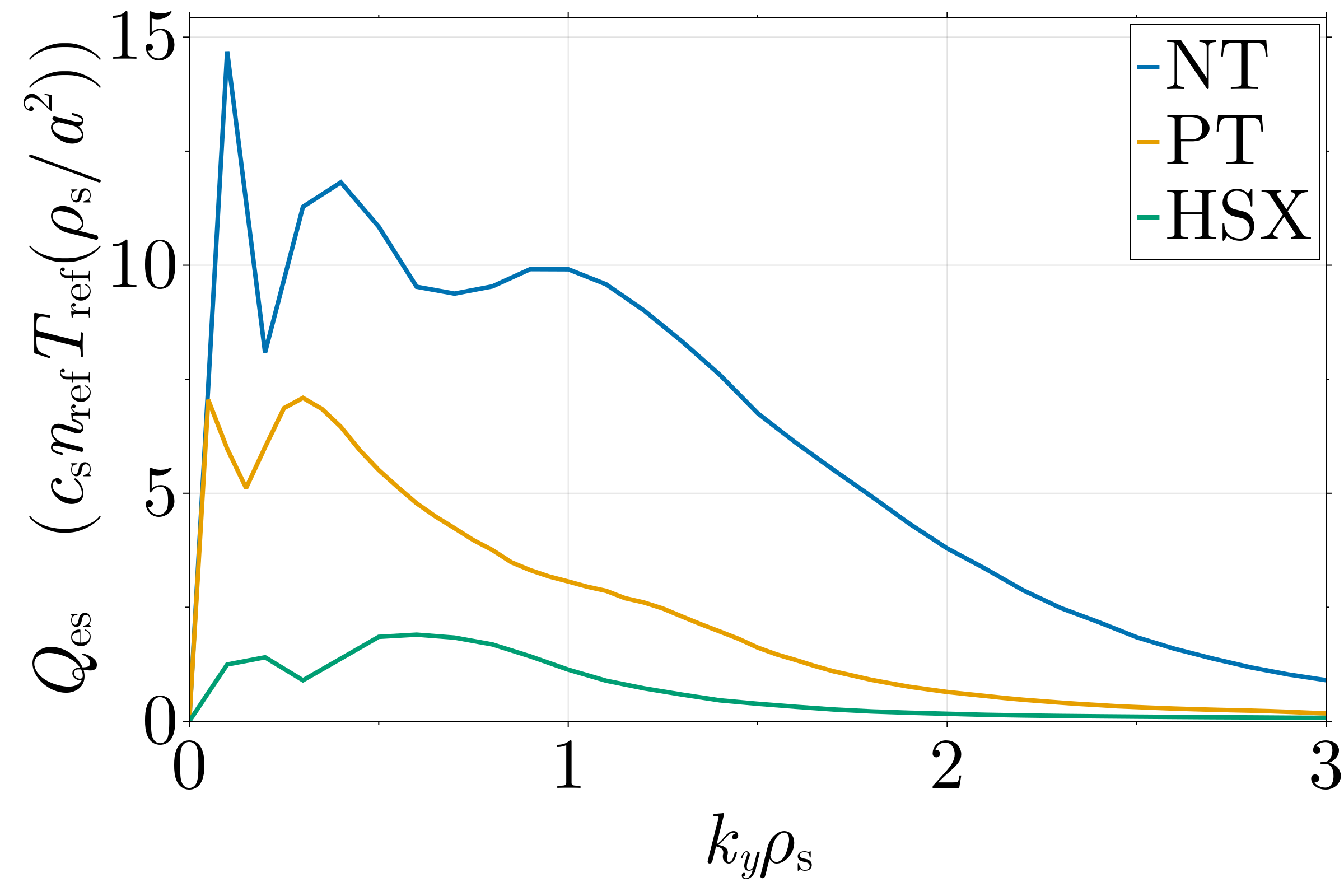}
		\caption{Nonlinear electron electrostatic heat flux with $\omega_n=4$ and $\omega_{T\text{e}}=0$ as a function of $k_y$ for NT (blue), PT (orange), and HSX (green). For NT, UIs are the dominant mode for $k_y\le0.7$ and TEM-UI hybrid modes at larger $k_y$. UIs are the dominant mode for $k_y\le0.6$ in PT. In HSX, TEMs drive the heat flux, which peaks at larger $k_y$ than the NT and PT cases.}
		\label{fig:Qes_ky_omn4}
	\end{figure}

	In all cases in Fig.~\ref{fig:Qes_ky_omn4}, flux spectra are very broad, with substantial flux at scales $k_y\le1$ and multi-peaked. The NT equilibrium had the largest heat flux. The peaks at $k_y=0.1$ and 0.4 were where UIs were the dominant instability, and the nonlinear cross phases were consistent with UIs at these wavenumbers. The most unstable mode at $k_y=1$ was on the TEM-UI hybrid branch but had more UI characteristics than TEM. The PT configuration had less heat flux than NT and peaked at $k_y=0.05$ and 0.3, where the dominant instability was a slab UI. The nonlinear cross phases were also consistent with UIs at these wavenumbers. HSX had the lowest heat flux but is driven by TEMs \cite{faber_gyrokinetic_2015}.
	
\begin{figure}[!b]
		\centering
		\includegraphics[width=0.5\textwidth]{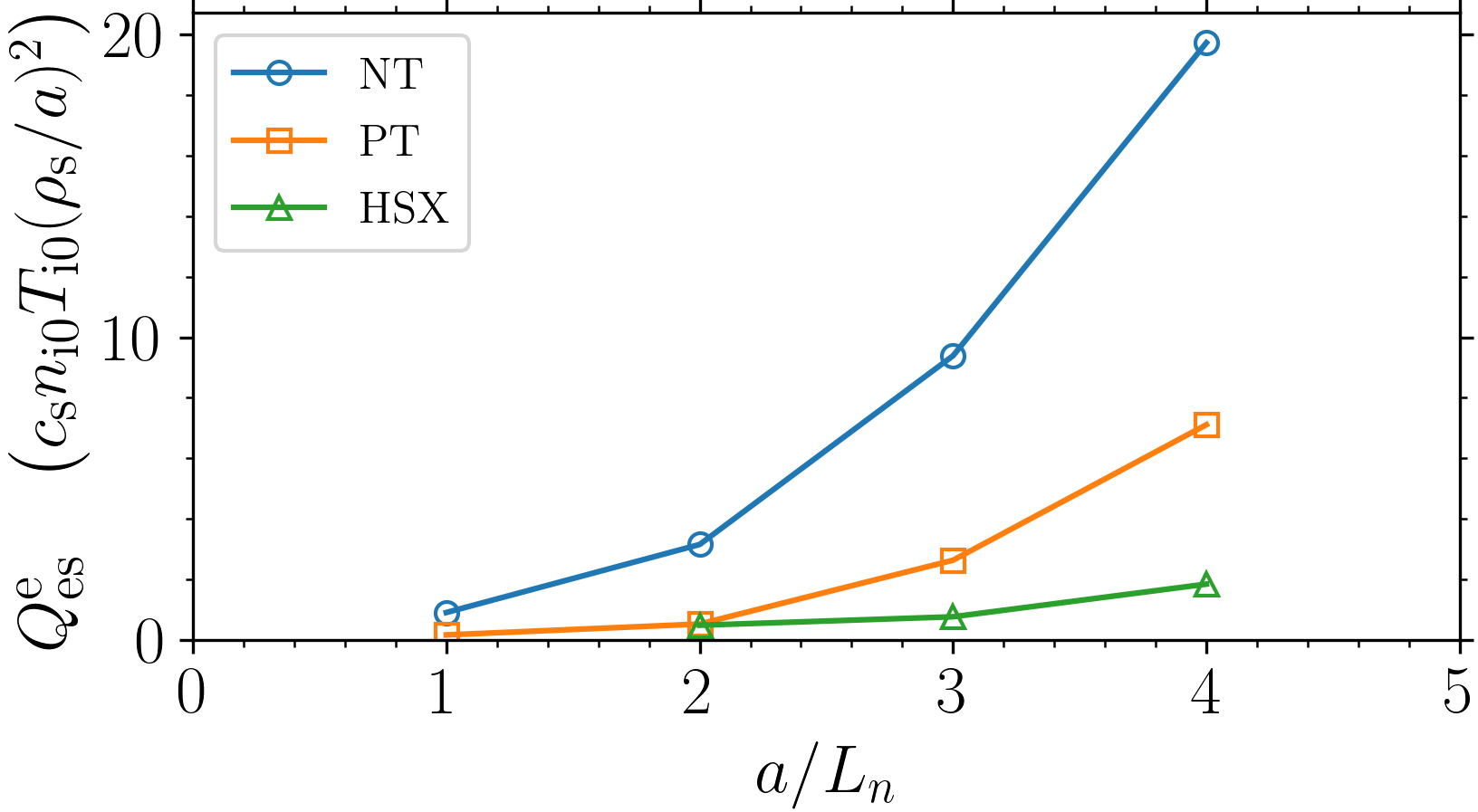}
		\caption{Nonlinear electron electrostatic heat flux with $\omega_{T\text{e}}=0$ as a function of $\omega_n$ for NT (blue), PT (orange), and HSX (green). UIs drive substantial heat flux in the NT and PT cases, while TEMs drive the heat flux in HSX.}
		\label{fig:Qes_omn4}
	\end{figure}

	The total electrostatic heat flux is computed with no electron temperature gradient for a range of density gradients. The electron electrostatic heat flux is plotted at each density gradient for the NT, PT, and HSX geometries in Fig.~\ref{fig:Qes_omn4}. At $\omega_n=4$, the heat flux for the NT configuration, driven by toroidal UIs, was more than twice as large as the slab-UI-driven heat flux in the PT geometry and over an order of magnitude larger than the TEM-driven heat flux in HSX. As the density gradient was decreased, there is a difference in the stiffness of the heat flux in each configuration. The heat fluxes in the NT and PT configurations decreased at a faster rate than HSX. At $\omega_n=2$, the PT configuration had a similar heat flux level to HSX. The heat flux of the NT configuration was similar to that of the PT geometry at $\omega_n=1$.

	When there was no density gradient drive and only electron temperature gradient drive, there were no significant fluctuations observed at scales with $k_y\le2$. The $\omega_n$ scenarios had heat fluxes driven at ion scales $k_y\rho_\text{s}\lesssim1$. The $\omega_{T\text{e}}$ scenarios for the reduced-TEM equilibria had ETGs as the dominant instability and had heat flux driven at length scales smaller than the ion scale. It is important to note that the nonlinear simulations for $\omega_n=0$, $\omega_{T\text{e}}=4$ do not have converged numerical resolutions. Additionally, nonlinear simulations with adiabatic ions for $\omega_n=0$, $\omega_{T\text{e}}=4$ were not substantially different from simulations with kinetic ions.


	\section{Electromagnetic Stabilization of the Universal Instability}
	\label{sec:beta}
\begin{figure}[!b]
		\centering
		\includegraphics[width=0.5\textwidth]{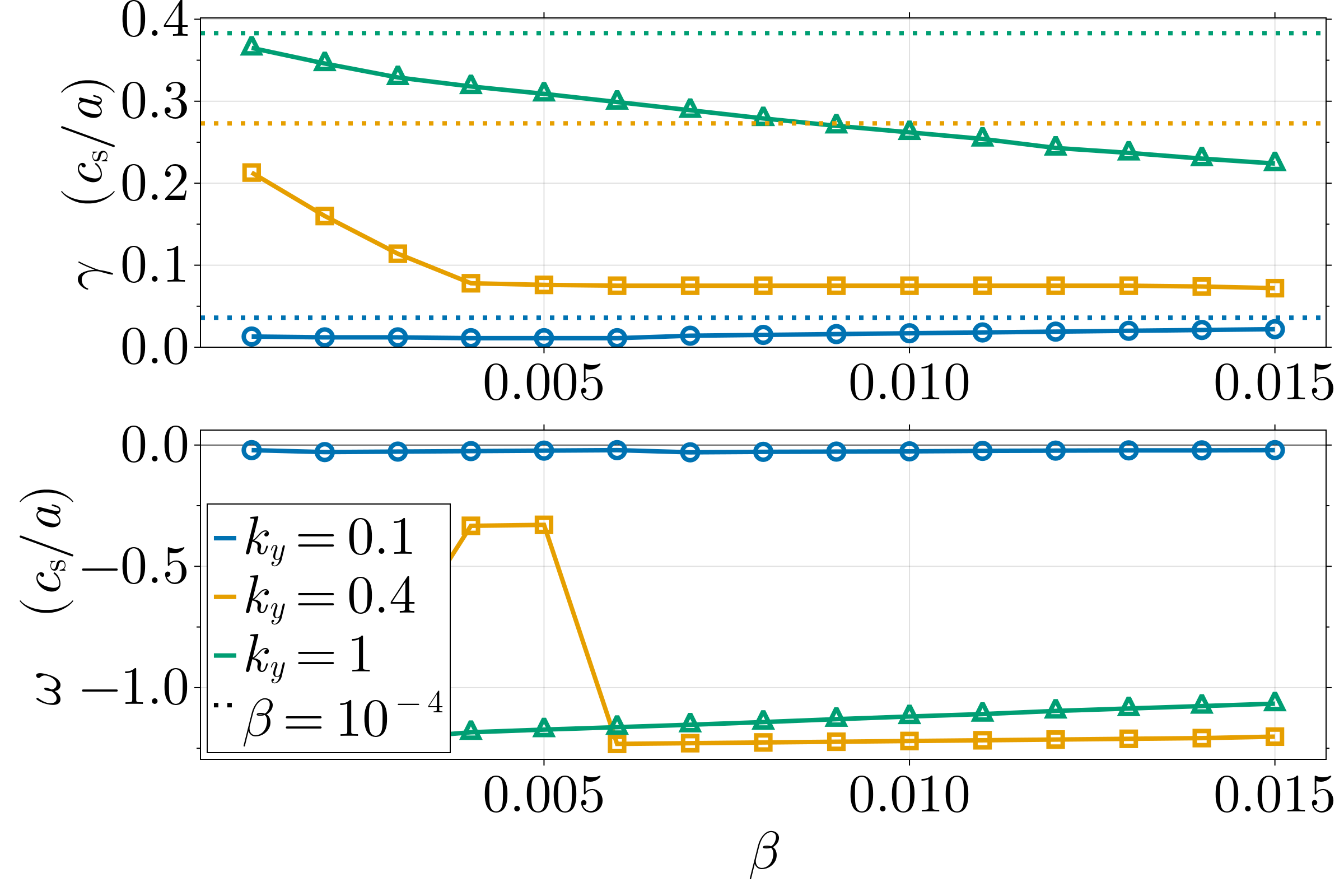}
		\caption{Growth rates (top) and frequencies (bottom) of the fastest growing modes with $k_x=0$ for the NT configuration with $\omega_n=0$, $\omega_{T\text{e}}=4$. The eigenvalues for $k_y=0.1$ (blue), $k_y=0.4$ (orange), and $k_y=1$ (green) are plotted as a function of $\beta$, with the growth rates at $\beta=10^{-4}$ are the horizontal dashed lines. The low-$k_y$ growth rates decrease with increasing $\beta$ until $\beta=4\times10^{-3}$, where they are more than half as large as those at $\beta=10^{-4}$. The growth rate at $k_y=1$ decreases for the entire range of $\beta$.}
		\label{fig:NT_linear_beta_scan}
	\end{figure}

	Because UIs are can be stabilized by electromagnetic effects in slab-like magnetic geometries\cite{mikhailovskaya_drift_1964,huba_finite-_1982,hastings_high-_1982}, this section discusses the effect of $\beta$ on linear stabilization of UIs and the reduction in heat flux in the NT and PT geometries. The linear and nonlinear simulations presented in this section use the same settings as the electrostatic cases---except for $\beta$---and were rechecked for numerical convergence. A density gradient of $\omega_n=4$ and electron temperature gradient of $\omega_{T\text{e}}=0$ is used. An initial-value solver scan over $\beta$ is performed for $k_y=0.1$, 0.4, and 1. The growth rates for the NT configuration, shown in Fig.~\ref{fig:NT_linear_beta_scan}, decrease from the initial $\beta=10^{-4}$. For $k_y=0.1$ and 0.4, the growth rates stopped decreasing at $\beta=4\times10^{-3}$, then begin increasing very slightly with increasing $\beta$. The fastest growing mode at $k_y=1$ is stabilized by increasing $\beta$ over the whole range scanned. The frequencies of these modes remain in the electron diamagnetic direction.

\begin{figure}[!ht]
		\centering
		\includegraphics[width=0.5\textwidth]{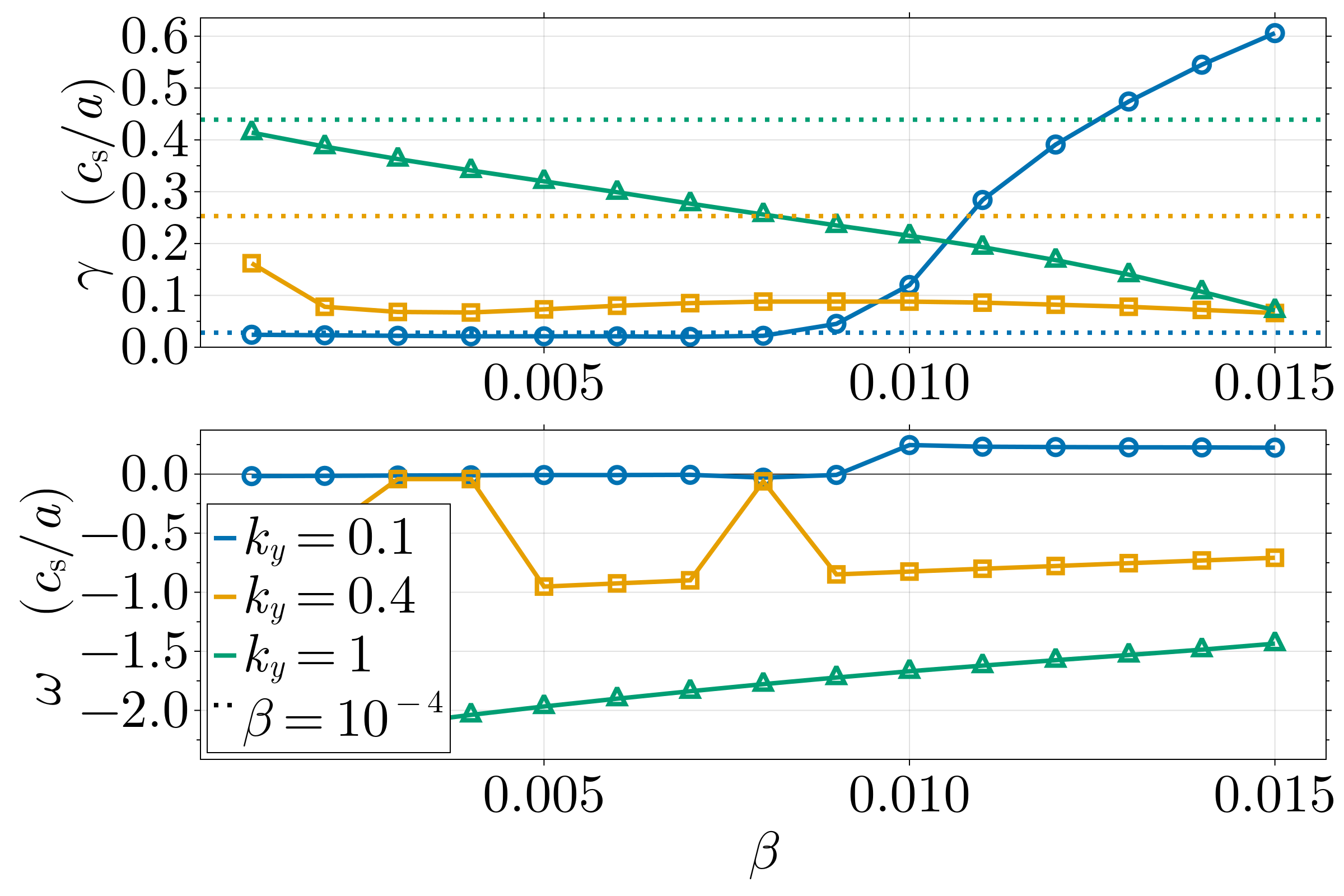}
		\caption{Growth rates (top) and frequencies (bottom) of the fastest growing modes with $k_x=0$ for the PT configuration with $\omega_n=0$, $\omega_{T\text{e}}=4$. The eigenvalues for $k_y=0.1$ (blue), $k_y=0.4$ (orange), and $k_y=1$ (green) are plotted as a function of $\beta$, with the growth rates at $\beta=10^{-4}$ are the horizontal dashed lines. The low-$k_y$ growth rates decrease with increasing $\beta$ until $\beta=4\times10^{-3}$. The mode with $k_y=0.4$ is substantially weakened. The fastest growing mode at $k_y=0.1$ is consistent with a KBM for $\beta\ge10^{-2}$, increasing in growth rate and is in the ion diamagnetic direction. The growth rate at $k_y=1$ decreases for the entire range of $\beta$.}
		\label{fig:PT_linear_beta_scan}
	\end{figure}

	The corresponding scan over $\beta$ for the PT case is shown in Fig.~\ref{fig:PT_linear_beta_scan}. There, the growth rates are reduced from the initial $\beta=10^{-4}$, where the fastest growing mode at $k_y=1$ is monotonically stabilized by increasing $\beta$ over the whole range scanned. For $k_y=0.1$ and 0.4, the growth rates stopped decreasing at $\beta=4\times10^{-3}$, then increase slightly with increasing $\beta$ until $\beta=10^{-2}$. At $\beta=10^{-2}$, the fastest growing mode at $k_y=0.1$ switches to a frequency in the ion diamagnetic direction, with rapidly increasing growth rates, consistent with a kinetic ballooning mode (KBM) \cite{hirose_ion_1995,hirose_kinetic_1996}. KBM-driven turbulence occurs below the ideal MHD ballooning limit\cite{pueschel_transport_2010} where ion drifts help destabilize the mode in the limit of low $k_\perp$ \cite{kotschenreuther_compressibility_1986,aleynikova_quantitative_2017}. Several stellarator studies have shown that KBMs can be destabilized at low $\beta$ and drive or otherwise cause a significant amount of turbulence \cite{aleynikova_kinetic_2018,mckinney_kinetic-ballooning-mode_2021,mulholland_enhanced_2023}. Because electromagnetic modes are not the focus of the present work, nonlinear simulations will be kept well below $\beta=10^{-2}$.
	
\begin{figure}[!b]
		\centering
		\includegraphics[width=0.5\textwidth]{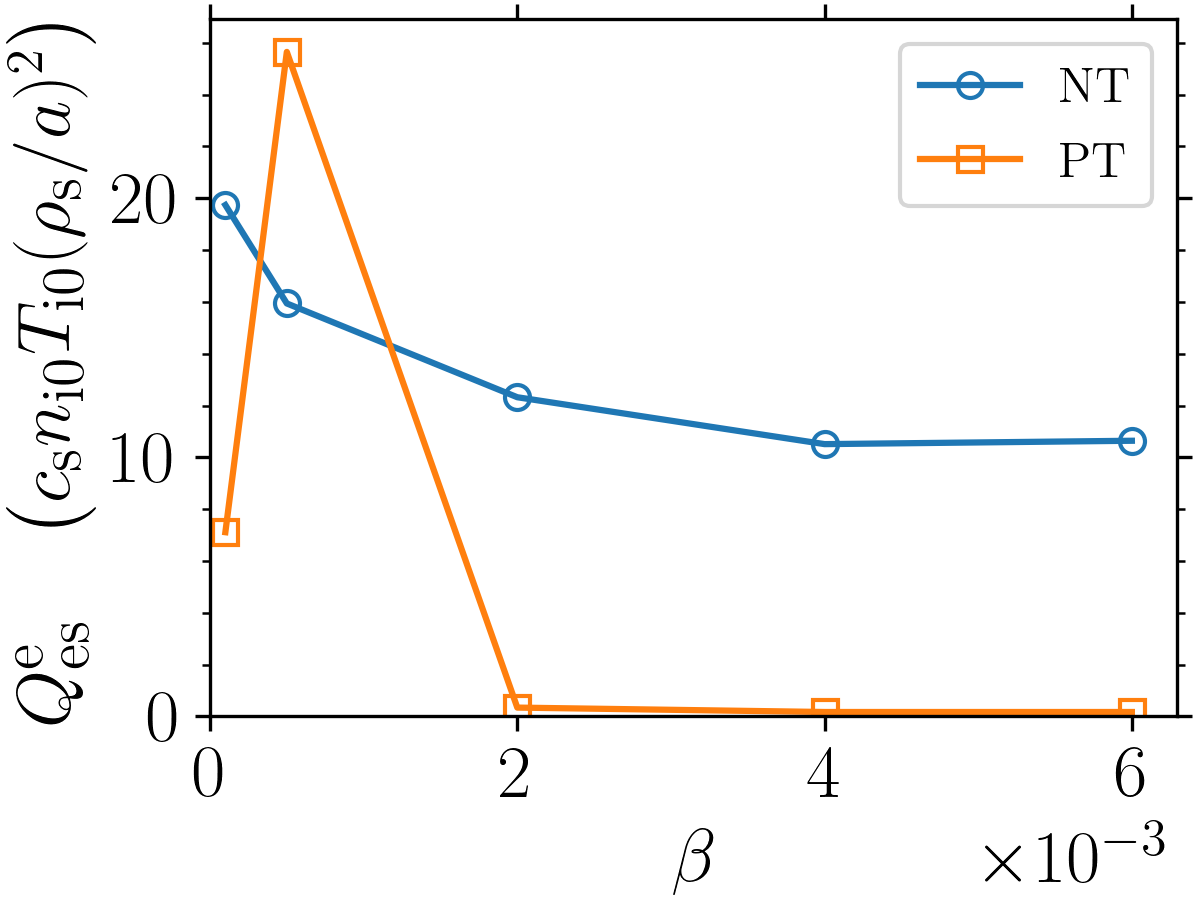}
		\caption{Nonlinear electron electrostatic heat flux with $\omega_{T\text{e}}=0$ as a function of $\beta$ for the NT (blue) and PT (orange) reduced-TEM configurations. Increasing $\beta$ to a fraction of a percent halves the heat flux in the NT geometry and almost completely suppresses the turbulence in the PT geometry. The increased heat flux in the PT configuration with $\beta=5\times10^{-4}$ is not well understood. Electromagnetic heat flux was at least an order of magnitude lower than electrostatic heat flux for all cases.}
		\label{fig:Qes_beta}
	\end{figure}

	Nonlinear heat flux as a function of $\beta$ for the two reduced-TEM configurations is presented in Fig.~\ref{fig:Qes_beta}. With an increase to $\beta=4\times10^{-3}$, the heat flux for the NT geometry is halved, while the PT geometry sees nearly complete suppression at $\beta=2\times10^{-3}$. The decease in heat fluxes is consistent with the stabilization of the dominant UI instabilities at low $k_y$. While the turbulence in the NT equilibrium was not completely suppressed, the linear stabilization and large reduction of the heat flux is evidence that toroidal UIs are also stabilized by $\beta$, despite current theory only covering slab-like UIs\cite{mikhailovskaya_drift_1964,huba_finite-_1982,hastings_high-_1982}. Although slab UIs in the PT equilibrium were monotonically stabilized by $\beta$, there was a sharp increase in heat flux at $\beta=5\times10^{-4}$. This datum is not well understood, but the phenomenon is numerically resolved, potentially suggesting a nonlinear effect as the cause. Nevertheless, increasing $\beta$ to $\mathcal{O}(10^{-3})$ results in a dramatic decrease in electrostatic heat flux in these UI-dominated scenarios. Additionally, the electromagnetic heat flux remained at least an order of magnitude lower than the electrostatic heat flux at each $\beta$.
	
	\section{Conclusions}
	\label{sec:concl}
	In this work, TEM-driven turbulence was suppressed in two 3D equilibria via optimization. A highly nonconvex optimization was performed on two 3D local MHD equilibria. These equilibria were optimized for quasihelical symmetry and for TEM turbulence via the available energy metric\cite{mackenbach_available_2022}. Additional penalty functions were included to keep the rotational transform away from low-order rational surfaces, the global shear low, the aspect ratio not too large, the flux-surface-averaged parallel current near zero, and to keep the flux surface shape amenable to practical numerical resolution requirements. One initial equilibrium was a helically rotating negative triangular shape and the other initial point had helically rotating positive triangularity. The optimized negative triangularity (NT) and positive triangularity (PT) cases successfully improved quasisymmetry and reduced available energy while keeping the other penalty functions low. While NT and PT initial configurations were chosen for the starting point of the optimization, there is no clear benefit of positive versus negative helical triangularity with respect to TEM stability or turbulence after optimization. The improvement of TEM stability and heat flux is likely a result of other geometry changes introduced by the optimization to achieve improved available energy.
	
	The reduced-TEM equilibria were compared to HSX, which is known to have TEM-driven turbulence. Linear characteristics of the three equilibria with only $\nabla n$ drive and with only $\nabla T_\text{e}$ drive were examined. Linear growth rates at low $k_y$ are similar between the three equilibria for both drive scenarios. However, in the $\nabla n$ case, cross phases, eigenmode structure, and the response of eigenmodes and eigenvalues to artificially removing particle trapping showed that the NT equilibrium was dominated by unstable toroidal UIs and the PT by slab UIs. In the $\nabla T_\text{e}$ case, linear cross phases and eigenmode structures pointed to ETGs.
	
	UIs were shown to drive substantial nonlinear heat flux in the $\nabla n$ scenario. Heat flux spectra showed that modes at low $k_y$ drove significant heat flux in the reduced-TEM equilibria. Nonlinear cross phases were consistent with UIs and agreed with linear cross phases at low $k_y$. The UI-driven heat flux in the NT scenario was twice as large as that in the PT equilibrium, and was almost ten times as large as the TEM-driven HSX heat fluxes. As the density gradient was decreased, the nonlinear heat fluxes for the NT and PT cases decreased much faster than HSX, reaching HSX levels by $\omega_n=2$ for the PT case and $\omega_n=1$ for the NT case. In the $\nabla T_\text{e}$ drive scenario, there were no significant heat fluxes or electrostatic potential amplitudes driven at ion scales.
	
	A scan over normalized electron pressure $\beta$ was performed for the NT and PT configurations for the $\nabla n$ drive scenario to investigate the electromagnetic stabilization of UIs. The growth rates of the UIs were significantly reduced at $k_y\ge0.4$ in both cases. For $k_y=0.1$, the growth rate decreased slightly, reaching a minimum at $\beta=4\times10^{-3}$, before increasing slightly. In the PT case, a KBM became the dominant instability at $\beta=10^{-2}$. Nonlinear electrostatic heat fluxes decreased by half in the NT case when $\beta$ was increased from $10^{-4}$ to $4\times10^{-3}$. In the PT case, the electrostatic heat flux were almost completely suppressed at $\beta=2\times10^{-3}$.

	The objective function used to target TEMs is simple and fast to compute, showing that high-fidelity models may not always be necessary to achieve the desired results in a stellarator optimization. However, when optimizing for one type of instability, one may arrive at a configuration with unfavorable stability or turbulence properties arising from another type of instability. In the present work, the results of the density gradient-driven scenarios showed that optimizing for TEMs lead can to unstable UIs that drive substantial amounts of heat flux. As a result, future optimization may need to simultaneously target UIs and TEMs to completely suppress ion-scale electrostatic turbulence in quasihelically symmetric configurations with substantial density gradients.
	
	A possible path forward to design a stellarator with strongly reduced electrostatic turbulence is to develop fast reduced metrics for ITGs and UIs and implement them along with the available energy metric for TEMs in an optimization. However, it is unclear if each instability can be suppressed or to what degree. Additionally, when considering electrostatic drift-wave-driven turbulence in the design of a future stellarator experiment or simulating an existing one, the operational parameters, such as $\beta$, should be considered, as they can substantially alter the nature of the turbulence.
	
	\section*{Acknowledgments}
	This work is supported by the US Department of Energy, Office of Science, Fusion Energy Services under grants no.~DE-SC0022257, DE-FG02-89ER53291, DE-FG02-86ER53218, DE-FG02-99ER54546, and DE-FG02-93ER54222; the Simons Foundation Award No. 1013655.; and by EUROfusion. This work has been carried out within the framework of the EUROfusion Consortium, funded by the European Union via the Euratom Research and Training Programme (Grant Agreement No 101052200 -- EUROfusion). Views and opinions expressed are however those of the author(s) only and do not necessarily reflect those of the European Union or the European Commission. Neither the European Union nor the European Commission can be held responsible for them. Computing resources were provided through the National Energy Research Scientific Computing Center, a DOE Office of Science User Facility, grant no.~DE-AC02-05CH11231.

\appendix
\section{Linear mode identification}
\label{sec:lin_mode_appx}
The linear mode identification procedure was applied to two scenarios: one with only $\nabla n$ drive and one with only $\nabla T_\text{e}$ drive. Following the procedure of Ref.~\onlinecite{costello_universal_2023}, also outlined in Sec.~\ref{sec:linear}, the dominant modes are identified. The linear cross phases, electrostatic potenetial structure relative to the magnetic geometry, and the response of the electrostatic potential and eigenvalues of the dominant instability are used to distinguish between TEMs and UIs when there is only a density gradient, and between TEMs and ETGs when there is only an electron temperature gradient. 

\subsection{Density Gradient Drive}
\begin{figure}[!b]
	\centering
	\includegraphics[width=0.5\textwidth]{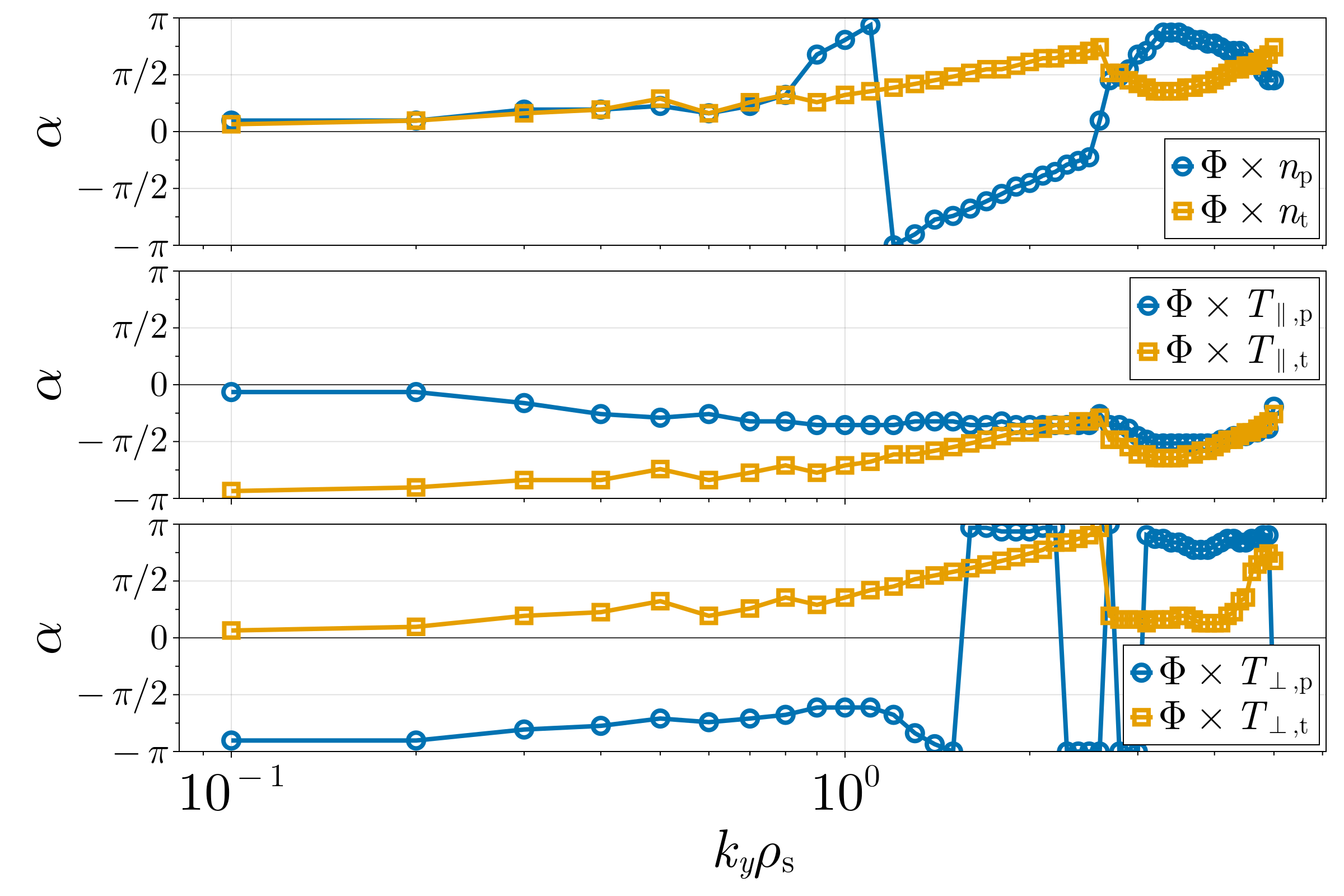}
	\caption{Cross phases in HSX with $\omega_n=4$, $\omega_{T\text{e}}=0$ of the dominant instability for electrostatic potential fluctuations $\mathrm{\Phi}$ with fluctuations of density (top) of passing (blue) and trapped (orange) electrons, parallel electron temperature (middle), and perpendicular electron temperature (bottom). Cross phases of $n_\text{t}$ near $\pi/2$ for $k_y\ge1$ indicate TEMs, while similar cross phases of $n_\text{t}$ and $n_\text{p}$ at $k_y\lesssim 1$ do not distinguish these modes as TEM or UI.}
	\label{fig:HSX_cross}
\end{figure}

\begin{figure}[h]
	\centering
	\includegraphics[width=0.5\textwidth]{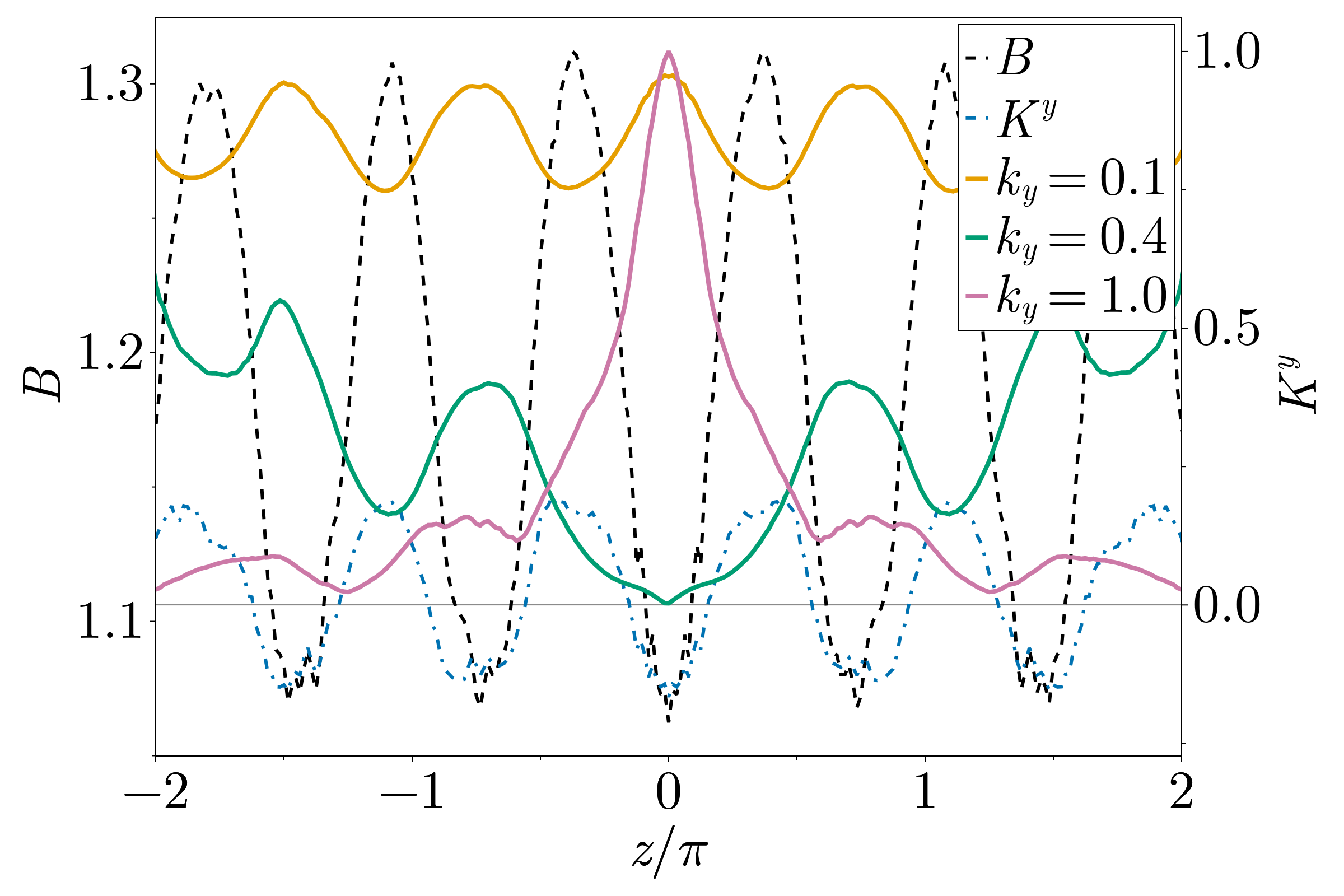}
	\caption{Amplitude of the electrostatic potential for dominant modes (solid) in HSX with $\omega_n=4$, $\omega_{T\text{e}}=0$ plotted with the magnetic field (dashed black, left axis), and curvature drive (blue dot dash, right axis) as a function of the parallel coordinate $z$. The normalized $\abs{\mathrm{\Phi}}$ are plotted for modes with $k_x=0$ and $k_y=0.1$ (yellow), $k_y=0.4$ (green), and $k_y=1$ (pink) (right axis). Each mode has strong localization in magnetic wells with destabilizing curvature, which is consistent with TEMs.}
	\label{fig:HSX_Phi}
\end{figure}

Beginning with HSX, the cross phases are presented in Fig.~\ref{fig:HSX_cross}. For $k_y\ge1$, the cross phase between $\mathrm{\Phi}$ and $n_\text{t}$ is $\sim\pi/2$, and the cross phase between $\mathrm{\Phi}$ and $n_\text{p}$ is inefficient, which is consistent with TEMs \cite{faber_gyrokinetic_2015}. At low $k_y$, both trapped and passing density cross phases are inefficient, a property both shared by UIs and TEMs \cite{faber_gyrokinetic_2015,costello_universal_2023}. Looking at the mode structure for low $k_y$ provides insight into classifying these modes. In Fig.~\ref{fig:HSX_Phi}, the magnetic field and curvature are plotted with the $k_x=0$ component of $\abs{\mathrm{\Phi}}$ for the most unstable mode at $k_y=0.1$, 0.4, and 1 along the field line for HSX. Each mode plotted is strongly localized in regions of magnetic wells with destabilizing curvature. The localization of these modes was consistent with TEMs. 

\begin{figure}[h]
	\centering
	\includegraphics[width=0.5\textwidth]{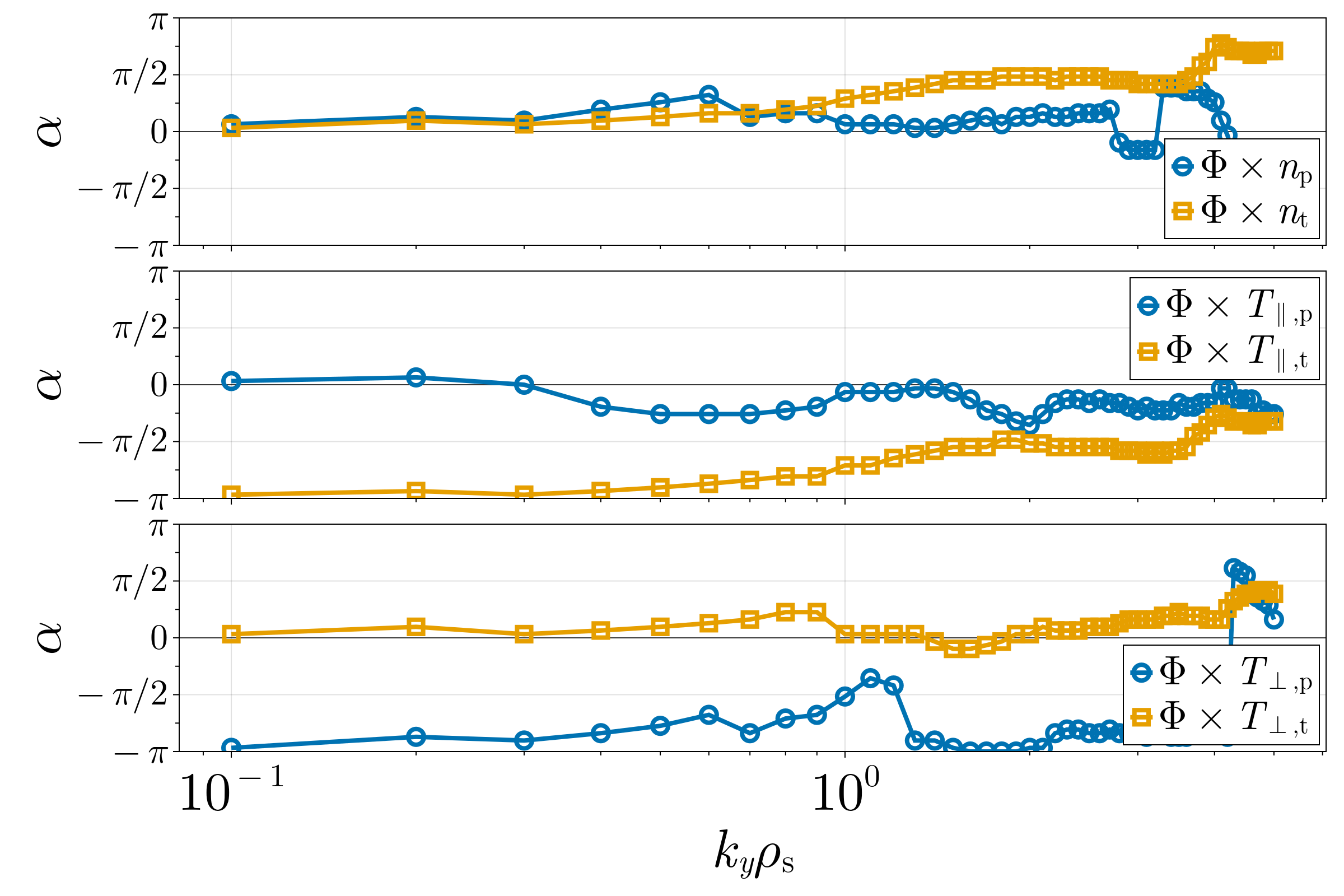}
	\caption{Cross phases in the NT reduced-TEM geometry with $\omega_n=4$, $\omega_{T\text{e}}=0$ of the dominant instability for electrostatic potential fluctuations $\mathrm{\Phi}$ with fluctuations of density (top) of passing (blue) and trapped (orange) electrons, parallel electron temperature (middle), and perpendicular electron temperature (bottom). Cross phases of $n_\text{t}$ are closer to $\pi/2$ than $n_\text{p}$ for $k_y\ge1$ indicating TEMs. For lower $k_y$, similar cross phases of $n_\text{t}$ and $n_\text{p}$ do not distinguish these modes as TEM or UI.}
	\label{fig:NT_cross}
\end{figure}

For the NT reduced-TEM equilibrium, the cross phases are shown in Fig.~\ref{fig:NT_cross}. The density cross phases indicate that for $k_y\ge1$, the modes are TEMs. The modes at lower $k_y$ can not be identified based on their cross phases alone, because both trapped and passing channels are similarly inefficient. Because there is a smooth increase in cross phase of $\mathrm{\Phi}$ and $n_\text{t}$ to $\pi/2$ with the decrease of $\mathrm{\Phi}$ and $n_\text{p}$ to zero, and the frequencies in Fig.~\ref{fig:omn4_gamma_omega} are on the same branch, the properties of these modes with $k_y\ge0.7$ were consistent with a TEM or TEM-UI hybrid mode.

\begin{figure}[h]
	\centering
	\includegraphics[width=0.5\textwidth]{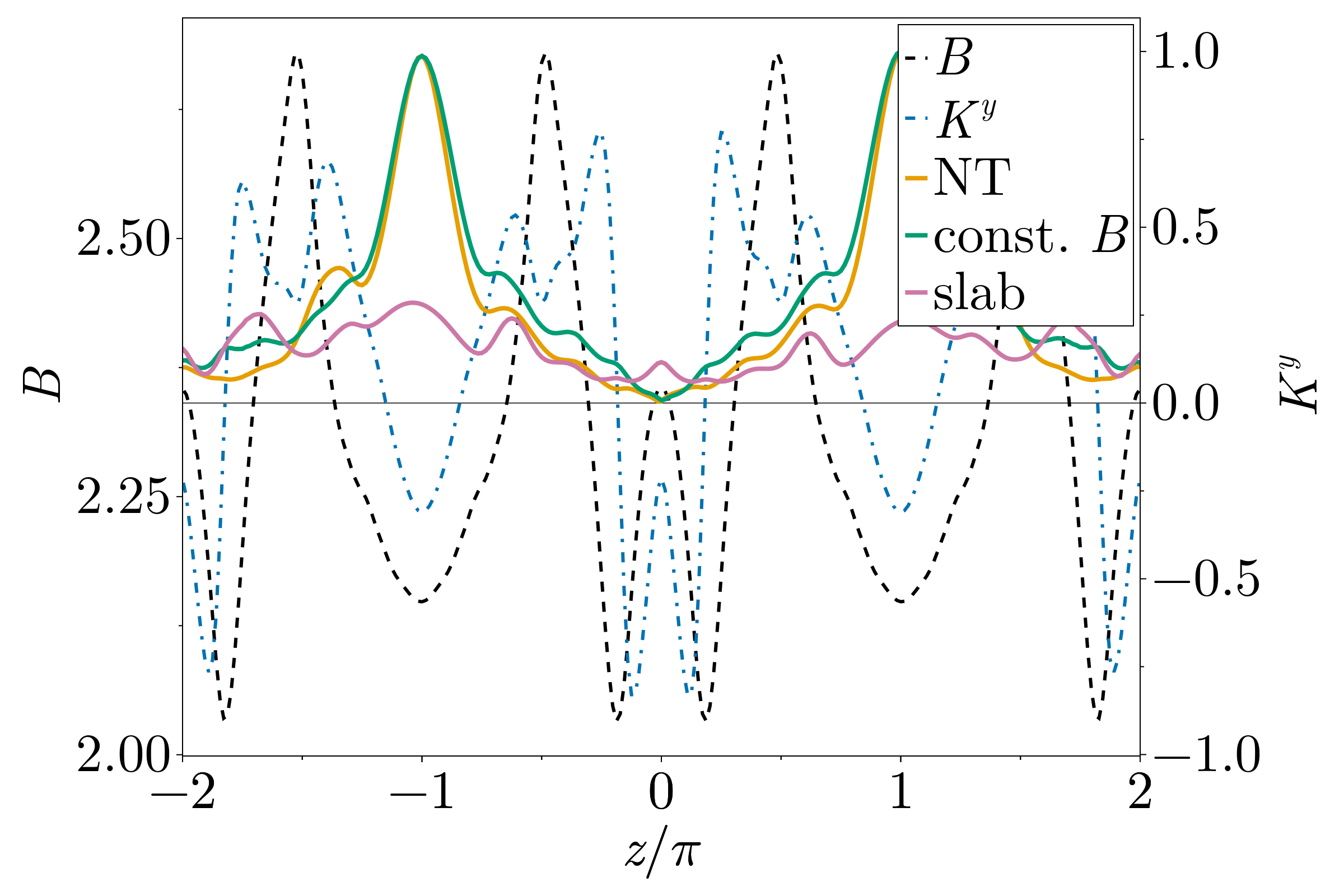}
	\caption{Amplitude of the electrostatic potential for dominant modes (solid) in the NT geometry with $\omega_n=4$, $\omega_{T\text{e}}=0$ for $k_x=0$, $k_y=0.5$ (right axis) normalized to its maximum value plotted with the magnetic field (dashed black, left axis) and curvature drive (blue dot dash, right axis) as a function of the parallel coordinate $z$. The mode with real geometry is in yellow, the constant-$B$ geometry is in green, and the slab-like geometry is in pink. The mode structure is similar between the real and constant-$B$ but not the slab geometries, indicating this mode is a toroidal UI.}
	\label{fig:NT_Phi_ky5e-1}
\end{figure}

\begin{figure}[h]
	\centering
	\includegraphics[width=0.5\textwidth]{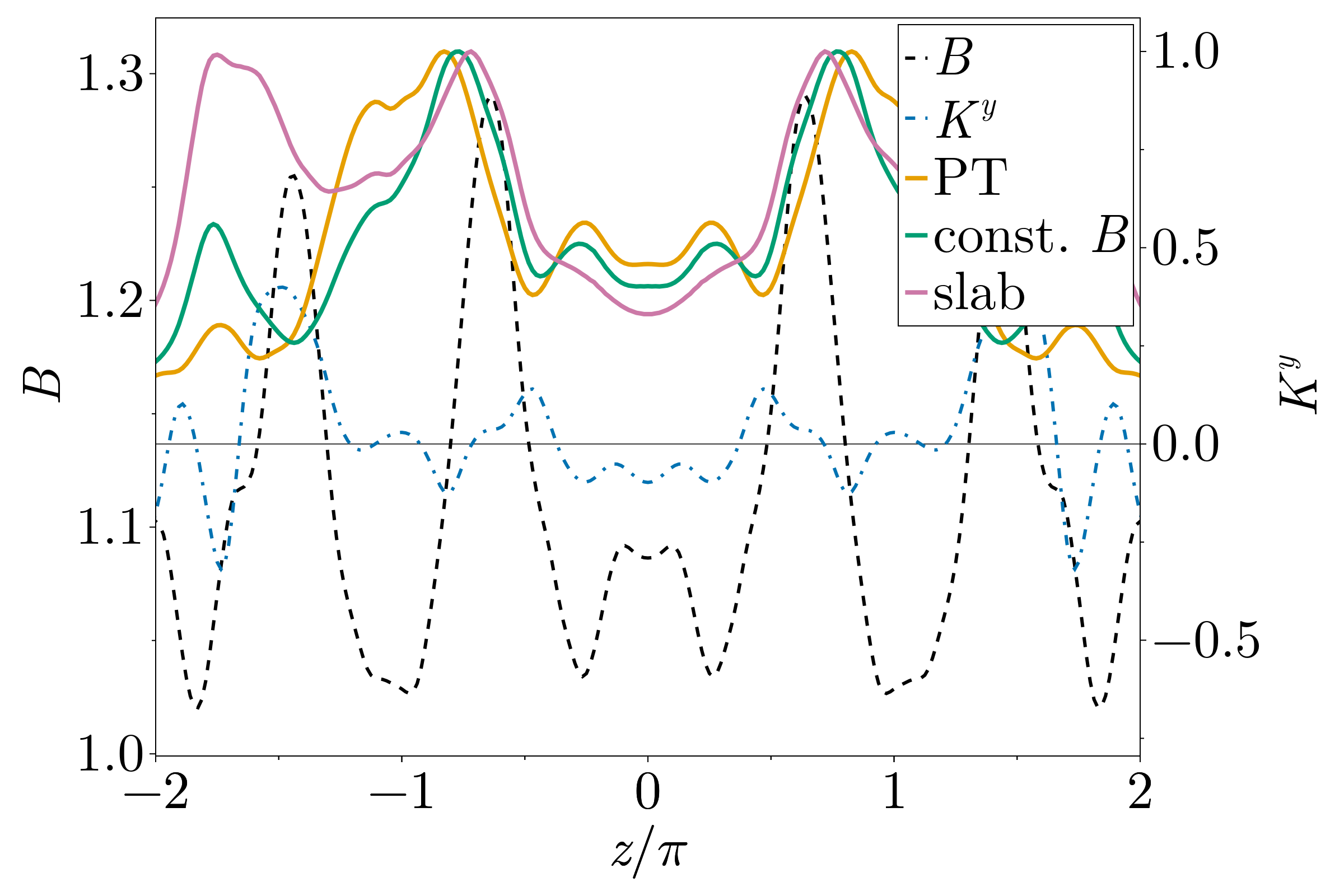}
	\caption{Amplitude of the electrostatic potential for dominant modes (solid) in the PT geometry with $\omega_n=4$, $\omega_{T\text{e}}=0$ for $k_x=0$, $k_y=0.5$ (right axis) normalized to its maximum value plotted with the magnetic field (dashed black, left axis) and curvature drive (blue dot dash, right axis) as a function of the parallel coordinate $z$. The mode with real geometry is in yellow, the constant-$B$ geometry is in green, and the slab-like geometry is in pink. The mode structure does not change when the trapping or when the curvature drive is additionally removed in slab geometry, which is consistent with a slab UI.}
	\label{fig:PT_Phi_5e-1}
\end{figure}

In Fig.~\ref{fig:NT_Phi_ky5e-1}, the parallel structure of the $k_x=0$ component of $\abs{\mathrm{\Phi}}$ for the dominant instability at $k_y=0.5$ for the physical NT, for the NT with constant-$B$, and for the NT slab-like geometries are plotted with $B_0$ and $\mathcal{K}^y$. While the mode is strongly localized in the magnetic wells centered near $z=\pm\pi$, it is important to note that the polarization term $g^{yy}$ is low in these wells too. Small values of $g^{yy}$ where curvature is destabilizing localizes toroidal branches of fluid-like modes at low $k_y$\cite{duff_effect_2022}. The destabilizing curvature in the wells centered at $z=\pm\pi$ is much larger than that of HSX and the reduced-TEM PT equilibrium (see, e.g., Fig.~\ref{fig:PT_Phi_5e-1}), due to the large helical radius of the magnetic axis, displayed in Fig.~\ref{fig:opt_xsection}. When the trapping is removed in the constant-$B$ geometry, the mode structure does not significantly change, consistent with a UI. When the curvature is also removed in the slab-like geometry, the mode structure changes significantly, meaning the fastest growing mode at $k_y=0.5$ for the NT equilibrium is consistent with a toroidal UI. 

\begin{figure}[h]
	\centering
	\includegraphics[width=0.5\textwidth]{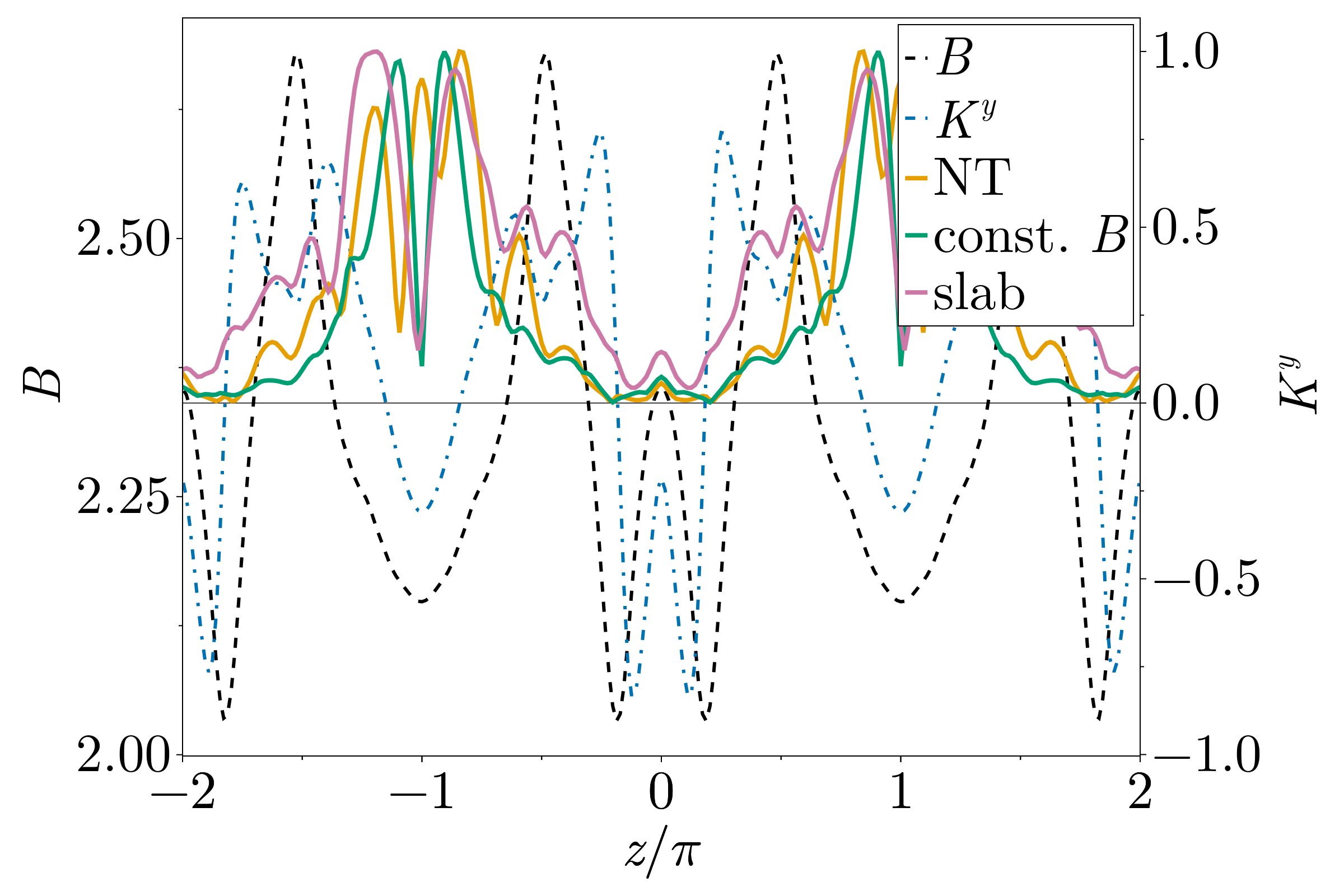}
	\caption{Amplitude of the electrostatic potential for dominant modes (solid) in the NT geometry with $\omega_n=4$, $\omega_{T\text{e}}=0$ for $k_x=0$, $k_y=2.2$ (right axis) normalized to its maximum value plotted with the magnetic field (dashed black, left axis) and curvature drive (blue dot dash, right axis) as a function of the parallel coordinate $z$. The mode with real geometry is in yellow, the constant-$B$ geometry is in green, and the slab-like geometry is in pink. The mode has strong TEM and UI properties.}
	\label{fig:NT_Phi_22}
\end{figure}

In the NT configuration, this toroidal UI response of the mode structure to the changes in geometry is consistent up to $k_y=2.2$. Because the cross phases suggests the modes with $k_y\gtrsim1.5$ are TEMs, and the mode structure response suggests the modes are toroidal UIs for $k_y\le2.1$, these modes for $1.5\lesssim k_y\le2.1$ are likely hybrid modes. At $k_y=2.2$, the frequency change in Fig.~\ref{fig:omn4_gamma_omega} for the NT configuration indicates the dominant instability is on a different branch. The same information as Fig.~\ref{fig:NT_Phi_ky5e-1} but for $k_y=2.2$ is presented in Fig.~\ref{fig:NT_Phi_22}. While much of the mode structure is similar for the physical, constant-$B$, and slab-like geometries, there is localization in $\mathrm{\Phi}$ at the bottom of the wells centered near $z=\pm\pi$. This localization in $\mathrm{\Phi}$ is consistent with a contribution from the trapped electron population because it is not present in the mode of the constant-$B$ or slab-like geometries. Therefore, the fastest growing mode at $k_y=2.2$ is also a TEM-UI hybrid mode. This hybridization occurrs in modes with $k_y\gtrsim0.7$ to varying degrees, becoming more TEM-like as $k_y$ increased.

\begin{figure}[h]
	\centering
	\includegraphics[width=0.5\textwidth]{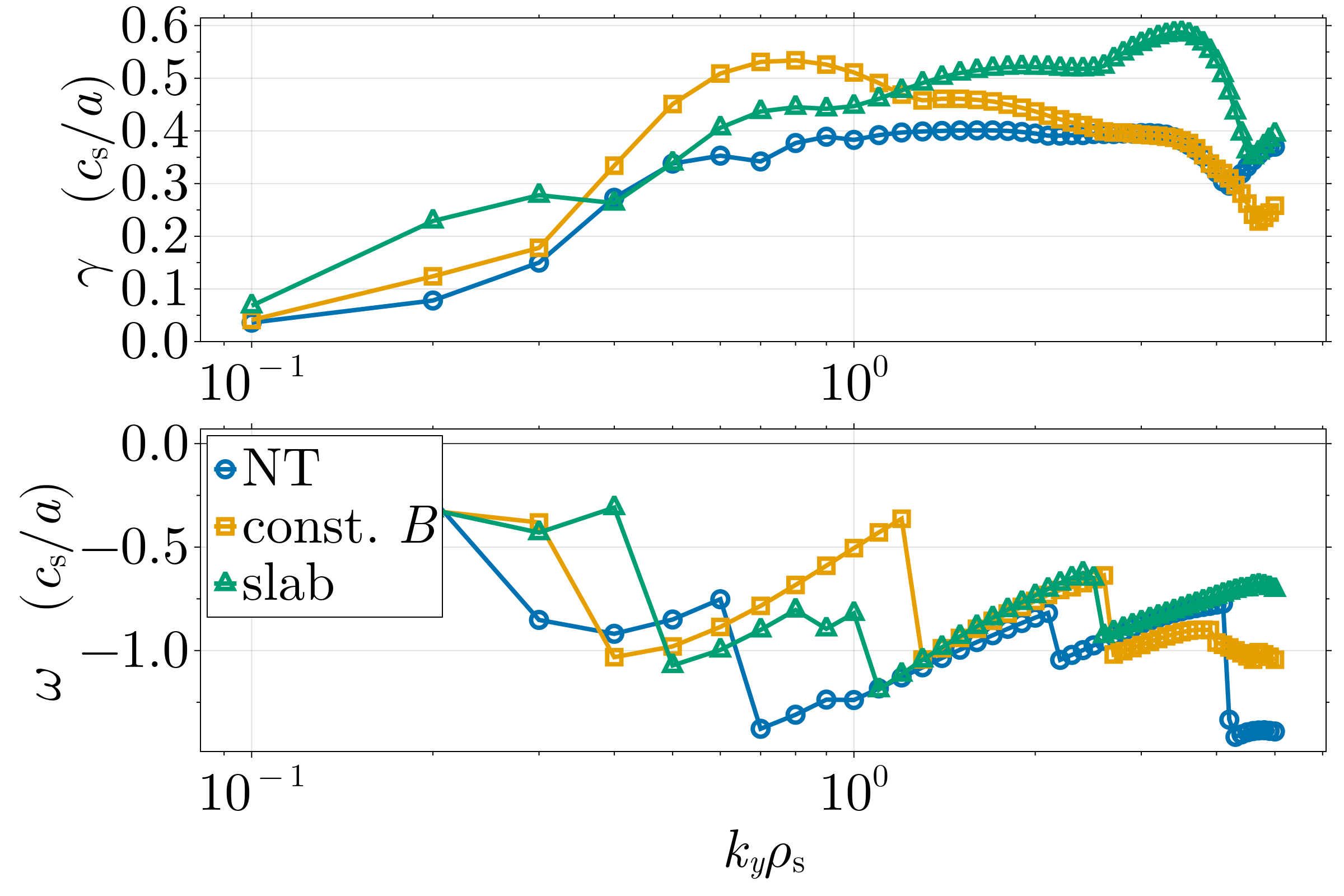}
	\caption{Growth rates (top) and frequencies (bottom) of the fastest growing modes with $k_x=0$ for the NT (blue) and constant-$B$ (orange), and slab-like (green) geometries for $\omega_n=4$, $\omega_{T\text{e}}=0$. The fastest growing modes in the NT configuration are toroidal UIs for $k_y\le0.7$, with growth rates increasing from the NT to constant-$B$ due to the removal of stabilizing non-resonant trapped electrons.}
	\label{fig:NT_omn4_gamma_omega_constB_slab}
\end{figure}

The growth rates and real frequencies of the NT equilibrium and its constant-$B$ and slab-like variations are presented in Fig.~\ref{fig:NT_omn4_gamma_omega_constB_slab}. The conclusion of toroidal UIs for $k_y\le0.7$ that was drawn from the cross phases and eigenfunction trends with geometry concurred with the eigenvalues. At these low wavenumbers, when the particle trapping effects is removed in the constant-$B$ geometry, the growth rate trends stayed similar. The slight increase in growth rates is possibly due to the removal of non-resonant trapped electrons, which play a stabilizing role \cite{helander_universal_2015,costello_universal_2023}. The frequencies are also similar. However, for $k_y=0.3$ and 0.4, the dominant instability switches UI branches before the modes in the constant-$B$ and slab geometries.

\begin{figure}[h]
	\centering
	\includegraphics[width=0.5\textwidth]{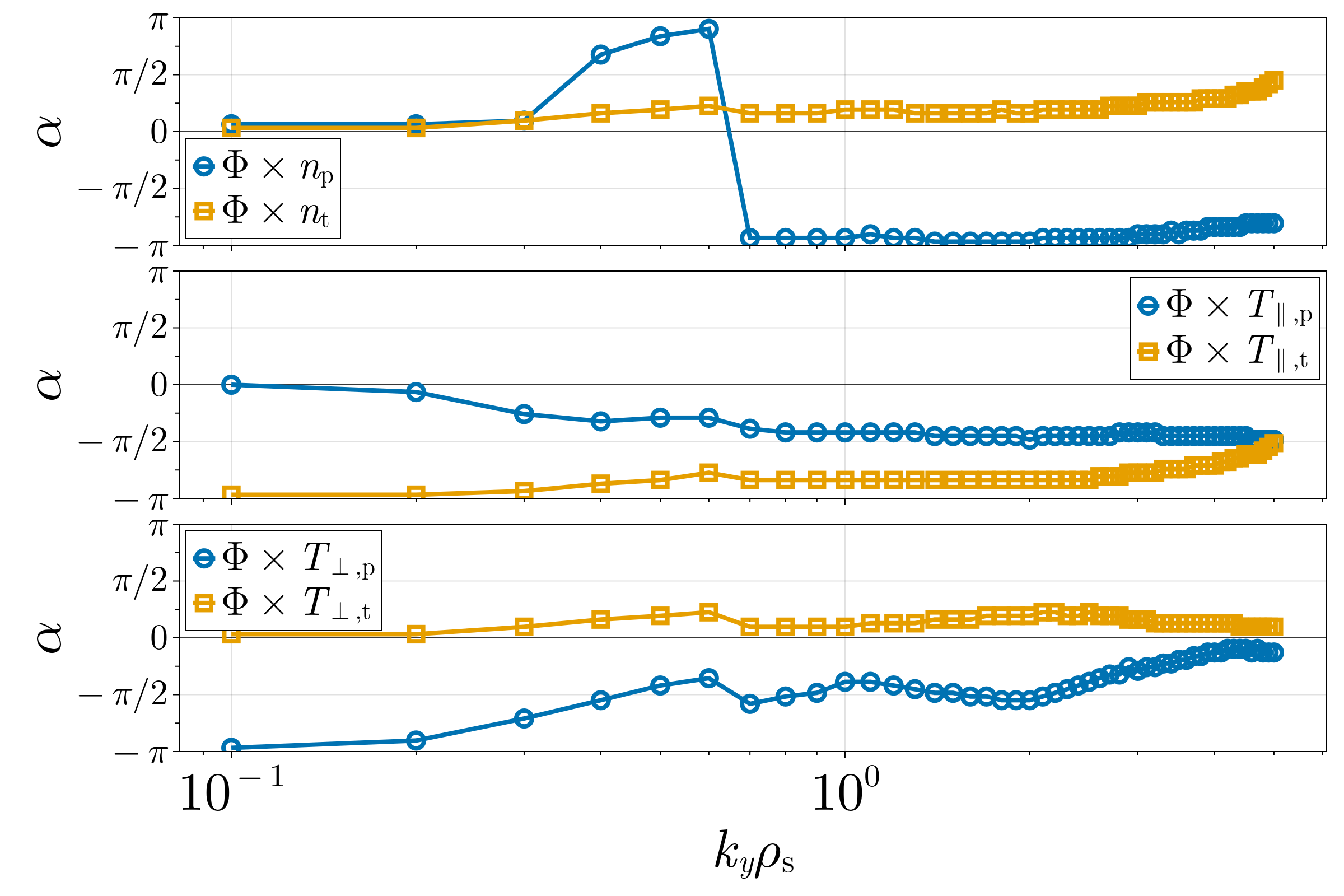}
	\caption{Cross phases in the PT reduced-TEM geometry with $\omega_n=4$, $\omega_{T\text{e}}=0$ of the dominant instability for electrostatic potential fluctuations $\mathrm{\Phi}$ with fluctuations of density (top) of passing (blue) and trapped (orange) electrons, parallel electron temperature (middle), and perpendicular electron temperature (bottom). Cross phases of $n_\text{t}$ are closer to $\pi/2$ than $n_\text{p}$ for $k_y\ge0.7$, indicating TEMs. For lower $k_y$, $n_\text{t}$ and $n_\text{p}$ are similarly out of phase with $\mathrm{\Phi}$ and cannot be used not distinguish these modes as TEM or UI.}
	\label{fig:PT_cross}
\end{figure}

For the PT case, the cross phases are presented in Fig.~\ref{fig:PT_cross}. At wavenumbers of $k_y\ge0.7$, the density fluctuations for the passing particles are out of phase with $\mathrm{\Phi}$, and $n_\text{t}$ is a moderately effective loss channel. For $k_y\le0.6$, $n_\text{p}$ is out of phase to a similar degree as $n_\text{t}$. From these cross phases, the most unstable modes at $k_y\ge0.7$ are TEMs, and the modes with $k_y\le0.6$ require an analysis of how the mode structure and eigenvalues respond to the geometry.

The parallel structure of the $k_x=0$ component of $\mathrm{\Phi}$ for the dominant instability at $k_y=0.5$ for the real PT, for thePT with constant-$B$, and for the PT slab geometries are plotted with $B_0$ and $\mathcal{K}^y$ are shown in Fig.~\ref{fig:PT_Phi_5e-1}. For $k_y=0.5$, the mode was a slab UI, because the mode structure did not change significantly from the real geometry to the slab-like geometry.

\begin{figure}[h]
	\centering
	\includegraphics[width=0.5\textwidth]{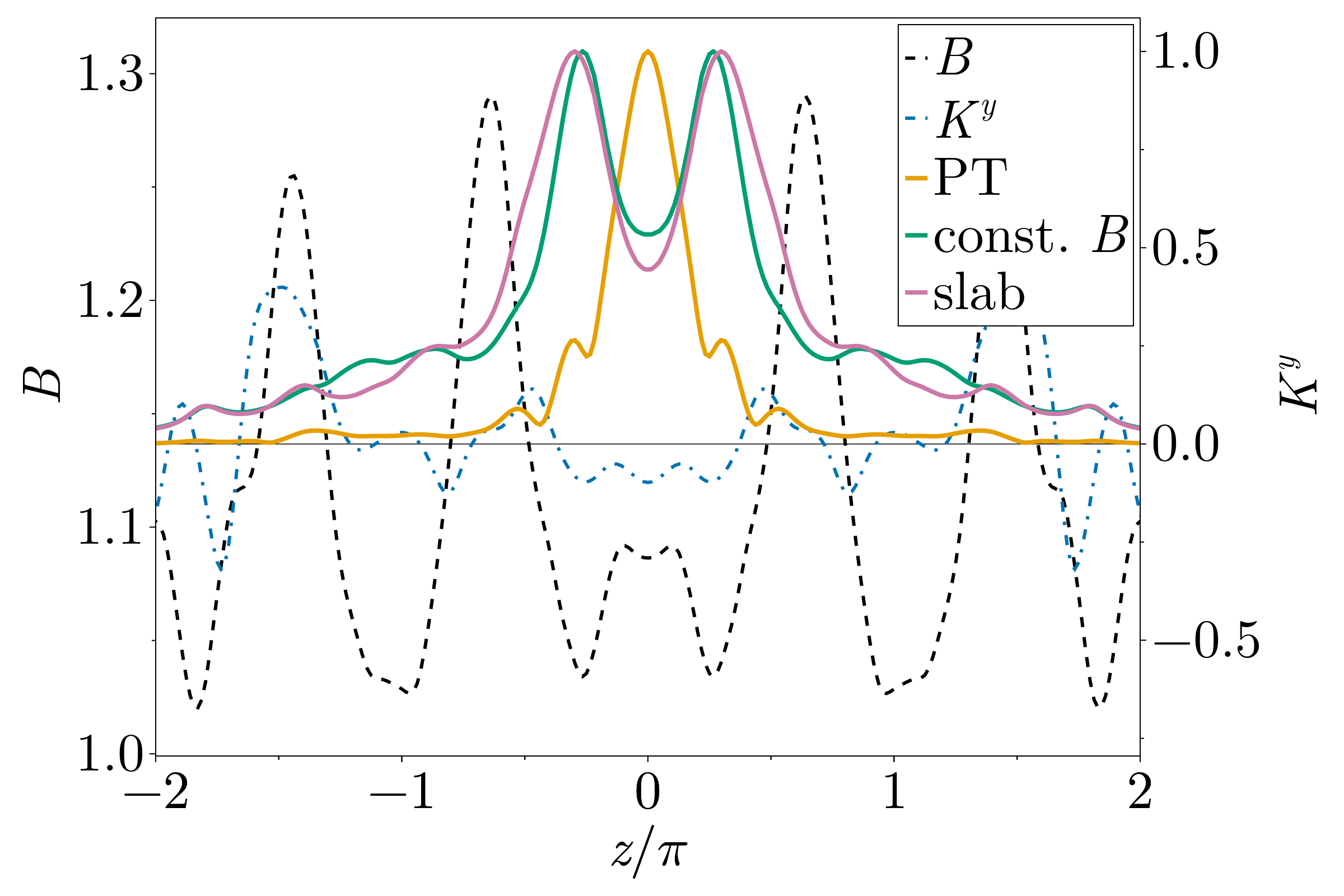}
	\caption{Amplitude of the electrostatic potential for dominant modes (solid) in the PT geometry with $\omega_n=4$, $\omega_{T\text{e}}=0$ for $k_x=0$, $k_y=0.7$ (right axis) normalized to its maximum value plotted with the magnetic field (dashed black, left axis) and curvature drive (blue dot dash, right axis) as a function of the parallel coordinate $z$. The mode with real geometry is in yellow, the constant-$B$ geometry is in green, and the slab-like geometry is in pink. The mode structure in the real geometry has strong localization in magnetic wells where curvature is destabilizing, and the mode structure changed significantly when the trapping was removed, indicating a TEM.}
	\label{fig:PT_Phi_6e-1}
\end{figure}

At $k_y=0.7$, there was a jump in frequency and cross phases from $k_y=0.6$, the values of which indicate a TEM. Figure~\ref{fig:PT_Phi_6e-1} shows the $k_x=0$ component of $\mathrm{\Phi}$ for the dominant instability at $k_y=0.7$ for the real PT, for the PT with constant-$B$, and for the PT slab-like geometries plotted with the magnetic field strength and curvature drive. The mode structure is well localized to the central magnetic well, where $\mathcal{K}^y$ is destabilizing. The mode structure also changes when trapping is removed. Combined with the cross phase information, this mode is consistent with a TEM. As $k_y$ increases, the most unstable modes also exhibit these TEM properties.

In Fig.~\ref{fig:NT_omn4_gamma_omega_constB_slab}, the growth rates and real frequencies of the PT equilibrium and its constant-$B$ and slab-like geometries are shown. The conclusion drawn from the cross phases and eigenfunction trends with geometry also agree with the conclusion of slab UIs for $k_y\le0.6$. The growth rates were very similar for these low wavenumber modes. The frequencies were also similar, however for $k_y=0.2$, the dominant instability switches UI branches before the modes in the slab-like geometry. Similarly, there is a branch switch of the dominant UI at $k_y=0.6$ in the physical geometry compared to the constant-$B$ and slab-like geometries.

\begin{figure}[h]
	\centering
	\includegraphics[width=0.5\textwidth]{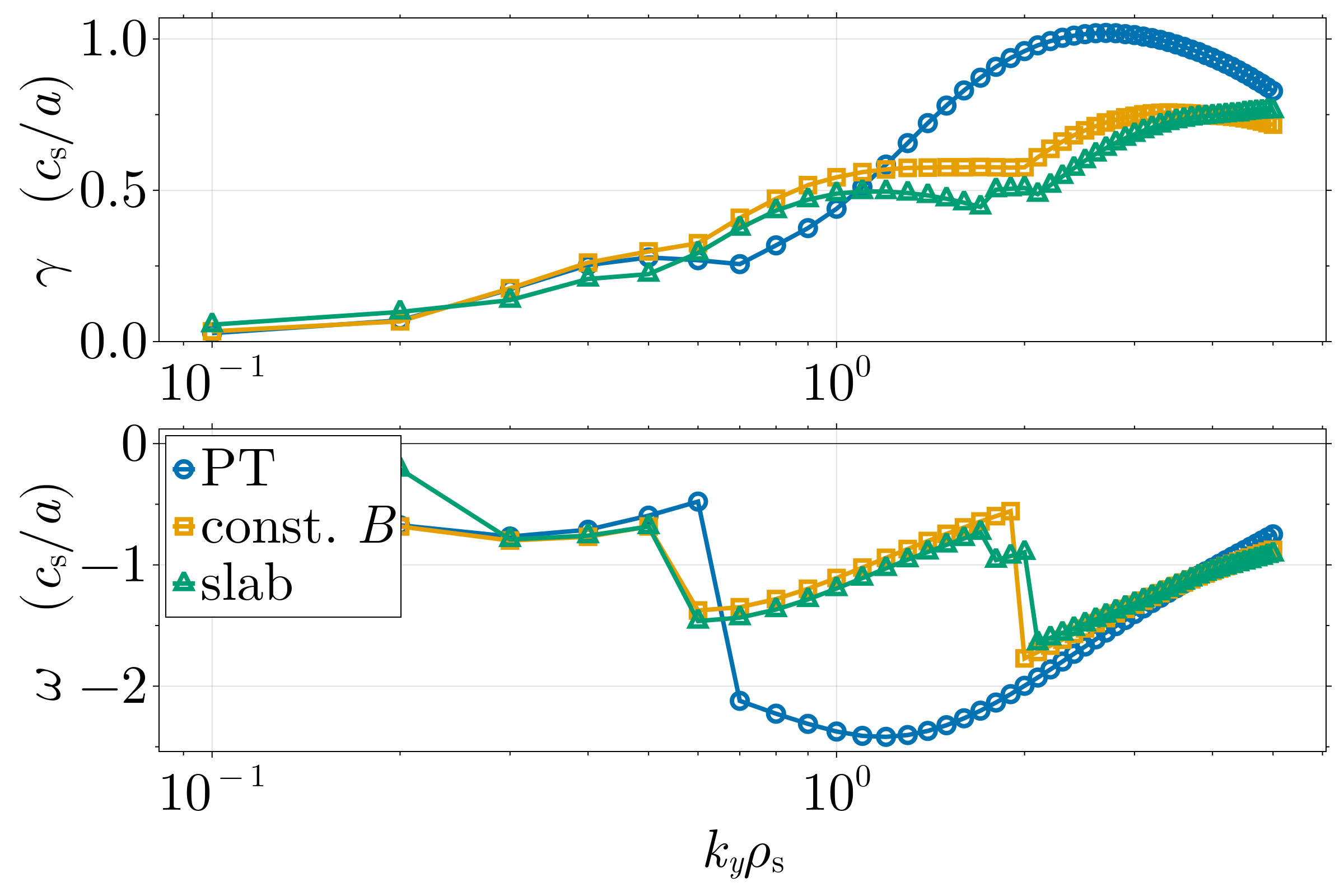}
	\caption{Growth rates (top) and frequencies (bottom) of the fastest growing modes with $k_x=0$ for the PT (blue) and constant-$B$ (orange), and slab-like (green) geometries for $\omega_n=4$, $\omega_{T\text{e}}=0$. The fastest growing modes in the NT configuration are slab UIs for $k_y\le0.7$, where growth rates are similar between the three geometries.}
	\label{fig:PT_omn4_gamma_omega_constB_slab}
\end{figure}

\subsection{Electron Temperature Gradient Drive}
\begin{figure}[h]
	\centering
	\includegraphics[width=0.5\textwidth]{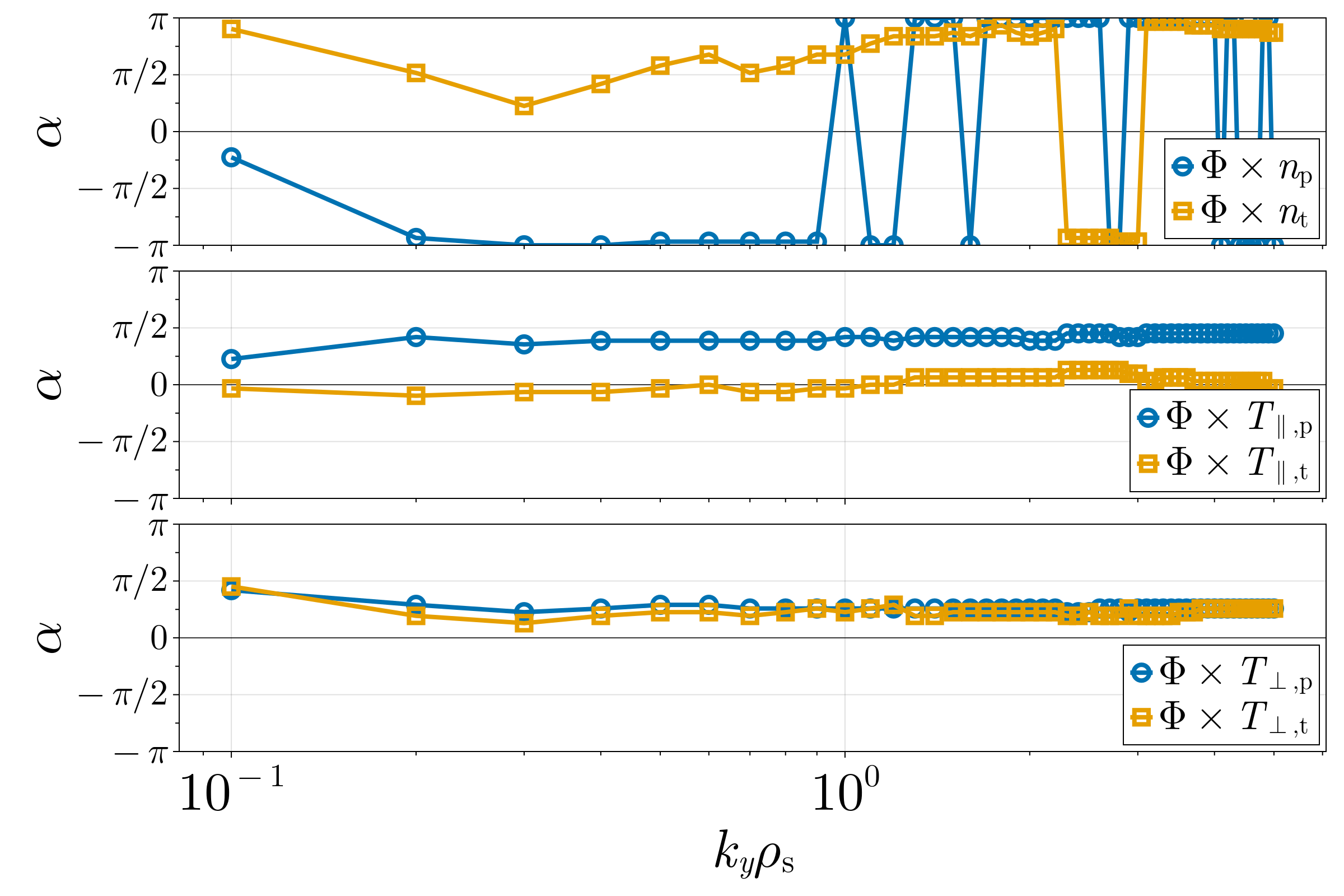}
	\caption{Cross phases in HSX with $\omega_n=0$, $\omega_{T\text{e}}=4$ of the dominant instability for electrostatic potential fluctuations $\mathrm{\Phi}$ with fluctuations of density (top) of passing (blue) and trapped (orange) electrons, parallel electron temperature (middle), and perpendicular electron temperature (bottom). Cross phases of $T_{\parallel,\text{p}}$ near $\pi/2$ for all $k_y$ indicate ETGs.}
	\label{fig:HSX_cross_Te}
\end{figure}

\begin{figure}[h]
	\centering
	\includegraphics[width=0.5\textwidth]{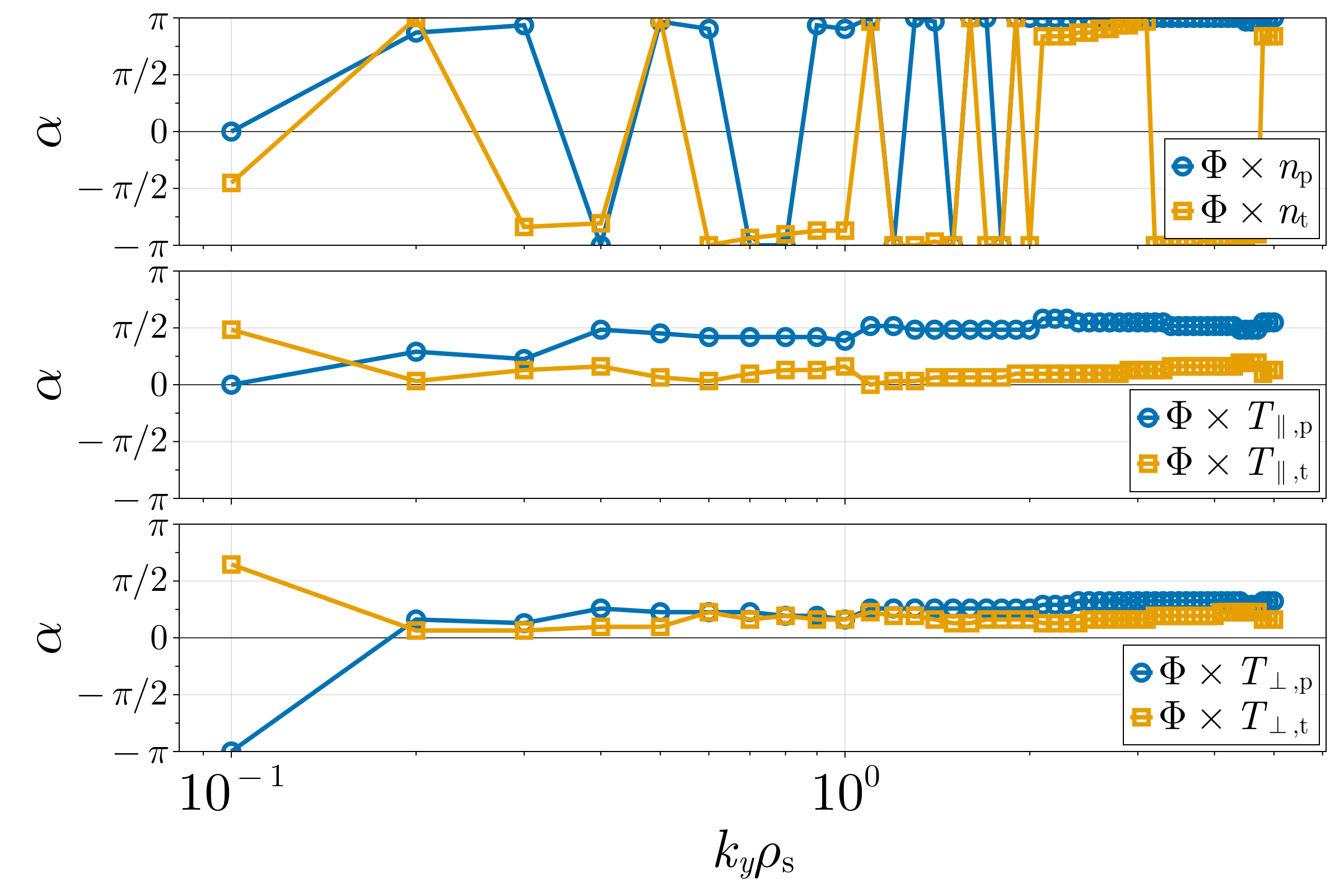}
	\caption{Cross phases in the reduced-TEM NT geometry with $\omega_n=0$, $\omega_{T\text{e}}=4$ of the dominant instability for electrostatic potential fluctuations $\mathrm{\Phi}$ with fluctuations of density (top) of passing (blue) and trapped (orange) electrons, parallel electron temperature (middle), and perpendicular electron temperature (bottom). Cross phases of $T_{\parallel,\text{p}}$ near $\pi/2$ for $k_y>0.1$ indicate ETGs, while the fastest growing mode at $k_y=0.1$ has cross phases of $T_{\parallel,\text{t}}$ and $T_{\perp,\text{t}}$ near $\pi/2$ indicating a TEM.}
	\label{fig:NT_cross_Te}
\end{figure}

\begin{figure}[h]
	\centering
	\includegraphics[width=0.5\textwidth]{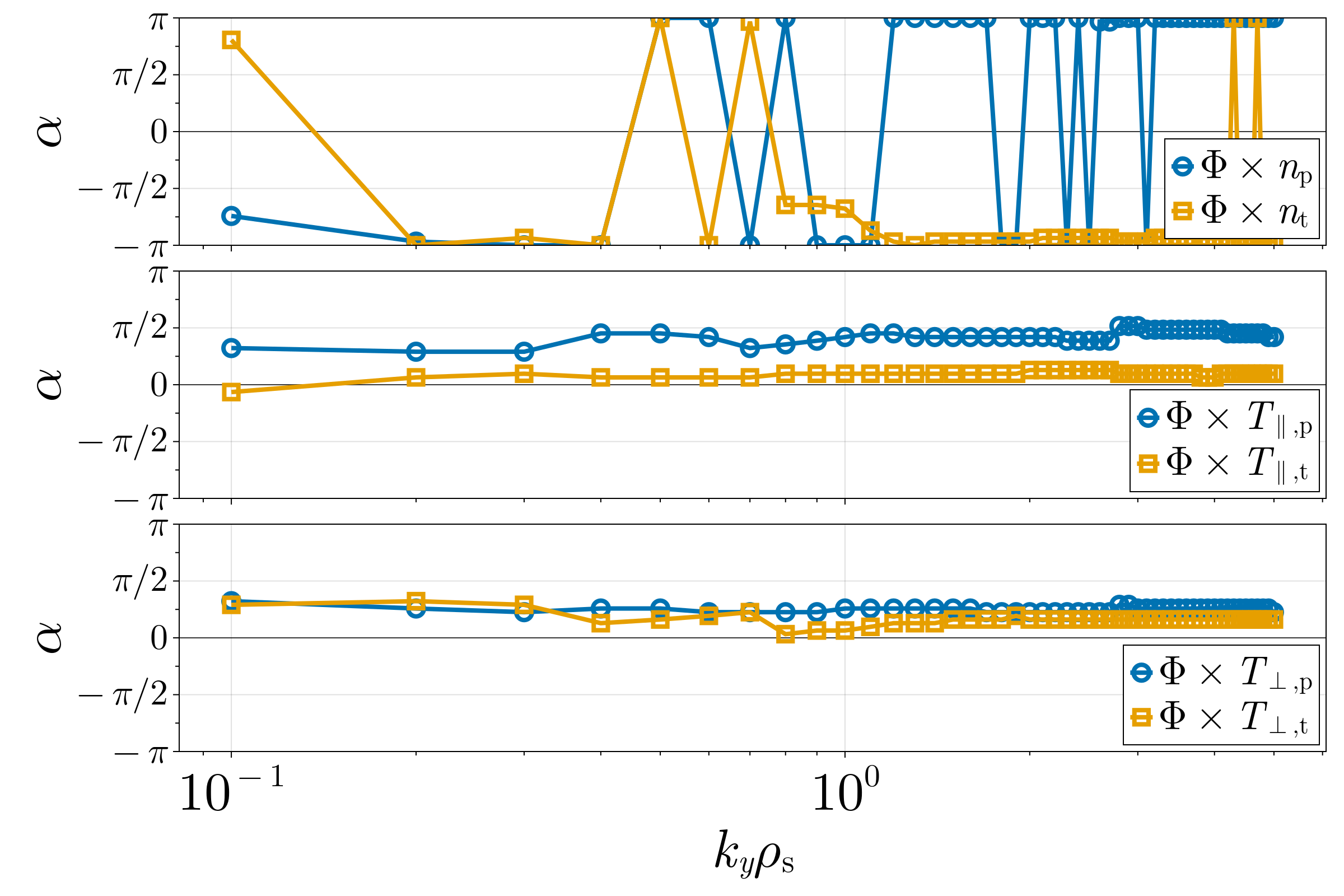}
	\caption{Cross phases in the reduced-TEM PT geometry with $\omega_n=0$, $\omega_{T\text{e}}=4$ of the dominant instability for electrostatic potential fluctuations $\mathrm{\Phi}$ with fluctuations of density (top) of passing (blue) and trapped (orange) electrons, parallel electron temperature (middle), and perpendicular electron temperature (bottom). Cross phases of $T_{\parallel,\text{p}}$ near $\pi/2$ for all $k_y$ indicate ETGs.}
	\label{fig:PT_cross_Te}
\end{figure}

In Fig.~\ref{fig:HSX_cross_Te}, the cross phases for HSX are shown for the case with $\omega_n=4$ and $\omega_{T\text{e}}=4$. The cross phase of $T_{\parallel,\text{p}}$ os near $\pi/2$ for all $k_y$, while all other temperature fluctuation cross phases are less efficient. Therefore, the cross phases indicate that modes are likely ETGs. In Fig.~\ref{fig:NT_cross_Te}, the $T_{\parallel,\text{p}}$ cross phase is the most efficient, except at $k_y=0.1$, so the dominant instability for $k_y>0.1$ for the NT case are consistent with ETGs. At $k_y=0.1$, both trapped electron temperature cross phases are near $\pi/2$, while the passing population cross phases are far from $\pi/2$, indicating a TEM. Similarly, the dominant instabilities in the PT configuration, whose cross phases are shown in Fig.~\ref{fig:PT_cross_Te}, are consistent with ETGs at all $k_y$ except for one datum. It is also important to note that while ETGs are commonly observed on electron scales, ion-scale ETG activity has been seen previously and can form hybrid modes with TEMs at low and intermediate $k_y$ \cite{pueschel_multi-scale_2020}.


\begin{figure}[H]
	\centering
	\includegraphics[width=0.5\textwidth]{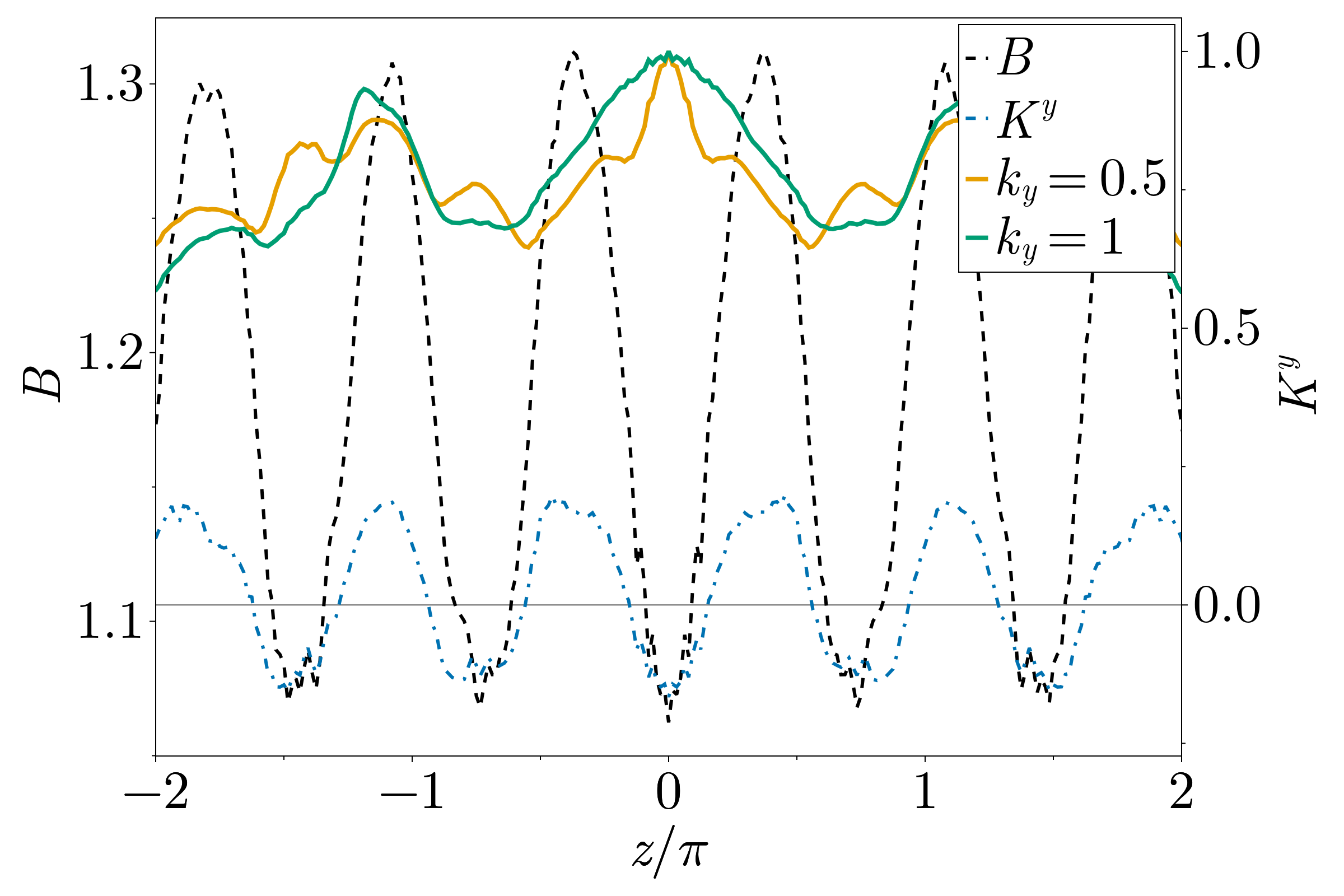}
	\caption{Amplitude of the electrostatic potential along the field line for the fastest growing mode at $k_x=0$, $k_y=0.5$ (yellow) and $k_y=1$ (green) (right axis), normalized to their maximum value, in HSX with $\omega_n=0$, $\omega_{T\text{e}}=4$. The magnetic field (dashed black, left axis) and curvature drive (blue dot dash, right axis) are plotted as a function of the parallel coordinate $z$. The $k_y=0.5$ mode with ETG-like cross phases is more slab-like than the $k_y=1$ mode, which has more localization close to the outboard midplane.}
	\label{fig:HSX_Phi_Te}
\end{figure}

As with the density gradient drive scenario, how $\mathrm{\Phi}$ localizes relative to magnetic wells with destabilizing curvature is used as supporting evidence to the mode identification given by the cross phases. In Fig.~\ref{fig:HSX_Phi_Te}, the amplitude of the $k_x=0$ component of the electrostatic potential is plotted along the field line with the magnetic field $B_0$ and curvature drive $\mathcal{K}^y$. The fastest growing mode with $k_y=0.5$, which was identified as a ETG by its cross phases, had localization of $\mathrm{\Phi}$ in magnetic wells with destabilizing curvature, which is attributed as part of the curvature drive of a slab-like ETG. The fastest growing mode for $k_y=1$ is also plotted in Fig.~\ref{fig:HSX_Phi_Te}, where $\mathrm{\Phi}$ is more localized near the outboard midplane, suggesting the mode is becoming more toroidal-like. 

\begin{figure}[H]
	\centering
	\includegraphics[width=0.5\textwidth]{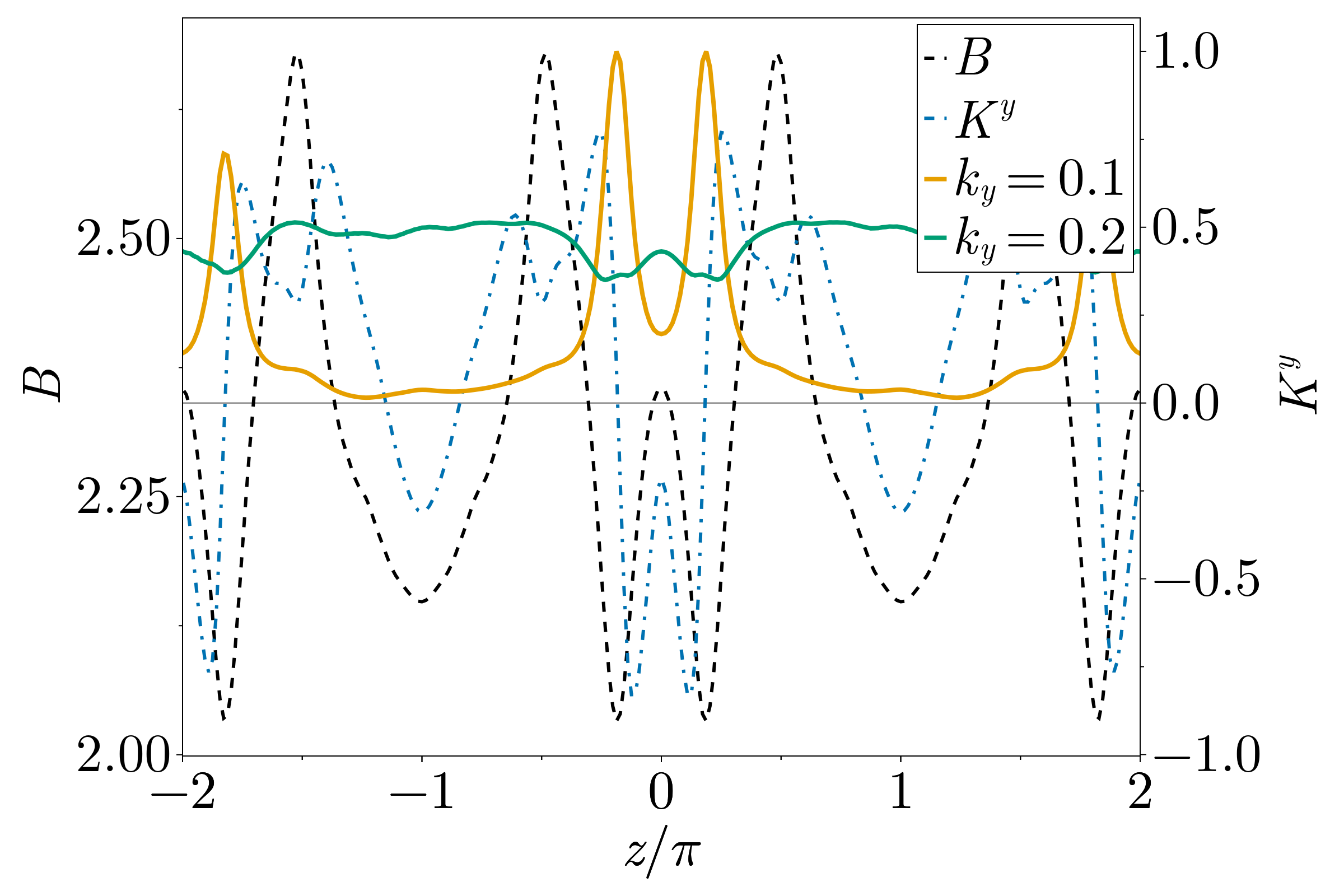}
	\caption{Amplitude of the electrostatic potential along the field line for the fastest growing mode at $k_x=0$, $k_y=0.1$ (yellow) and $k_y=0.2$ (green) (right axis), normalized to their maximum value, in the NT geometry with $\omega_n=0$, $\omega_{T\text{e}}=4$. The magnetic field (dashed black, left axis) and curvature drive (blue dot dash, right axis) are plotted as a function of the parallel coordinate $z$. The $k_y=0.1$ mode has localization in magnetic wells with destabilizing curvature consistent with a TEM. The mode at $k_y=0.2$ does not show localization in magnetic wells with destabilizing curvature, indicating an ETG.}
	\label{fig:NT_Phi_Te}
\end{figure}

Likewise, Fig.~\ref{fig:NT_Phi_Te} plots the $k_y=0.1$ and $k_y=0.2$ modes for the NT configuration with $B_0$ and $\mathcal{K}^y$. At $k_y=0.1$, the fastest growing mode, which had cross phases consistent with a TEM has $\mathrm{\Phi}$ localized in magnetic wells with destabilizing curvature. For $k_y=0.2$, the mode was not localized, and this, coupled with the ETG-like cross phases, suggests a slab ETG. This is the case for all $k_y$ in the NT configuration, confirming that ETGs were the dominant instability. For all $k_y\ge0.2$, the fastest growing mode structure is consistent with an ETG. 

\begin{figure}[H]
	\centering
	\includegraphics[width=0.5\textwidth]{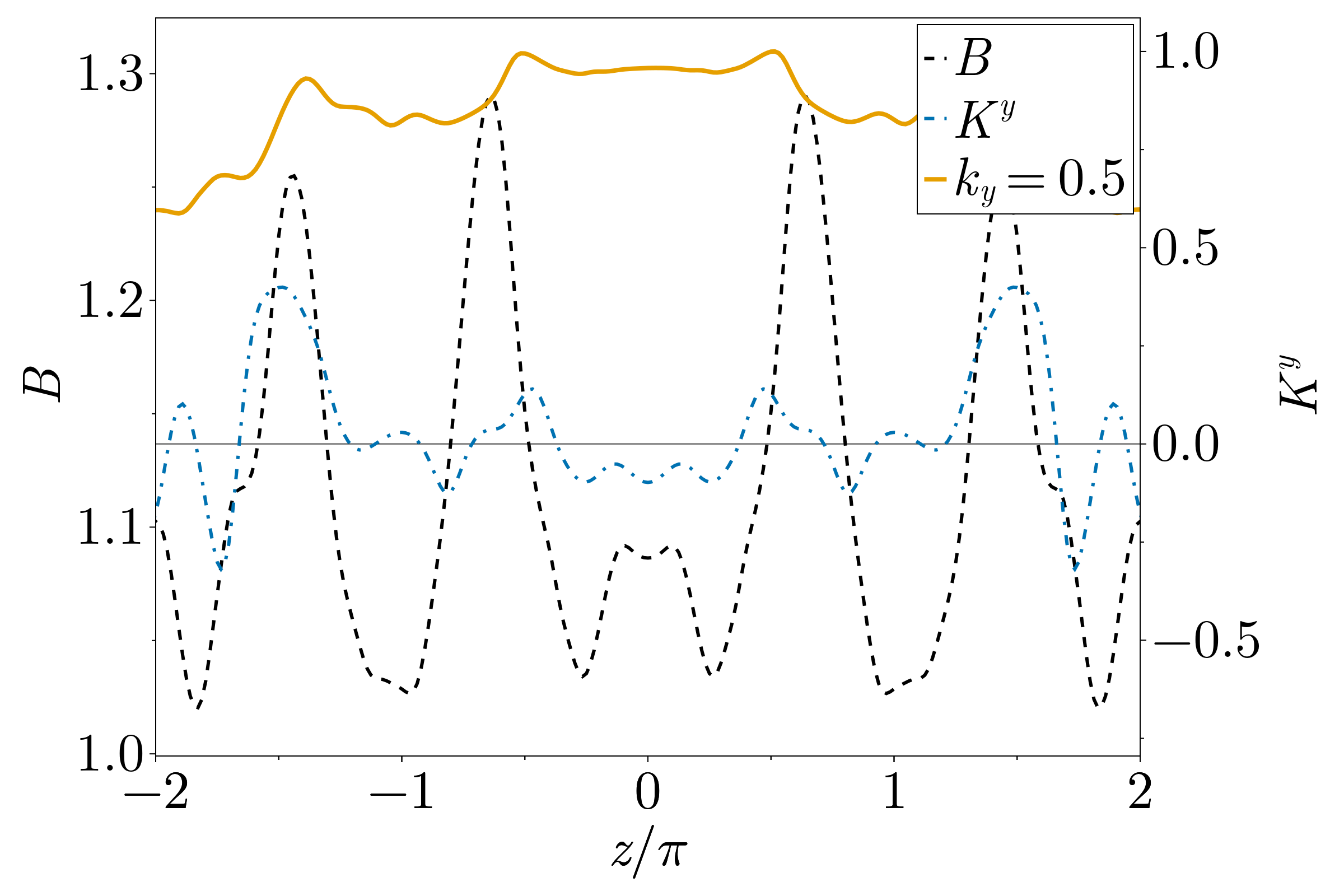}
	\caption{Amplitude of the electrostatic potential along the field line for the fastest growing mode at $k_x=0$, $k_y=0.5$ (yellow) normalized to its maximum value, in the PT geometry with $\omega_n=0$, $\omega_{T\text{e}}=4$. The magnetic field (dashed black, left axis) and curvature drive (blue dot dash, right axis) are plotted as a function of the parallel coordinate $z$. The $k_y=0.5$ mode does not show localization in magnetic wells with destabilizing curvature, indicating an ETG.}
	\label{fig:PT_Phi_Te}
\end{figure}

The same analysis as the NT configuration was applied to the PT configuration, with the $k_y=0.5$ mode plotted with $B_0$ and $\mathcal{K}^y$ in Fig.~\ref{fig:PT_Phi_Te}. For all the dominant instabilities investigated in the PT case, the mode structures are not consistent with TEMs. Based on the mode localization trends with magnetic geometry and cross phases, the reduced-TEM configurations have ETGs as the dominant instability, even at ion scales, except for the NT case at $k_y=0.1$.

	\newpage
	\bibliographystyle{apsrev4-2}
	\bibliography{Duff_StellOpt_2024}
\end{document}